\documentclass[
		twoside,openright,titlepage,numbers=noenddot,headinclude,
	 	footinclude=true,cleardoublepage=empty,
		dottedtoc, 
		BCOR=5mm,paper=a4,fontsize=11pt, 
		british  
		]{scrreprt} 
      
\PassOptionsToPackage{utf8}{inputenc} 
\usepackage{inputenc}

\PassOptionsToPackage{eulerchapternumbers,listings, pdfspacing, subfig,beramono,eulermath,parts}{classicthesis}

\newcommand{\myTitle}{A primer of group theory \newline for Loop Quantum Gravity \newline and Spin-foams}
\newcommand{\myName}{Pierre Martin-Dussaud\xspace}

\newcommand{\ie}{i.\,e.\,}

\newcommand{\eg}{e.\,g.\,}

\newcounter{dummy} 
\providecommand{\mLyX}{L\kern-.1667em\lower.25em\hbox{Y}\kern-.125emX\@}

\AtBeginDocument{} 
\DeclareRobustCommand{\C}{$\mathbb{C}$} 

\newcommand*{\SU}{\textsc{SU(2)}\xspace}%
\newcommand*{\U}{\textsc{U(1)}\xspace}%
\newcommand*{\SL}{\textsc{SL$_2$(\C)}\xspace}
\newcommand*{\slc}{\textfrak{s:l$_2$(\C)}\xspace}
\newcommand*{\su}{\textfrak{s:u(2)}\xspace}
\newcommand*{\SO}{\textsc{SO(3)}\xspace}
\newcommand*{\so}{\textfrak{s:o(3)}\xspace}
\newcommand*{\GL}{\textsc{GL$_2$(\C)}\xspace}
\newcommand*{\gl}{\textfrak{gl$_2$(\C)}\xspace}

\newcommand*\cleartoleftpage{%
  \clearpage
  \ifodd\value{page}\hbox{}\newpage\fi
}

\providecommand{\rm}{}\renewcommand{\rm}{\mathrm}
\renewcommand{\hbar}{\hslash}

\newcommand{\Wthree}[6]{\begin{pmatrix} #1 & #2 & #3 \\ #4 & #5 & #6 \end{pmatrix}}
\newcommand{\Wfour}[9]{\begin{pmatrix} #1 & #2 & #3 & #4 \\ #5 & #6 & #7 & #8 \end{pmatrix}^{(#9)}}

\newenvironment{reminder}
{\par\addvspace{\medskipamount}\noindent\scriptsize$\clubsuit$ \textsc{Reminder.}}
{\par\addvspace{\medskipamount}}

\newenvironment{proof}
{\par\addvspace{\medskipamount}\noindent\scriptsize$\blacktriangleright$ \textsc{Proof.}}
{$\Box$\par\addvspace{\medskipamount}}

\newenvironment{notabene}
{\par\addvspace{\medskipamount}\noindent\scriptsize$\bigstar$ \textsc{Nota Bene.}}
{\par\addvspace{\medskipamount}}

\newenvironment{physics}
{\par\addvspace{\medskipamount}\noindent\scriptsize$\varheart$ \textsc{Physics.}}
{\par\addvspace{\medskipamount}}

\usepackage{physics} 
\usepackage[percent]{overpic} 
\usepackage{amssymb}
\usepackage{lastpage}
\usepackage{yfonts}
\makeatletter
\newsavebox{\obj@fbbox}\newsavebox{\obj@fbboxii}
\newcommand{\obj@fbtext}[1]{\leavevmode\sbox{\obj@fbbox}{#1}%
  \copy\obj@fbbox\kern-\wd\obj@fbbox\kern0.3pt\raise0.2pt\copy\obj@fbbox}
\newcommand{\obj@fbmath}[1]{\mathpalette\obj@fbm@{#1}}
\newcommand{\obj@fbm@}[2]{\sbox{\obj@fbboxii}{$\m@th#1#2$}%
  \copy\obj@fbboxii\kern-\wd\obj@fbboxii\kern0.3pt\raise0.2pt\copy\obj@fbboxii}
\newcommand{\objdeflabel}{%
  \normalfont\bfseries\boldmath
  \let\textsc\@firstofone
  \let\obj@mathbb\mathbb \renewcommand{\mathbb}[1]{\obj@fbmath{\obj@mathbb{##1}}}%
  \let\obj@textfrak\textfrak \renewcommand{\textfrak}[1]{\obj@fbtext{\obj@textfrak{##1}}}%
}
\newcommand{\defobj}[1][]{%
  \par\addvspace{0.5\baselineskip}%
  \noindent{\objdeflabel #1}%
  \@ifnextchar,{}{\nobreakspace}}
\makeatother

\DeclareSymbolFont{extraup}{U}{zavm}{m}{n}
\DeclareMathSymbol{\varheart}{\mathalpha}{extraup}{86}

\usepackage{lipsum} 

\PassOptionsToPackage{british}{babel}  
\usepackage{babel}

\PassOptionsToPackage{%
backend=bibtex8,bibencoding=ascii,%
language=auto,%
style=numeric-comp,%
sorting=nyt, 
maxbibnames=10, 
backref=true,
natbib=true, 
isbn=false}{biblatex}
\usepackage{biblatex}

\usepackage{csquotes}
 
\MakeOuterQuote{"}

\PassOptionsToPackage{fleqn}{amsmath} 
 \usepackage{amsmath}

\PassOptionsToPackage{T1}{fontenc} 
\usepackage{fontenc}

\usepackage{textcomp} 

\usepackage{scrhack} 

\usepackage{xspace} 

\usepackage{mparhack} 

\PassOptionsToPackage{smaller}{acronym} 
\usepackage{acronym} 

\makeatletter
\AtBeginDocument{%
  \renewcommand*{\AC@hyperlink}[2]{%
    \begingroup
      \hypersetup{hidelinks}%
      \hyperlink{#1}{#2}%
    \endgroup
  }%
}

\PassOptionsToPackage{pdftex}{graphicx}
\usepackage{graphicx} 

\usepackage{tabularx} 
\usepackage{caption}
\usepackage{subfig}  

\usepackage{listings} 
\PassOptionsToPackage{pdftex,hyperfootnotes=false,pdfpagelabels}{hyperref}
\usepackage{hyperref}  
\hypersetup{
colorlinks=true, linktocpage=true, pdfstartpage=3, pdfstartview=FitV,
breaklinks=true, pdfpagemode=UseNone, pageanchor=true, pdfpagemode=UseOutlines,%
plainpages=false, bookmarksnumbered, bookmarksopen=true, bookmarksopenlevel=1,%
hypertexnames=true, pdfhighlight=/O,
urlcolor=webbrown, linkcolor=RoyalBlue, citecolor=webgreen, 
pdftitle={\myTitle},
pdfauthor={\textcopyright\ \myName},
pdfsubject={},
pdfkeywords={},
pdfcreator={pdfLaTeX},
pdfproducer={LaTeX with hyperref and classicthesis}
}

\makeatletter
\@ifpackageloaded{babel}
{
\addto\extrasamerican{

}
\addto\extrasngerman{

}
}{\relax}
\makeatother

\usepackage{classicthesis}

\usepackage{etoolbox}
\pretocmd{\chapter}{\cleardoublepage}{}{}

\titleformat{\section}      {\normalfont\large\bfseries}     {\thesection}      {1em}{}
\titleformat{\subsection}   {\normalfont\normalsize\bfseries}{\thesubsection}   {1em}{}
\titleformat{\subsubsection}{\normalfont\normalsize\bfseries}{\thesubsubsection}{1em}{}
\titleformat{\paragraph}[runin]{\normalfont\normalsize\bfseries}{\theparagraph}{0pt}{}

\usepackage{tikzit}

\tikzstyle{Green Node}=[fill={zx_green}, inner sep=0mm, minimum size=2mm, draw=black, shape=circle, tikzit fill=green, font={\footnotesize\boldmath}]
\tikzstyle{Red Node}=[fill={zx_red}, inner sep=0mm, minimum size=2mm, draw=black, shape=circle, tikzit fill=red, font={\footnotesize\boldmath}]
\tikzstyle{H}=[fill=yellow, draw=black, shape=rectangle, inner sep=0.6mm, minimum height=1.5mm, minimum width=1.5mm]
\tikzstyle{new style 0}=[fill={rgb,255: red,255; green,0; blue,4}, draw=black, shape=rectangle]
\tikzstyle{new style 1}=[fill=green, draw=black, shape=rectangle]
\tikzstyle{swirl}=[fill=white, draw=black, shape=circle]
\tikzstyle{new style 2}=[fill=black, draw=black, shape=circle]
\tikzstyle{gphase}=[rounded rectangle, rounded rectangle arc length=90, fill={zx_green}, inner sep=2pt]
\tikzstyle{rphase}=[rounded rectangle, rounded rectangle arc length=90, fill={zx_red}, inner sep=2pt]
\tikzstyle{box}=[shape=rectangle, text height=1.5ex, text depth=0.25ex, yshift=0.5mm, fill=white, draw=black, minimum height=5mm, yshift=-0.5mm, minimum width=5mm, font={\small}]
\tikzstyle{Z dot}=[Green Node, tikzit fill=green]
\tikzstyle{Z phase dot}=[minimum size=5mm, font={\footnotesize\boldmath}, shape=rectangle, rounded corners=2mm, inner sep=0.2mm, outer sep=-2mm, scale=0.8, tikzit shape=circle, draw=black, fill={zx_green}, tikzit draw=blue, tikzit fill=green]
\tikzstyle{X phase dot}=[Z phase dot, tikzit shape=circle, fill={zx_red}, font={\footnotesize\boldmath}, tikzit draw=blue, tikzit fill=red]
\tikzstyle{X dot}=[Red Node, tikzit fill=red]
\tikzstyle{hadamard}=[H, tikzit shape=rectangle, tikzit fill=yellow]
\tikzstyle{vertex}=[inner sep=0mm, minimum size=1mm, shape=circle, draw=black, fill=black]
\tikzstyle{vertex set}=[inner sep=0mm, minimum size=1mm, shape=circle, draw=black, fill=white, font={\footnotesize\boldmath}]
\tikzstyle{wide box}=[fill=white, draw=black, shape=rectangle, minimum width=10 mm]

\tikzstyle{Edge}=[-]
\tikzstyle{new edge style 0}=[draw=black, {|->}]
\tikzstyle{new edge style 1}=[-, draw={rgb,255: red,191; green,0; blue,64}]
\tikzstyle{brace edge}=[-, tikzit draw=blue, decorate, decoration={brace,amplitude=1mm,raise=-1mm}]
\tikzstyle{arrow}=[->]
\tikzstyle{blue wire}=[-, tikzit draw=blue, draw=blue]
\tikzstyle{dashed gray}=[-, dashed, tikzit fill={rgb,255: red,191; green,191; blue,191}, fill={rgb,255: red,191; green,191; blue,191}, draw={rgb,255: red,128; green,128; blue,128}]

\input{ZX.tikzdefs}

\begin{document}

\frenchspacing 

\raggedbottom 

\selectlanguage{british} 

\pagenumbering{roman} 

\pagestyle{plain} 


\begin{titlepage}

\begin{addmargin}[-1cm]{-3cm}
\begin{center}
\large

\hfill
\vfill

\begingroup
\color{Maroon}\spacedallcaps{A primer of group theory} \\ 
\color{Maroon}\spacedallcaps{for Loop Quantum Gravity} \\
\color{Maroon}\spacedallcaps{and Spin-foams}\\ 
\bigskip 
\endgroup

\spacedlowsmallcaps{\myName} 

\vfill

\label{fig:matteo}
\vspace*{\fill}
\includegraphics[width = 1 \textwidth]{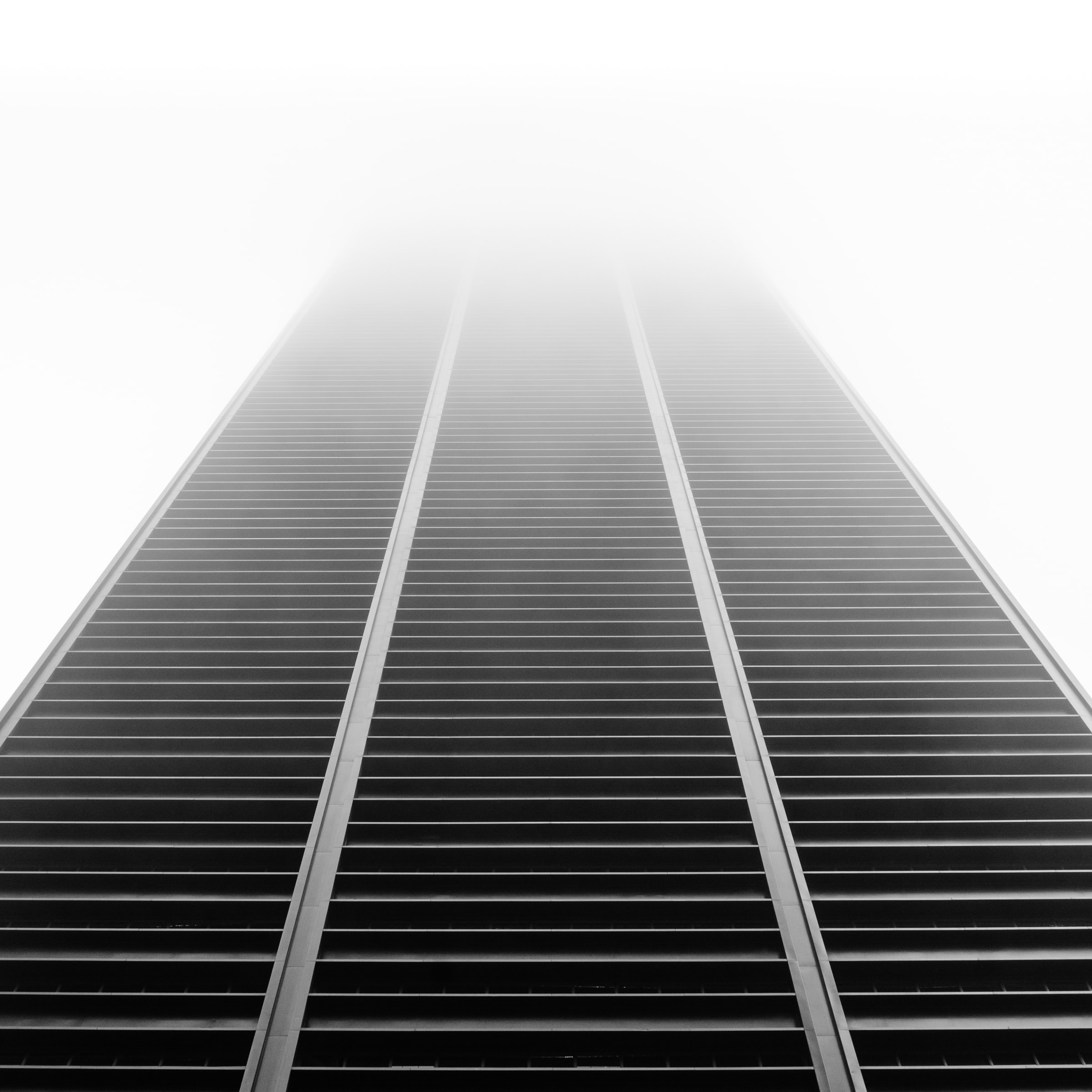}
\vspace*{\fill}

\vfill

\end{center}
\end{addmargin}

\end{titlepage}

\thispagestyle{empty}

\noindent\spacedlowsmallcaps{Nota Bene}\\
\sloppy
This is a \textit{lively} paper. I regularly release updated versions with corrections and complements. Please contribute by sending me spotted typos and personal comments: 
\begin{center}
\texttt{martindussaud[at]protonmail[dot]com}
\end{center}

\vspace{\baselineskip}

\noindent\spacedlowsmallcaps{Acknowledgements}\\
\sloppy
I would like to thank Alexander Thomas, Carlo Rovelli, Daniel Martinez, Fabio D'Ambrosio, Farshid Soltani, Giorgio Sarno, Jacob Drori, Richard East and Simone Speziale for their comments, suggestions, encouragements and proofreadings. I also want to thank Alexandra Elbakyan for her help to access the scientific literature. This revised version was reviewed by Claude Opus.

\hfill

\vfill

\noindent
Cite this article as
\begin{quote}
Martin-Dussaud, P. \textit{A primer of group theory for Loop Quantum Gravity and spin-foams}. Gen Relativ Gravit \textbf{51}, 110 (2019). DOI:\href{https://doi.org/10.1007/s10714-019-2583-5}{10.1007/s10714-019-2583-5}. arXiv:\href{https://arxiv.org/abs/1902.08439}{1902.08439}
\end{quote}

\cleardoublepage 

\begingroup

\refstepcounter{dummy}

\pdfbookmark[1]{\contentsname}{tableofcontents} 

\setcounter{tocdepth}{2} 

\setcounter{secnumdepth}{3} 

\manualmark

\begingroup 
\let\clearpage\relax
\let\cleardoublepage\relax
\let\cleardoublepage\relax
\tableofcontents 
\endgroup

\automark[section]{chapter}
\renewcommand{\chaptermark}[1]{\markboth{\spacedlowsmallcaps{#1}}{\spacedlowsmallcaps{#1}}}
\renewcommand{\sectionmark}[1]{\markright{\thesection\enspace\spacedlowsmallcaps{#1}}}

\cleardoublepage

\endgroup

\begingroup 
\let\clearpage\relax
\let\cleardoublepage\relax
\let\cleardoublepage\relax

\refstepcounter{dummy}

\addcontentsline{toc}{chapter}{Acronyms} 


\chapter*{Acronyms}

\begin{acronym}[UML]
\acro{CSCO}{Complete Set of Commuting Observables}
\acro{EPRL}{Engle-Pereira-Rovelli-Livine}
\acro{GGV}{Gel'fand-Graev-Vilenkin}
\acro{GMS}{Gel'fand-Minlos-Shapiro}
\acro{LQG}{Loop Quantum Gravity}
\acro{RHS}{Right Hand Side}
\end{acronym}

\begingroup 
\let\clearpage\relax
\let\cleardoublepage\relax
\let\cleardoublepage\relax

\refstepcounter{dummy}
\addcontentsline{toc}{chapter}{Notations} 

\chapter*{Notations}

\begin{tabular}{ll}
$\overset{\text{def}}=$ & Equal by definition\\
$\propto$ & Proportional to \\
$\sim$ & Of the order of, scaling as \\
$\cong$ & Isomorphic to \\
$\approx$ & Approximately equal to \\
$\det M$ & Determinant of $M$\\
$\Tr M$ & Trace of $M$ \\
$Sp(M)$ & Spectrum of $M$ \\
$M^\dagger$ & Hermitian conjugate of $M$ \\
$M^T$ & Transpose of $M$ \\
$\epsilon_{ijk}$ & Levi-Civita symbol \\
$\sigma_\mu$ & Pauli matrices $(\sigma_0 = \mathbb{1},\sigma_1,\sigma_2,\sigma_3)$ \\
$\bar \sigma^\mu$ & Pauli matrices $(\sigma_0 = \mathbb{1},\sigma_1,\sigma_2,\sigma_3)$, see chapter \ref{ch:warmup} \\
$\eta_{\mu \nu}$ & Minkowski metric tensor. Convention: $\eta_{\mu \nu} = \text{diag}(-,+,+,+)$ \\
$\binom{n}{p}$ & Binomial coefficient $\frac{n!}{p!(n-p)!}$ \\
$]a,b[$ & Open interval from $a$ to $b$, sometimes denoted $(a,b)$ \\
$\pmb z$ & $(z_0,z_1) \in \mathbb{C}^2$ \\
$\bar z$, $z^*$ & Complex conjugate of $z$ \\
$\dd \Omega^2$ & The usual metric on the $2$-sphere, $\dd \Omega^2 = \dd \theta^2 + \sin^2\theta \, \dd \phi^2$\\
$G$ & Newton constant, $G \approx 6.7 \times 10^{-11} ~\mathrm{m^3 \, kg^{-1} \, s^{-2}}$ \\
$\hbar$ & Reduced Planck constant, $\hbar \approx 1.1 \times 10^{-34} ~\mathrm{J \, s}$ \\
$c$ & Speed of light in vacuum, $c \approx 3.0 \times 10^{8} ~\mathrm{m \, s^{-1}}$\\
$\gamma$ & Immirzi parameter \\
$l_P$ & Planck length, $l_P \approx 1.6 \times 10^{-35} ~\mathrm{m}$ \\
\end{tabular}

\endgroup


\endgroup 

\cleardoublepage 

\refstepcounter{dummy}
\addcontentsline{toc}{chapter}{Introduction} 

\begingroup 
\let\clearpage\relax
\let\cleardoublepage\relax
\let\cleardoublepage\relax

\chapter*{Introduction}

\begin{quote}
\textit{The aim of a theory really is, to a great extent, that of systematically organizing past experience in such a way that the next generation, our students and their students and so on, will be able to absorb the essential aspects in as painless a way as possible, and this is the only way in which you can go on cumulatively building up any kind of scientific activity without eventually coming to a dead end.} \\
\hspace*{\fill} --- M. F. Atiyah in \cite{Atiyah1974}.
\end{quote}

\sloppy
The central role of group theory in physics has been largely revealed in the modern theories of the 20th century. Quantum gravity is no exception. The two main groups of interest for quantum gravity are \SL and its subgroup \SU. This may seem natural since \SL is, in some sense, the "quantum version" of the restricted Lorentz group \textsc{SO$^+$(3,1)}, which is an important symmetry group of Minkowski spacetime, and similarly, \SU is the "quantum version" of the group of space rotations; but the reason why these groups come out in quantum gravity is actually more subtle.

Even though many monographs devoted to this theory exist, the various tools needed (\eg representation theory, harmonic analysis, recoupling theory...) are often scattered across books, each with its own conventions and notations. This was the initial motivation for the compilation of the present document. Now it serves three main purposes:
\begin{enumerate}
\item A concise introduction for students to the essential mathematical tools of \ac{LQG}. It bridges a gap between the level of students at the end of a master programme, and the minimum level required to start doing research in \ac{LQG}. In case the pace is too fast, paragraphs have been inserted for a quick refresher. They are written in small font size, and introduced by "{\scriptsize $\clubsuit$ \textsc{Reminder}}". Instead of introducing new formulae out of nowhere, we emphasise the motivation for introducing them. Proofs are given when helpful for understanding, but are sometimes only sketched. They are written in small font size, introduced by "{\scriptsize $\blacktriangleright$  \textsc{Proof}}", so that they can also be skipped easily.
\item A convenient compendium for researchers. Instead of having each formula in a different heavy, old book, the most useful ones are gathered in a short toolbox. 
\item A translation hub between the conventions of the main references. For many notions, each author tends to use their own notations, which makes it difficult to switch easily from one reference to another. We have made some choices ourselves, but we show explicitly how they relate to various major references: such discussions are given in small font size, introduced by "$\star$ {\scriptsize \textsc{Nota Bene}}". We see this attempt as a step towards a more widely accepted use of common notations. In particular, we give the conventions of the Wolfram Language, which are helpful for implementing numerical computations.
\end{enumerate}

Although most of the technical content is not new, the overall compilation is. We also offer some new derivations of results, simpler than what can be found elsewhere, or sometimes not written anywhere else. A commented bibliography is provided at the end to give a panorama of the existing literature and to help readers looking for more details.

These notes are aimed both at physicists, caring about their tools being mathematically well grounded, and at mathematicians, curious about how some of their familiar abstract structures can reveal the beauty of quantum gravity. For the latter, we have included specific short paragraphs in small font size, introduced by "{\scriptsize $\varheart$ \textsc{Physics}}", that provide general ideas and references on how the mathematics has been exploited by theoretical physicists, especially in quantum gravity.

The plan is the following:
\begin{itemize}
	\item[] Ch. \ref{ch:equal} discusses the foundational notion of equality.
	\item[] Ch. \ref{ch:warmup} wraps up all the basics of \SU and \SL, the two Lie groups of main interest for quantum gravity.
	\item[] Ch. \ref{ch:geometric} lights up some geometrical aspects of spheres.
	\item[] Ch. \ref{ch:representation-SU2} catalogues various possible realisations of \SU-irreps used in the literature.
	\item[] Ch. \ref{ch:recoupling-SU2} condenses the main results of the recoupling theory of \SU.
	\item[] Ch. \ref{ch:harmonic} shows how functions over \SU can be decomposed in harmonics.
	\item[] Ch. \ref{ch:representation-SL2C} renders all the flourish of the representations of \SL.
	\item[] Ch. \ref{ch:recoupling-SL2C} attempts to generalise recoupling theory to \SL.
	\item[] Ch. \ref{ch:loops} summarises \ac{LQG} and spin-foams in a nutshell.
	\item[] Ch. \ref{ch:commented} offers a commented selection of reference textbooks to go further. 
\end{itemize}

\vfill

\endgroup

\pagenumbering{arabic} 

\refstepcounter{dummy}

\pagestyle{scrheadings} 

\chapter{Equality}
\label{ch:equal}

Since its axiomatic formalisation, notably carried out by Zermelo and Fraenkel, modern mathematics is based on the notion of \textit{set}. According to the usual bourbakian reconstruction, sets are endowed with \textit{structures}, which turn them into \textit{spaces}. However, this way does not allow us to recover exactly the intuitive notion of \textit{mathematical object}. Indeed, different spaces can describe the same mathematical object. In fact, different descriptions of the same object are related by \textit{isomorphisms}, which are bijections that preserve the structures in both directions (the inverse map has to preserve them too).

A simple example is provided by \textit{the circle}. For physicists, there is no doubt about what a circle is. The mathematicians, who are both praised and mocked for giving precise definitions, do surely have a good definition of a circle. However, a fair sample of mathematicians will not give you one, but two definitions of a circle. 
The real mathematicians define it as a submanifold of the plane $\mathbb{R}^2$:
\begin{equation}
S^1 \overset{\text{def}}=  \left \{ (x,y) \in \mathbb{R}^2 \mid x^2 + y^2 = 1 \right \}.
\end{equation}
The complex mathematicians define it as a subgroup of $\mathbb{C}^*$:
\begin{equation}
\U \overset{\text{def}}= \left \{\lambda \in \mathbb{C} \mid |\lambda| =1 \right \},
\end{equation}
$S^1$ and \U are different sets. They are even different spaces, as the former is a differentiable manifold, and the latter a group. Nevertheless, both deserve the name of "circle" and they can be regarded as the "same thing" through the following isomorphism
\begin{equation}
f : \left\{ \begin{array}{lll}
S^1 & \rightarrow & \U \\
(x,y) & \mapsto & x + i y
\end{array} \right.
\end{equation}
It is an isomorphism, as it enables to translate the group and the manifold structures from one set to another. It is much more than being only a bijection. A bijection is quite a weak requirement, as it only preserves the cardinal of sets, so that the circle is also in bijection for instance with the disk. Mathematicians have fancy names to distinguish all the kinds of isomorphism:
\begin{itemize}
\item Bijection, between sets;
\item Homeomorphism, between topological spaces;
\item Isometry, between metric spaces;
\item Isomorphism (or group isomorphism), between groups;
\item Diffeomorphism, between differentiable manifolds.
\end{itemize}
In all cases, we denote 
\begin{equation}
A \cong B
\end{equation}
to signify the existence of an isomorphism between $A$ and $B$, which kind of isomorphism should be clear from the context.

Since $S^1$ and \U share all the same structures, it is tempting to say that we should regard them as really the same object, and write $S^1 = \U$. This idea is reminiscent of Leibniz's definition of equality:
\begin{quote}
\textit{$x=y$, if and only if, $x$ and $y$ have all the same properties.}
\end{quote}
Unfortunately, his definition is too fuzzy to be useful, as $x$ and $y$ can \textit{never} share \textit{all} the same properties, just because, for instance, "$x$" and "$y$" are not written alike. The domain of properties has to be restricted to get a consistent definition of equality. As a result, even in mathematics, the meaning of the symbol "$=$" is not as sharp as people usually believe, it is always some kind of $\approx$. In our case, a mathematician refuses to write $S^1 = \U$ as long as they refuse to write $\mathbb{R}^2 = \mathbb{C}$. But the difference between "$=$" and "$\cong$" is inessential. When many structures are shared, it is fine to write simply $A=B$.

Our brain is a champion for performing this kind of identification, but a computer trying to do the same would often run into "typing" errors. Category theory provides an appropriate language for speaking about these identifications. By contrast, the theory of sets suffers from a certain formalist rigidity due to its requirement to define all objects based on the notion of set.

Some other people would probably say that $S^1$ and \U are actually two incarnations of a third object, which is, \textit{really}, the circle. In this view, the circle belongs to the platonic world of Ideas, and only contingent witnesses of it are seen \textit{in real life}. Such a discussion would have been a delight for medieval scholasticism, but is useless for proving theorems about circles.

Physicists are usually more flexible with notations and definitions, as long as "it is clear what it means". As Feynman says
\begin{quote}
\textit{We cannot define anything precisely. If we attempt to, we get into that paralysis of thought that comes to philosophers, who sit opposite each other, one saying to the other, "You don't know what you are talking about!". The second one says, "What do you mean by know? What do you mean by talking? What do you mean by you?" (\cite{feynman1964}, lecture 8)}
\end{quote}
My opinion is more qualified. The flexibility with definitions and notations can be a strength when it makes us agile to juggle with concepts, but it is a weakness when it blurs the beauty of details. It took me a while to understand this simple lesson, maybe because it is not often said explicitly. So I thought someone would appreciate to read it here someday.
\chapter{Warmup}
\label{ch:warmup}

This chapter is a melting pot of the basic algebraic mathematical tools that will be later used extensively. It also fixes many of the notations. If you already feel warmed up, you would do well to skip this chapter. If you have never seen these notions in your life, you would do better to first learn them from an introductory book. Good ones are for instance Knapp \cite{knapp1986}, Hall \cite{hall2015}, and Bernard-Laszlo-Renard \cite{bernard2012}.

\section{Basics of \texorpdfstring{\SL}{SL(2,C)}}

\defobj[$\mathcal{M}_2(\mathbb{C})$]is the algebra of $2 \times 2$ complex matrices. It is an \textit{algebra} because it is a vector space (with addition of matrices) endowed with a bilinear product (the usual matrix product).
\defobj[\GL]is a \textit{linear group} defined by 
\begin{equation}
\GL \overset{\text{def}}= \left \{ M \in \mathcal{M}_2(\mathbb{C}) \mid \det M \neq 0 \right \}.
\end{equation}
It is a $4$-dimensional complex \textit{Lie group}, \ie both a $4$-dimensional complex differentiable manifold and a group whose multiplication and inversion are analytic maps.
\defobj[\gl]denotes the \textit{Lie algebra} of \GL, \ie the tangent space over the identity $\mathbb{1} \in \GL$. It is actually isomorphic to $\mathcal{M}_2(\mathbb{C})$, when it is endowed with the Lie product $\left[M,N\right] = MN-NM$.
\defobj[\SL]is a \textit{special linear group}, defined by 
\begin{equation}
\SL \overset{\text{def}}= \left \{ M \in \GL \mid \det M =1 \right \}.
\end{equation}
It is a $3$-dimensional complex Lie subgroup of \GL. Topologically, \SL is not compact but it is \textit{simply connected}, \ie it is path-connected and any loop can be contracted to a point.
\defobj[\slc]is the Lie algebra of \SL. One can show that
\begin{equation}
\slc =\left \{ M \in \mathcal{M}_2(\mathbb{C}) \mid \Tr M = 0 \right \}.
\end{equation}
It is a $3$-dimensional complex Lie subalgebra of \gl.
\defobj[$\sigma_1, \sigma_2, \sigma_3$]are the \textit{Pauli matrices}, defined by
\begin{align}
\sigma_1 \overset{\text{def}}=\begin{pmatrix}
0& 1 \\ 1 & 0
\end{pmatrix}, &&
\sigma_2 \overset{\text{def}}=\begin{pmatrix}
0&-i\\i&0
\end{pmatrix}, &&
\sigma_3 \overset{\text{def}}= \begin{pmatrix}
1 & 0 \\ 0 & -1
\end{pmatrix}.
\end{align}
They form a basis of \slc. \marginpar{Here and everywhere else, Einstein notation is understood over repeated indices.}Interestingly, they satisfy
\begin{equation}
[\sigma_i,\sigma_j] = 2i \, \epsilon_{ijk} \, \sigma_k.
\end{equation}
For convenience, the identity $\mathbb{1}$ is often denoted $\sigma_0$, and the index $\mu \in \{0,1,2,3\}$ is used to denote the enlarged set of Pauli matrices $\sigma_\mu$. Then, the $\sigma_\mu$ provide a basis of the \textit{complex} vector space $\mathcal{M}_2(\mathbb{C})$: any $a \in \mathcal{M}_2(\mathbb{C})$ can be written uniquely
\begin{equation} \label{eq:decomposition Pauli}
a = \sum_{\mu=0}^3 a_\mu \sigma_\mu \quad \text{with} \quad a_0,a_1,a_2,a_3 \in \mathbb{C}.
\end{equation}
Note that in this basis, the determinant reads 
\begin{equation}
\det a = a_0^2 - a_1^2 - a_2^2 - a_3^2.
\end{equation}
\defobj[$H_2(\mathbb{C})$] is the \textit{real} vector space of $2 \times 2$ \textit{hermitian} matrices, defined by
\begin{equation}
H_2(\mathbb{C}) \overset{\text{def}}= \{ M \in \mathcal{M}_2(\mathbb{C}) \mid M^\dagger = M \}.
\end{equation}
A basis is also given by the Pauli matrices: any $h \in H_2(\mathbb{C})$ can be written uniquely as
\begin{equation} \label{eq:hermitian decomposition Pauli}
h = \sum_{\mu=0}^3 h_\mu \sigma_\mu \quad \text{with} \quad h_0,h_1,h_2,h_3 \in \mathbb{R}.
\end{equation}
\defobj[$H_2^{++}(\mathbb{C})$], the set of $2 \times 2$ hermitian \textit{positive-definite} matrices, is
\begin{equation}
H_2^{++}(\mathbb{C}) \overset{\text{def}}= \left \{ M \in H_2(\mathbb{C}) \mid \forall \lambda \in Sp(M), \quad \lambda > 0 \right \},
\end{equation}
with $Sp(M)$, the \textit{spectrum} of $M$, \ie the set of its eigenvalues.

	\section{Spacetime symmetries}

\defobj[$\mathbb{M}$]is the spacetime of special relativity, called \textit{Minkowski spacetime}. Mathematically, it is the vector space $\mathbb{R}^4$, endowed with a \textit{lorentzian inner product}, whose signature is either $(-,+,+,+)$ (general relativists convention) or $(+,-,-,-)$ (particle physicists convention). In this primer, we stick to the convention of general relativity and we denote the metric tensor $\eta_{\mu \nu} = \text{diag} (-,+,+,+)$. It is used to lower indices while its inverse $\eta^{\mu \nu} = \text{diag} (-,+,+,+)$ is used to raise them. For instance, $\sigma^\mu = \eta^{\mu \nu} \sigma_\nu = (-\sigma_0, \sigma_1,\sigma_2, \sigma_3)$.
\defobj[$\mathbb{P}$]is the group of all isometries (distance-preserving transformations) of $\mathbb{M}$, called the \textit{Poincaré group} (or sometimes the \textit{inhomogeneous Lorentz group}).
\defobj[\textsc{O(3,1)}]is the linear subgroup of isometries that leave the origin fixed, called the \textit{Lorentz group} (or sometimes the \textit{homogeneous Lorentz group}), and sometimes also denoted \textsc{O(1,3)}. The Poincaré group $\mathbb{P}$ can be decomposed as a semi-direct product $\mathbb{P} = \textsc{O(3,1)} \ltimes \mathbb{R}^4$. \textsc{O(3,1)} is composed of four connected components related to each other by the operators of parity (space-reversal) and time-reversal.
\defobj[\textsc{SO$^+$(3,1)}] is the connected component to the identity in \textsc{O(3,1)}. It forms a subgroup made of transformations that preserves the orientation and the direction of time. It is called the \textit{proper orthochronous Lorentz group}, or the \textit{restricted Lorentz group}.	
	
As a real vector space, Minkowski spacetime $\mathbb{M}$ is isomorphic to $H_2(\mathbb{C})$, with the map
\begin{equation}\label{eq:hermitian Minkowski}
 X^\mu = (t,x,y,z) \mapsto h  = X^\mu \sigma_\mu = \begin{pmatrix}
t + z & x -iy \\
x+iy & t-z
\end{pmatrix}.
\end{equation}
For convenience, denote $\bar \sigma^\mu = (\sigma_0, \sigma_1,\sigma_2,\sigma_3)$. Then, the inverse map is given by
\begin{equation}
h \mapsto  X^\mu = \frac 12 \Tr h \bar \sigma^\mu,
\end{equation}
and the pseudo-scalar product
\begin{equation}
\begin{split}
X^\mu X'_\mu &= - tt' + xx' + yy' + zz' \\
&= \frac 14 \Tr \left(  hh' - h\sigma_1h'\sigma_1- h\sigma_2h'\sigma_2 - h \sigma_3 h' \sigma_3 \right) \\
&= - \frac{\det h}{2} \Tr (h^{-1} h')
\end{split}
\end{equation}
As a result, the pseudo-norm of $\mathbb{M}$ is mapped to the determinant over $H_2(\mathbb{C})$:
\begin{equation}
X^\mu X_\mu  = - \det h.
\end{equation}
From the latter property, we see that the action of $a \in \SL$ upon $h \in H_2(\mathbb{C})$, given by
\begin{equation}\label{eq:SL2C action hermitian}
h \mapsto a h a^\dagger,
\end{equation}
defines a linear isometry on $\mathbb{M}$. Thus, we deduce the homomorphism $\Lambda: \SL \to \textsc{SO$^+$(3,1)}$,
\begin{equation}
[\Lambda(a)]^\mu_{\ \ \nu} = \frac{1}{2} \Tr \left( a \sigma_\nu a^\dagger \bar \sigma^\mu \right).
\end{equation}
The kernel of $\Lambda$ is $Z_2 \overset{\text{def}}= \{\mathbb{1} , -\mathbb{1} \}$, the $2$-element group. Thus we have the following \textit{short exact sequence} of Lie groups:\marginpar{A short exact sequence is such that the image of a homomorphism is the kernel of the next one.}
\begin{equation}
1 \longrightarrow Z_2 \longrightarrow \SL \overset{\Lambda}\longrightarrow SO^+(3,1) \longrightarrow 1,
\end{equation}
where the map $Z_2 \to \SL$ is the inclusion of $Z_2$ into \SL. In particular, we have the following isomorphism of groups
\begin{equation}\label{eq:SO SL}
\SL / Z_2 \cong \textsc{SO$^+$(3,1)}.
\end{equation}
\SL is said to be the double cover, or the universal cover, of \textsc{SO$^+$(3,1)}. For that reason it is sometimes called the \textit{Lorentz spin group}. This gives a first glimpse on the role of \SL in fundamental physics.

\section{Subgroups of \texorpdfstring{\SL}{SL(2,C)}}
\label{sec:subgroups}

\begin{figure}[h!]
\centering
\begin{overpic}[scale=0.5]{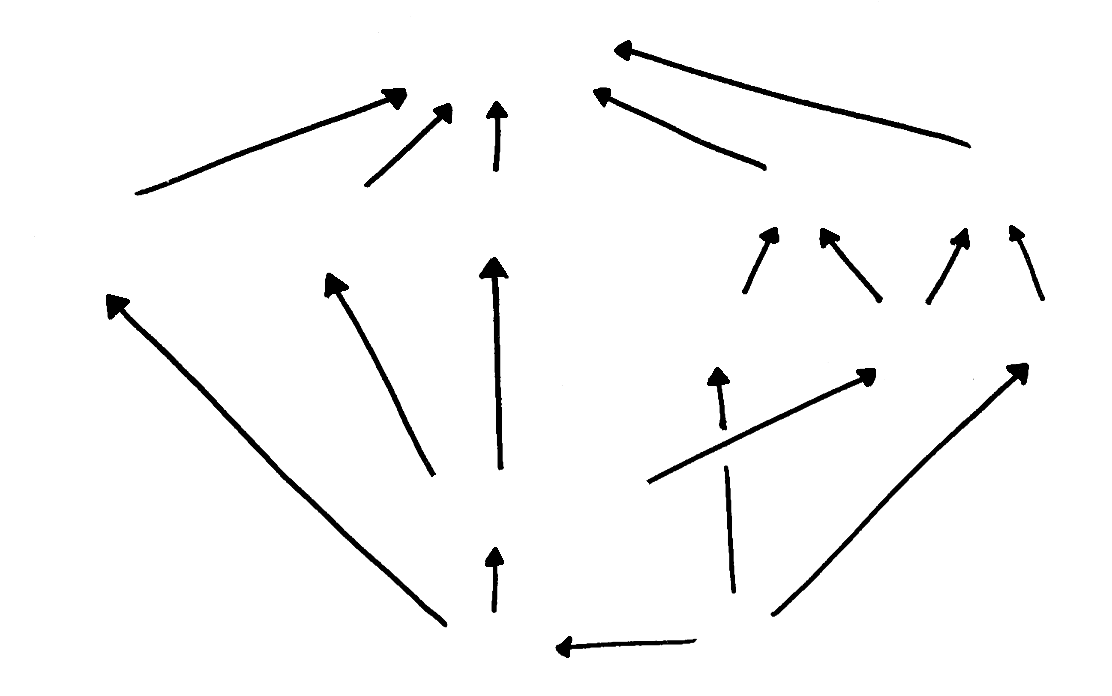}
\put (65,5) {$\mathbb{1}$}
\put (42,3) {$Z_2$}
\put (0,40) {\textsc{SL$_2$($\mathbb{R}$)}}
\put (21,40) {\textsc{SU(1,1)}}
\put (33,15) {S($\U \oplus \U$)}
\put (40,42) {\SU}
\put (38,57) {\SL}
\put (70,43) {$K^+$}
\put (87,43) {$K^-$}
\put (62,31) {$Z^+$}
\put (93,31) {$Z^-$}
\put (80,30) {$D$}
\end{overpic}
\caption[Hasse diagram of the subgroups of \SL. The arrows represent inclusions.]{Hasse diagram of the subgroups of \SL. The arrows represent inclusions.}
\label{fig:graph}
\end{figure}

There are many subgroups of \SL. We describe below the main ones. The figure \ref{fig:graph} shows the relations of inclusion between them.

\defobj[\SU], the unitary special group, is defined by:
\begin{equation}\label{eq:def SU(2)}
\SU \overset{\text{def}}= \left \{ u \in \SL \mid u^\dagger u = \mathbb{1} \right \}.
\end{equation}
Any $u \in \SU$ can be uniquely written as
\begin{multline}
u = u_0 e + i \sum_{k=1}^3 u_k \sigma_k \quad \text{with} \quad u_0,u_1,u_2,u_3 \in \mathbb{R} \\
\text{and} \quad \sum_{k=0}^3 u_k^2 = 1.
\end{multline}
Through the isomorphism \eqref{eq:hermitian Minkowski} and the action \eqref{eq:SL2C action hermitian}, the definition \eqref{eq:def SU(2)} enables us to see \SU as the stabiliser (also called \textit{little group} or \textit{isotropy group}) of the unit time vector $(1,0,0,0)$. Physically, it means that \SU only acts over the space, and not in the time direction. Choosing another time direction, related to $(1,0,0,0)$ by a boost $\Lambda$, would have defined another stabilizer, isomorphic to \SU, which makes physicists sometimes talk of \textit{a} \SU, as if there were several.
\defobj[\textsc{SU(1,1)}]is defined by:
\begin{equation}
\textsc{SU(1,1)} \overset{\text{def}}= \left \{ v \in \SL \mid v^\dagger \sigma_3 v= \sigma_3 \right \}.
\end{equation}
Any $v \in \textsc{SU(1,1)}$ can be uniquely written as:
\begin{multline}
v = v_0 e + v_1 \sigma_1 + v_2 \sigma_2 + iv_3 \sigma_3 \\
\text{with} \quad v_0,v_1,v_2,v_3 \in \mathbb{R} \\
\text{and} \quad v_0^2 - v_1^2 - v_2^2 + v_3^2 = 1.
\end{multline}
Similarly to the \SU case, \textsc{SU(1,1)} can be understood by its action in Minkowski spacetime as the stabiliser of $(0,0,0,1)$.
\defobj[\textsc{SL$_2$($\mathbb{R}$)}], the real special linear group, is defined by
\begin{equation}
\textsc{SL$_2$($\mathbb{R}$)} \overset{\text{def}}= \left \{ a \in \mathcal{M}_2(\mathbb{R}) \mid \det a = 1 \right \},
\end{equation}
and interestingly it is also
\begin{equation}
\textsc{SL$_2$($\mathbb{R}$)} = \left \{a \in \SL \mid a^\dagger \sigma_2 a = \sigma_2 \right \}.
\end{equation}
Any $a \in \textsc{SL$_2$($\mathbb{R}$)}$ can be uniquely written as:
\begin{multline}
a = a_0 e + a_1 \sigma_1 + ia_2 \sigma_2 + a_3 \sigma_3 \\
\text{with} \quad a_0,a_1,a_2,a_3 \in \mathbb{R} \\ 
\text{and} \quad a_0^2 - a_1^2 + a_2^2 - a_3^2 = 1.
\end{multline}
Again \textsc{SL$_2$($\mathbb{R}$)} can be understood by its action in Minkowski spacetime as the stabiliser of $(0,0,1,0)$.
\begin{notabene}
Following the previous sequence, it would be fair to expect the next subgroup to be the one defined by\begin{equation}
\left \{b \in \SL \mid b^\dagger \sigma_1 b = \sigma_1 \right \}.
\end{equation}
To the knowledge of the author, this group bears no standard name as an explicit subgroup. It is, however, nothing genuinely new. The Pauli matrices $\sigma_1$, $\sigma_2$ and $\sigma_3$ are mapped onto one another by rotations, that is by conjugation $u\, \sigma_i\, u^\dagger$ with $u \in \SU$. Consequently the three groups defined by $b^\dagger \sigma_i b = \sigma_i$ are all \textit{conjugate in} \SL, and therefore isomorphic. Recall that two subgroups $H, H' \subset G$ are said to be \textit{conjugate in} $G$ when there exists $g \in G$ such that $H' = g H g^{-1}$; the map $h \mapsto g h g^{-1}$ then provides an isomorphism between them. In particular, the group above is isomorphic to both \textsc{SU(1,1)} and \textsc{SL$_2$($\mathbb{R}$)}, thereby recovering the well-known isomorphism $\textsc{SU(1,1)} \cong \textsc{SL$_2$($\mathbb{R}$)}$.
\end{notabene}
\defobj[$K_+$ and $K_-$], the upper and lower triangular groups, are defined by:
\begin{equation}
\begin{split}
&K_+ \overset{\text{def}}= \left \{ \begin{pmatrix}
\lambda^{-1} & \mu \\
0 & \lambda 
\end{pmatrix} \mid \lambda \in \mathbb{C}^* \quad \text{and} \quad \mu \in \mathbb{C} \right \} 
\\
&K_- \overset{\text{def}}= \left \{ \begin{pmatrix}
\lambda^{-1} & 0 \\
\mu & \lambda 
\end{pmatrix} \mid \lambda \in \mathbb{C}^* \quad \text{and} \quad \mu \in \mathbb{C} \right \} .
\end{split}
\end{equation}
They are also called the \textit{Borel subgroups} or the \textit{parabolic subgroups}.
\defobj[$Z_+$ and $Z_-$], are defined by
\begin{equation}
\begin{split}
&Z_+ \overset{\text{def}}= \left \{ \begin{pmatrix}
1 & z \\
0 & 1 
\end{pmatrix} \mid z \in \mathbb{C} \right \} 
\\
&Z_- \overset{\text{def}}= \left \{ \begin{pmatrix}
1 & 0 \\
z & 1 
\end{pmatrix} \mid z \in \mathbb{C} \right \}.
\end{split}
\end{equation}
\defobj[$D$], the diagonal group is defined by
\begin{equation}
D \overset{\text{def}}= \left \{ \begin{pmatrix}
\delta & 0 \\ 0 & \delta^{-1} 
\end{pmatrix}
\mid \delta \in \mathbb{C}^* \right \}.
\end{equation}
\defobj[\textsc{S($\U \oplus \U$)}], defined by
\begin{equation}
\textsc{S($\U \oplus \U$)} \overset{\text{def}}= \left \{ \begin{pmatrix}
e^{i\theta} & 0 \\ 0 & e^{-i \theta} 
\end{pmatrix}
\mid \theta \in \mathbb{R} \right \}.
\end{equation}
\marginpar{The notation $\textsc{S}(.)$ means \textsf{special}, \ie of determinant $1$.}
It is the maximal torus (\ie the biggest compact, connected, abelian Lie subgroup) of \SU. We have obviously
\begin{equation}
\textsc{S($\U \oplus \U$)} \cong \U.
\end{equation}
\defobj[$Z_2$], the 2-element group,
\begin{equation}
Z_2 \overset{\text{def}}=\left \{ \mathbb{1},-\mathbb{1} \right \}.
\end{equation}
It is also the center of \SL, \ie the subset of \SL which commutes with all of \SL.
\marginpar{A subgroup $H \subset G$ is said to be \textsf{normal} if $g h g^{-1} \in H$, for all $g \in G$ and $h \in H$.}Since it is a normal subgroup (as any center of any group), the quotient $\SL/Z_2$ is also a group, which can be shown to be isomorphic to the restricted Lorentz group \textsc{SO$^+$(3,1)}, as was already said in equation \eqref{eq:SO SL}.

\section{Decomposition of \texorpdfstring{\SL}{SL(2,C)}}
\label{sec:decomposing SL}
	
The structural properties of a matrix group can be grasped through the study of its decompositions. We are going to present four of them for \SL. 

\begin{reminder}
The spectral theorem states that any hermitian matrix is diagonalisable. More precisely, $H$ is hermitian if and only if there exists a unitary matrix $U$ and a diagonal matrix $D$ with real coefficients such that\begin{align*}
H = U^\dagger D U.
\end{align*}
The coefficients of $D$ are the eigenvalues of $H$. The same result holds if $H$ is not hermitian but only \textit{normal}, \ie $[H,H^\dagger]=0$, to the only difference that the eigenvalues are not necessarily real in this case.
\end{reminder}

\vspace{\baselineskip}
\noindent
$\blacksquare$ \textbf{Polar decomposition.}\marginpar{\scriptsize
$\varheart$ \textsc{Physics.} The polar decomposition has been used notably by Thiemann and Winkler in their analysis of the coherent states of quantum gravity (see \cite{Thiemann:2000bw, Thiemann:2000bx, Thiemann:2000by, Thiemann:2000aa}).}
For all $M \in \GL$, there exists a unique unitary matrix $U \in \textsc{U(2)}$ and a unique positive-definite hermitian matrix $H \in H_2^{++}(\mathbb{C})$ such that:
\begin{equation}
M = HU.
\end{equation}
Remarks:
\begin{enumerate}
\item The order does not matter, and the theorem would also be true with $M = UH$.
\item If $M \in \SL$, then $U \in \SU$ and $\det H = 1$.
\item It is called "polar" because it is a generalisation of the polar decomposition of complex numbers $z = r e^{i\theta}$. It can be generalised further to any \textsc{GL$_n$(\C)}.
\end{enumerate}
\begin{proof}
The polar decomposition actually works for the set $H_n^{++}(\mathbb{C})$ of $n \times n$ positive-definite hermitian matrices. Recall that any $A \in H_n^{++}(\mathbb{C})$ has a unique square root $\sqrt{A} \in H_n^{++}(\mathbb{C})$, \ie a unique positive-definite hermitian matrix whose square is $A$.

Existence. Let $M \in GL_n(\mathbb{C})$ be the matrix to be decomposed, then $MM^\dagger$ is hermitian and positive. Let $H = \sqrt{MM^\dagger} \in H_n^{++}(\mathbb{C})$, so that $MM^\dagger = H^2$. We check finally that $H^{-1}M$ is unitary.

Uniqueness. If $M=HU$ with $U$ unitary and $H$ definite-positive hermitian, then $M^\dagger=U^{-1}H$, so $MM^\dagger=H^2$. But $MM^\dagger$ is positive hermitian and has therefore a unique positive hermitian square root, so that $H$ is this square root and $U$ is equal to $H^{-1}M$.
\end{proof}

\vspace{\baselineskip}
\noindent
$\blacksquare$ \textbf{Cartan decomposition.}\marginpar{\scriptsize
$\varheart$ \textsc{Physics.} This decomposition has been used notably for the asymptotics of spin-foams amplitude \cite{speziale2017}, and also for twisted geometries in \cite{langvik2016}.}
For all $g \in \SL$, there exists $u,v \in \SU$ and $r \in \mathbb{R}_+$ such that:
\begin{equation}
g = u \, e^{r \sigma_3 /2} \, v^{-1}.
\end{equation}
Remarks:
\begin{enumerate}
\item The number $r$ is called \textit{the rapidity of the boost along the axis $z$}.
\item This theorem can be generalized to the case of $SL_n(\mathbb{C})$. 
\item The rapidity $r$ is uniquely determined but $u$ and $v$ are not. The other possible choices are $(u \, e^{i \theta \sigma_3} , v \, e^{i \theta \sigma_3})$, with $\theta \in \mathbb{R}$.
\item The polar decomposition of \SL is a particular case of the Cartan decomposition where $v^{-1} = \mathbb{1}$. This requirement makes it unique.
\end{enumerate}
\begin{proof}
The proof is essentially the same as previously.
Existence. Let $g \in \SL$ be decomposed. $g^\dagger g$ is positive-definite hermitian. With the spectral theorem, we have $v \in U(2)$ and $d$ a real diagonal matrix with strictly positive coefficients such that $g^\dagger g = v^\dagger d v$. If $\det v= e^{i \theta}$, then $u = e^{-i\theta/2} v \in \SU$ and $g^\dagger g = u^\dagger d u$. Since $\det d = 1$, one can write $d = e^{r \sigma_3}$. Note that
\begin{equation}
e^{r\sigma_3/2} = \begin{pmatrix}
e^{r/2}&0\\0&e^{-r/2}
\end{pmatrix},
\end{equation} Then we show $g u e^{r \sigma_3/2} \in \SU$.
\end{proof}

\vspace{\baselineskip}
\noindent
$\blacksquare$ \textbf{Gauss decomposition.}
Let $g \in \SL$ such that $g_{22} \neq 0$. There exists a unique triplet $(z_+,d,z_-) \in Z_+ \times D \times Z_-$ such that
\begin{equation}\label{eq:Gauss decomposition}
g=z_+dz_-
\end{equation}
\begin{proof} Explicit computation.
If $g = \begin{pmatrix}
\alpha & \beta \\ \gamma & \delta
\end{pmatrix} \in \SL$, then one can write:
\begin{equation}
g = \begin{pmatrix}
1 & \beta \delta^{-1} \\ 0 & 1 
\end{pmatrix}
\begin{pmatrix}
\delta^{-1} & 0 \\ 0 & \delta
\end{pmatrix}
\begin{pmatrix}
1 & 0 \\ \gamma \delta^{-1} & 1
\end{pmatrix}
\end{equation}
\end{proof}

\vspace{\baselineskip}
\noindent
$\blacksquare$ \textbf{Iwasawa decomposition.}\marginpar{\scriptsize
$\varheart$ \textsc{Physics.} 
This decomposition has been used notably for the study of covariant twisted geometries \cite{livine2011a} and for hyperbolic discrete geometries \cite{bonzom2014b}.}
For any matrix $M \in \SL$, there exists a unique triplet $(Z,D,U) \in Z_+ \times D_{\mathbb{R}_+} \times \SU$ such that
\begin{equation}
M = ZDU = 
\begin{pmatrix}
1 & z \\ 0 & 1
\end{pmatrix}
\begin{pmatrix}
\lambda^{-1} & 0 \\ 0 & \lambda
\end{pmatrix}
\begin{pmatrix}
\alpha & - \beta^* \\ \beta & \alpha^*
\end{pmatrix}
\end{equation}
with $(z,\lambda,\alpha, \beta) \in \mathbb{C} \times \mathbb{R}^*_+ \times \mathbb{C}^2$.

	\section{Basics of \texorpdfstring{\SU}{SU(2)}}
	\label{sec:basics-SU(2)}

Let us now focus on the special unitary subgroup
\begin{equation}
\SU \overset{\text{def}}= \left \{ u \in \SL \mid u^\dagger u = \mathbb{1} \right \}.
\end{equation}
It is a $3$-dimensional real Lie subgroup of the $6$-dimensional real Lie group \SL. Any $u \in \SU$ can be uniquely written as
\begin{equation}
u = \begin{pmatrix}
	\alpha & - \beta^* \\
	\beta & \alpha^*
	\end{pmatrix}
	\quad \text{with} \quad (\alpha, \beta) \in \mathbb{C}^2, |\alpha|^2 + |\beta|^2 = 1,
\end{equation}
or equivalently
\begin{multline}
u = \begin{pmatrix}
	a+ib & -c+id \\
	c+id & a-ib
	\end{pmatrix}\\
	\text{with} \quad (a,b,c,d) \in \mathbb{R}^4 \\
	\text{and} \quad a^2 + b^2 + c^2 + d^2 = 1.
\end{multline}
The latter expression shows that \SU is diffeomorphic to $S^3$, the unit sphere of $\mathbb{R}^4$. Therefore it is connected, simply connected and compact.
The center of \SU is $Z_2$ and the quotient $\SU/Z_2$ is a group, which happens to be isomorphic to \SO (see section \ref{sec:rotation}). The real Lie algebra of \SU is
\begin{equation}
\su = \left \{ M \in \mathcal{M}_2(\mathbb{C}) \mid M^\dagger = - M \ \text{and} \  \Tr M = 0 \right \}.
\end{equation}
It is a real vector space, of which a basis is given by $( i \sigma_1, i \sigma_2,i\sigma_3 )$.
Since \SU is a compact and connected Lie group, any element of \SU can be written (non uniquely) as the exponential of an element of the associated Lie algebra \su (it is a general theorem for compact connected Lie groups).
\begin{physics}
The group \SU is central in quantum physics. First, it appears for the theory of the angular momentum (spin). Historically, it was also used as an approximate symmetry group for the isospin that relates protons and neutrons. Then it reappeared to describe the electro-weak interaction. In LQG, \SU comes with the holonomies, which are obtained by exponentiation of the Ashtekar variables, used for the quantization (see chapter \ref{ch:loops}).
\end{physics}

\vspace{\baselineskip}
\noindent
$\blacksquare$ \textbf{Exponential decomposition.}
If $u \in \SU$, there exists a (non-unique) $\vec{\alpha} \in \mathbb{R}^3$ such that 
\begin{equation}\label{eq:exponential decomposition}
u = e^{i \vec{\alpha} \cdot \vec{\sigma}} = \cos \| \vec \alpha \| \mathbb{1} + i \sin \| \vec \alpha \| \frac{\vec \alpha \cdot \vec \sigma}{\| \vec \alpha \|}.
\end{equation}
\begin{proof}
Let $M = \begin{pmatrix}
a & -b^* \\ b & a^*
\end{pmatrix} \in \SU$. The equality is easy to check for $r = \arccos(\Re a)$ and $\alpha_1 = \frac{r}{\sin r} \Im b$, $\alpha_2 = - \frac{r}{\sin r} \Re b$, $\alpha_3 = \frac{r}{\sin r} \Im a$.
\end{proof}

\vspace{\baselineskip}
\noindent
$\blacksquare$ \textbf{Euler angles decomposition.}
For all $u \in \SU$, there exists  $\alpha, \beta, \gamma \in \mathbb{R}$ (called Euler angles) such that:
\begin{equation} \label{eq:Euler decomposition}
u = e^{-\frac{i\alpha}{2} \sigma_3} e^{-\frac{i \beta}{2} \sigma_2} e^{-\frac{i \gamma}{2} \sigma_3}
\end{equation}
\marginpar{We use the French notation for open intervals, that is $]a,b[$ instead of the more familiar $(a,b)$ on the other side of the pond.}The choice can be made unique by restricting the domain of definition of the angles, for instance $\alpha \in ]-2\pi,2\pi [, \beta \in [0,\pi] \ \text{and} \ \gamma \in [|\alpha| , 4\pi- |\alpha|[$.
\begin{proof}
Explicit computation. The \ac{RHS} gives 
\begin{equation}
\begin{pmatrix}
e^{-\frac{i(\alpha + \gamma)}{2}} \cos \beta/2 & - e^{\frac{i ( \gamma - \alpha)}{2}} \sin \beta /2\\
e^{- \frac{i(\gamma - \alpha)}{2}} \sin \beta/2 & e^{\frac{i(\alpha + \gamma)}{2}} \cos \beta /2
\end{pmatrix}
\end{equation}
and for any $u \in \SU$, it is clearly possible to find $\alpha, \beta, \gamma \in \mathbb{R}$ to write $u$ in this form. 
Note that there are other conventions for the definition of Euler angles. The definition we have chosen is the one of Varshalovich (\cite{varshalovich1987} p. 27) and Sakurai (\cite{sakurai2011} p. 177). Rühl (\cite{ruhl1970} p. 43) and the Wolfram Language have chosen the convention $u = e^{\frac{i\alpha}{2} \sigma_3} e^{\frac{i \beta}{2} \sigma_2} e^{\frac{i \gamma}{2} \sigma_3}$ instead.
\end{proof}

\section{The rotations \texorpdfstring{\SO}{SO(3)}}
\label{sec:rotation}

As we have said in section \ref{sec:subgroups}, the action of \SU over Minkowski spacetime, given by equation \eqref{eq:SL2C action hermitian}, preserves the time direction. More precisely, an element $u \in \SU$ is acting over $\mathbb{M}$ through \eqref{eq:SL2C action hermitian} as a matrix $\Lambda(u) \in SO^+(3,1)$ given by
{\scriptsize
\begin{equation}
\Lambda(u) = \left(
\begin{array}{c | c c c}
1 & 0 & 0 & 0 \\
\hline 
0 & & & \\
0 & & R(u) & \\
0 & & & 
\end{array}
\right)
\end{equation}
}
with
\begin{equation}
[R(u)]_{ij} = \frac{1}{2} \Tr \left(u \sigma_j u^\dagger \sigma_i \right).
\end{equation}
If $u$ is written as
\begin{equation}
u = \begin{pmatrix}
\alpha & - \beta^* \\
\beta & \alpha^*
\end{pmatrix}
\end{equation}
then $R(u)$ reads 
{\scriptsize
\begin{equation}\label{eq:homomorphism}
R(u) = \begin{pmatrix}
\frac{1}{2} ( \alpha^2 + \alpha^{*2} - \beta^2 - \beta^{*2} ) &  \frac{i}{2} ( -\alpha^2 + \alpha^{*2} + \beta^2 - \beta^{*2} ) & \alpha \beta^* + \alpha^* \beta \\
\frac{i}{2} ( \alpha^2 - \alpha^{*2} + \beta^2 - \beta^{*2} )  & \frac{1}{2} ( \alpha^2 + \alpha^{*2} + \beta^2 + \beta^{*2} ) & i (\alpha \beta^* - \alpha^* \beta) \\
- \alpha \beta - \alpha^* \beta^* & i(\alpha \beta - \alpha^* \beta^*) & \alpha \alpha^* - \beta \beta^*
\end{pmatrix}.
\end{equation}}
The map $R$ is a $2$-to-$1$ onto homomorphism from \SU to the group of rotations of $\mathbb{R}^3$, denoted \SO and defined as
\begin{equation}
\SO \overset{\text{def}}= \left \{ M \in \mathcal{M}_3(\mathbb{R}) \mid M^T M = \mathbb{1} \ \text{and} \  \det M = 1 \right \}.
\end{equation}
\SO is connected, but not simply connected. Topologically, it is homeomorphic to the sphere $S^3$ with the antipodal points being identified. This can be understood through the property that $R(u) = R(-u)$, or through the following isomorphism of groups
\begin{equation}\label{eq:SO=SU}
\SO \cong \SU / Z_2.
\end{equation}
The Lie algebra of \SO is
\begin{equation}
\so = \left \{ M \in \mathcal{M}_3(\mathbb{R}) \mid M^T + M = 0 \ \text{and} \  \Tr M = 0 \right \}.
\end{equation}
The isomorphism \eqref{eq:SO=SU} implies the following isomorphism of Lie algebra
\begin{equation}
\so \cong \su.
\end{equation}
Remark that the action defined by $R$ is equal to the adjoint representation of \SU (see equation \eqref{eq:adjoint_representation}) when written in the basis $i \sigma_j \in \su$.

As it is written in equation \eqref{eq:homomorphism}, it is hard to understand the geometrical meaning of the coefficients of \SU. It becomes simpler if one uses the exponential decomposition \eqref{eq:exponential decomposition}: $R(e^{i \theta \vec n \cdot \sigma})$, with $\vec n \in S^2$ and $\theta \in \mathbb{R}$, is the rotation around the axis of $\vec n$, of an angle $- 2 \theta$. In particular, the action of $R$ on the Euler decomposition \eqref{eq:Euler decomposition}, yields 
\begin{multline}
R (u) = r_z(\alpha) r_y(\beta) r_z(\gamma) \\
\text{with} \ 
r_z(\phi) = \begin{pmatrix}
\cos \phi & - \sin \phi & 0 \\
\sin \phi & \cos \phi & 0 \\
0 & 0 & 1  
\end{pmatrix} \\
\text{and} \quad r_y(\beta) = \begin{pmatrix}
\cos \beta & 0 & \sin \beta \\
0 & 1 & 0 \\
- \sin \beta &  0 & \cos \beta
\end{pmatrix}.
\end{multline}
where $(\alpha, \beta, \gamma)$ are any choice of Euler angles. The unicity of the decomposition can be obtained for instance with the restriction $\alpha \in ]-\pi,\pi [, \beta \in [0,\pi] \ \text{and} \ \gamma \in [|\alpha| , 2\pi- |\alpha|[$.

\section{Measure and integration}

\paragraph{Haar measure.} A Borel set in \SU is any subset of \SU obtained from open sets through countable union, countable intersection, or taking the complement. All Borel sets form a $\sigma$-algebra\footnote{A $\sigma$-algebra on a set $X$ is a collection of subsets of $X$ that contains $X$ itself and is closed under complementation and countable unions (and hence countable intersections too). The prefix ``$\sigma$'' signals closure under \emph{countable}, rather than merely finite, operations.} called the Borel $\sigma$-algebra $\mathcal{B}(\SU)$. A Borel \SU-measure $\mu$ is a non-negative function over $\mathcal{B}(\SU)$ for which $\mu(\emptyset)=0$, and which is countably additive (the measure of a disjoint union is the sum of the measures of each set). A Borel measure is said to be quasi-regular if it is both
\begin{enumerate}
\item Outer regular: $ \mu(S) = \inf \{\mu(U) \mid S \subseteq U, U \ \text{open}\}$;
\item Inner regular: $\mu(S) = \sup \{\mu(K) \mid K \subseteq S, K \ \text{compact} \}$.
\end{enumerate}
It can be shown that there exists a unique quasi-regular Borel measure $\mu$ over \SU satisfying
\begin{enumerate}
\setlength\itemsep{0 \baselineskip}
\item invariance: $\mu(S) = \mu(gS) = \mu(Sg)$ for all Borel sets $S$ and all $g \in \SU$;
\item normalisation: $\mu(\SU) = 1$.
\end{enumerate}
It is called the (two-sided normalised) \textit{Haar measure} of \SU. The Haar measure enables the definition of integrals of functions $f$ over \SU:
\begin{equation}
\int_{\SU} f(u) \, \dd \mu(u) \quad \text{also denoted} \quad \int_{\SU} f(u) \, \dd u.
\end{equation}

\paragraph{The Hilbert space $L^2(\SU)$.} The space of complex functions over \SU satisfying 
\begin{equation}
\int_{\SU} |f(u)|^2 du < \infty,
\end{equation}
is denoted $L^2(\SU)$. It is an infinite-dimensional Hilbert space with the scalar product
\begin{equation}
(f_1,f_2) \overset{\text{def}}= \int_{\SU} f_1^*(u) f_2(u) \dd u.
\end{equation}

\paragraph{Measure over \texorpdfstring{\SL}{SL(2,C)}.}
In \acs{GGV} (\cite{gelfand1966} pp. 214--215), an invariant measure $\dd a$ over \SL is defined, so that for any $g \in \SL$, we have
\begin{equation}
\dd a = \dd (ga) = \dd (ag) = \dd (a^{-1}).
\end{equation}
In terms of the matrix components $a_{ij}$, it is given explicitly by 
\begin{equation}
\dd a = \left( \frac{i}{2} \right)^3 |a_{12}|^{-2} \, \dd a_{11} \dd \overline{a_{11}} \, \dd a_{12} \dd \overline{a_{12}} \, \dd a_{22} \dd \overline{a_{22}} .
\end{equation}
In Rühl \cite{ruhl1970}, the invariant measure is given in terms of the coefficients in the decomposition \eqref{eq:decomposition Pauli}, by
\begin{equation}
\dd a = \frac{1}{\pi^4} \delta \left( a_0^2 - \sum_{k=1}^3 a_k^2 - 1 \right) \dd a_0 \dd \overline{a_0}  \, \dd a_1 \dd \overline{a_1} \, \dd a_2 \dd \overline{a_2} \, \dd a_3 \dd \overline{a_3}.
\end{equation}
It is normalised so that the induced measure over \SU is the same Haar measure defined in the previous section. In Rühl (\cite{ruhl1970} p. 285), it is shown that using the Cartan decomposition $a = u e^{r \sigma_3/2} v^{-1}$, we have:
\begin{equation}
\dd \mu(a) = \frac{1}{4 \pi} \sinh^2r \, \dd r \, \dd u \, \dd v.
\end{equation}

\section{Representations}

\paragraph{Representation of groups.}
A good way to understand the structural properties of a group is to look for its action on vector spaces. By "action", we mean specifically a linear action that preserves the group product: it is called a \textit{representation}. In physics, notably in quantum mechanics, we often focus on representations over Hilbert spaces.

Let $G$ be a locally compact group, and $GL(\mathcal{H})$ the group of bounded linear operators over a Hilbert space $\mathcal{H}$ that admit a bounded inverse. A (bounded continuous) \textit{representation} $\rho$ of $G$ over $\mathcal{H}$ is a homomorphism $\rho : G \to GL(\mathcal{H})$, such that the action map $G \times \mathcal{H} \to \mathcal{H}$ is continuous. In the case of a finite-dimensional Hilbert space $\mathcal{H}$, $GL(\mathcal{H})$ is just the space of invertible linear maps, and a representation is any linear action of $G$ over $\mathcal{H}$. It is called \textit{unitary} if it preserves the scalar product.

\paragraph{Representation of Lie algebras.}
There are also representations of Lie algebras, which are linear actions preserving the Lie bracket. Any representation of a Lie group defines, by differentiation, a representation of its Lie algebra. Precisely, if $\rho: G \rightarrow GL(\mathcal{H})$ is a representation of $G$, the \textit{differential of $\rho$}, is the linear map $D\rho: \mathfrak{g} \rightarrow \textfrak{gl}(\mathcal{H})$ defined for all $X \in \mathfrak{g}$ by:
\begin{equation}\label{eq:diff}
(D\rho)(X) \overset{\text{def}}= \left. \frac{d}{dt}\rho(e^{tX}) \right|_{t=0}.
\end{equation}
Moreover, for all $X \in \mathfrak{g}$,
\begin{equation}\label{eq:exp}
\rho(e^X) = e^{D\rho(X)}.
\end{equation}
One can show that:
\begin{enumerate}
\item If $F \subset \mathcal{H}$ is stable for $\rho$, then $F$ is also stable for $D \rho$.
\item If $D\rho$ is irreducible, then $\rho$ is also irreducible.
\item If $G$ is connected, the converses of (1) and (2) are also true.
\end{enumerate}
Conversely, given a Lie algebra $\mathfrak{g}$, there is no unique Lie group associated to it, but there is a unique simply connected one $G$, whose Lie algebra is $\mathfrak{g}$. Then, given any morphism of Lie algebra $\phi$, there exists a morphism of a Lie group $\rho$ such that $\phi = D\rho$. Thus, a representation of $\mathfrak{g}$ will infer a representation on each of its associated Lie groups.

\paragraph{Irrep.}
A representation is \textit{irreducible} if it admits no closed stable subspace other than $\left\{ 0 \right\}$ and $\mathcal{H}$. For brevity, we commonly say "irrep" instead of "irreducible representation". 
They can be seen as the building blocks of the other representations. From two representations, one can build others using notably the direct sum and the tensor product. If $V$ and $W$ are two vector spaces of representation of a group $G$ and its algebra $\mathfrak{g}$, we define a representation over the direct sum $V \oplus W$ by
\begin{equation}
\begin{split}
&\forall g \in G, \quad  g \cdot (v + w) = g \cdot v + g \cdot w \\
&\forall X \in \mathfrak{g}, \quad  X \cdot (v + w) = X \cdot v + X \cdot w .
\end{split}
\end{equation}
We also define a representation over the tensor product $V \otimes W$
\begin{equation}\label{eq:tensor_representation}
\begin{split}
&\forall g \in G, \quad  g \cdot (v \otimes w) = (g \cdot v) \otimes (g\cdot w) \\
&\forall X \in \mathfrak{g}, \quad X \cdot (v \otimes w) =  (X \cdot v) \otimes w + v \otimes (X \cdot w). 
\end{split}
\end{equation}

\section{Intertwiners}
\label{sec:intertwiners}

If $V$ and $W$ are two vector spaces of representation of a group $G$ and its algebra $\mathfrak{g}$, an \textit{intertwiner} (or \textit{equivariant map} or \textit{intertwining operator}) is a linear map $T : V \rightarrow W$ satisfying:
\begin{equation}
T( g \cdot v) = g \cdot T(v). 
\end{equation}
\marginpar{In the language of category theory, an intertwiner is nothing but a \textit{natural transformation} between two \textit{functors}, each functor being a representation of the group.}The space of intertwiners, denoted $\text{Hom}_G(V,W)$, is a subspace of the vector space of linear maps $\text{Hom}(V,W)$. A useful result is the following isomorphism
\begin{equation}\label{eq:hom-inv}
\text{Hom}_G(V,W) \cong \text{Inv}_G(V \otimes W^*),
\end{equation}
where $W^*$ is the dual space of $W$ and
\begin{equation}
\text{Inv}_G (E)  \overset{\text{def}}= \left\{ \psi \in E \mid \forall g \in G, \ g \cdot \psi = \psi \right\}.
\end{equation}
Two representations are \textit{equivalent} if there is an invertible intertwiner between them. An invertible intertwiner is a way to \textit{identify} two representations, as if there were only a change of notation between them. It is common to alleviate the notations by making the intertwiner implicit, and using instead the symbol of congruence "$\cong$", which should be understood as "equal from the perspective of the group representation". For instance, anticipating on section \ref{sec:angular momentum realisation}, we denote $\ket{jm} \cong v_{j-m}$ instead of $\ket{jm} = T ( v_{j-m})$, and similarly for operators, we write $ J_+ \cong e$, rather than $ J_+ = T \circ e \circ T^{-1}$. Thus, two equivalent representations will often be presented as two \textit{realisations} of the same representation. The symbol $\cong$ is not a strict equality "$=$" in the mathematical sense since it only identifies \textit{some} of the structures on the two sides of the equation. See chapter \ref{ch:equal} for a discussion about what being equal or isomorphic means.

\paragraph{Schur's Lemma.} If $T : V \to W$ is an intertwiner between two finite irreps of $G$, then either $T=0$, or $T$ is bijective. Moreover, if the irreps are complex and $T$ is bijective, then for any other bijective intertwiner $T'$ there exists $\lambda \in \mathbb{C}$ such that $T' = \lambda T$. This lemma is useful notably to prove the following and very important theorem.

\paragraph{Peter-Weyl's theorem.}
An important case is when the group $G$ is compact (\eg \SU, but not $SL_2(\mathbb{C})$). In this case, we have the following properties:
\begin{enumerate}
\item Any complex finite representation of $G$ can be endowed with a hermitian product which makes the representation unitary.
\item Any unitary irrep of $G$ is finite-dimensional.
\item Any unitary representation can be decomposed into a direct sum of irreps.
\end{enumerate}
Theses results justify notably that focusing on unitary irreps of \SU, as we do in chapter \ref{ch:representation-SU2}, is sufficient to describe all possible finite or unitary representations of \SU. Finally, the compactness of $G$ enables to define the space of square-integrable functions $L^2(G)$ with the Haar measure, and then
\begin{enumerate}
\setcounter{enumi}{3}
\item The linear span of all matrix coefficients of all finite unitary irreps of $G$ is dense in $L^2(G)$.
\end{enumerate}
A proof can be found in Knapp (\cite{knapp1986} pp. 17--20).

\section{Induced representations}
\label{sec:induced}

There is a well-known method to build a representation of group, \textit{induced} from a representation of one of its subgroups. We present below two possible formal definitions of the method (see the book \cite{maurin1997} for details). In section \ref{sec:principal series}, we will apply the method to construct the principal series of irreps of \SL.

Consider a group $G$ and $K$ one of its subgroup. Say $\rho$ is a representation of $K$ over a vector space $V$. We are going to build a representation of $G$ using $\rho$. We first define a vector space $\mathcal{H}$, then a group homomorphism $U: G \rightarrow GL(\mathcal{H})$. There are two equivalent ways to proceed.
\begin{enumerate}
\item Let $\mathcal{H}$ be the vector space of functions $f:G \rightarrow V$ such that 
\begin{equation}
\forall g \in G, \quad \forall k \in K, \quad f(gk) = \rho(k)f(g).
\end{equation}
For all $g \in G$, we define the linear map $U(g) : \mathcal{H} \rightarrow \mathcal{H}$ by 
\begin{equation}
\forall f \in \mathcal{H}, \quad \forall x \in G, \quad  U(g)f(x) = f(g^{-1}x).
\end{equation}
\item Denote the quotient $M\overset{\text{def}}=G/K$. Let $P(M,K)$ be a $K$-principal bundle over $M$. Denote $P \times_\rho V$ the associated vector bundle of base $M$. It has $V$ as fibre. Let
\begin{equation}
\mathcal{H} = \Gamma \left( P \times_\rho V \right),
\end{equation}
the set of sections of $P \times_\rho V$. For all $g \in G$, we define the linear map $U(g)$ by
\begin{equation}
\forall f \in \mathcal{H}, \quad \forall x \in G/K, \quad (U(g) f )(x) = f(g^{-1}x).
\end{equation}
\end{enumerate}
In both cases, $(U, \mathcal{H})$ is the representation of $G$ \textit{induced} from the representation $(\rho,V)$ of the subgroup $K$. 

As an example, consider the trivial subgroup $\{e\}$ of a Lie group $G$, and its trivial representation over \C. The induced representation is then given by the Hilbert space $L^2(G)$, endowed with a left-invariant (resp. right-invariant) measure, and the linear action $g \cdot f(h) = f(g^{-1} h)$ (resp. $g \cdot f(h) = f(hg)$). It is also called the left (resp. right) regular representation.
\chapter{Bundles over Spheres}
\label{ch:geometric}

In this chapter, we present three ways to think about spheres and we introduce two essential bundles over the sphere. All these geometrical tools are omnipresent in theoretical physics and especially in quantum gravity.

\section{Variations upon a sphere}

In chapter \ref{ch:equal}, we have seen how subtle the definition of a circle may be. Here, we do it again with three equivalent descriptions of the sphere:
\begin{enumerate}\setlength\itemsep{0em}
\item The submanifold $S^2$;
\item The Riemann sphere $\bar{\mathbb{C}}$;
\item The complex projective line $\mathbb{C}P^1$.
\end{enumerate}

\subsection{The sphere $S^2$}

Define the sphere $S^2$  as 
\begin{equation}
S^2 \overset{\text{def}}= \left \{ (x,y,z) \in \mathbb{R}^3 \mid x^2 + y^2 + z^2 = 1 \right \}.
\end{equation}
It is a topological space with the induced topology of $\mathbb{R}^3$, meaning the open subsets of $S^2$ are the intersection of $S^2$ with the open sets of $\mathbb{R}^3$. $S^2$ is also a $2$-dimensional differentiable manifold. It can be parametrised with the spherical coordinates $(\theta, \phi)$ as
\begin{equation}
S^2 = \left \{ (\sin \theta \cos \phi, \sin \theta \sin \phi , \cos \theta) \mid \theta \in \left[ 0,\pi \right] \ \text{and} \ \phi \in \left[0, 2 \pi \right[ \right \}.
\end{equation}
$S^2$ is endowed with a metric induced from the euclidian metric of $\mathbb{R}^3$
\begin{equation}
\dd s^2 = \dd x^2 + \dd y^2 + \dd z^2.
\end{equation}
In spherical coordinates, it reads
\begin{equation}
\dd s^2 = \dd \theta^2 + \sin^2 \theta \, \dd \phi^2 .
\end{equation}
The associated surface element, used to integrate functions over the sphere, is
\begin{equation}
\dd A = \sin \theta \, \dd \theta \, \dd \phi .
\end{equation}

\subsection{The Riemann sphere}

A topological space is locally compact if every point admits a compact neighbourhood. Such a space $X$ can be compactified by adding a single point to it. The resulting compact space is denoted $\bar{X}$, and called the \textit{Alexandroff extension}. The Alexandroff extension of $\mathbb{C}$ is the \textit{Riemann sphere}, denoted $\bar{\mathbb{C}} = \mathbb{C} \cup \{ \infty \}$. It is a Riemann surface, that is a uni-dimensional complex manifold, and it is diffeomorphic to $S^2$:
\begin{equation}
\bar{\mathbb{C}} \cong S^2.
\end{equation}
Different diffeomorphisms are used in the literature. Here we give the \textit{stereographic projection from the south pole}, from $S^2$ to $\bar{\mathbb{C}}$, that reads, 
\begin{equation}
(x,y,z) \mapsto \zeta = \frac{-x+iy}{1+z},
\end{equation}
or, in spherical coordinates,
\begin{equation}
(\theta,\phi) \mapsto \zeta = - \tan \frac \theta 2 e^{-i\phi}.
\end{equation}
The inverse is  
\begin{equation}\label{eq:inverse-stereographic}
\zeta \mapsto \begin{pmatrix}
x \\ y \\ z
\end{pmatrix} = \frac{1}{1 + |\zeta|^2} \begin{pmatrix}
-\zeta - \zeta^* \\ i(\zeta^* - \zeta) \\ 1- |\zeta|^2
\end{pmatrix}.
\end{equation}
The metric now takes the form
\begin{equation}
\dd s^2 = \frac{4 \dd \zeta \, \dd \zeta^*}{(1 + |\zeta|^2)^2},
\end{equation}
with $\dd \zeta =  \dd \Re (\zeta) + i \, \dd \Im (\zeta)$.
The metric enables to measure lengths and areas. $\bar{\mathbb{C}}$ is actually a \textit{Kähler manifold}, so that its metric can be locally written as the second derivative of a potential, in our case, $F(\zeta) = \log (1 + | \zeta |^2)$, and 
\begin{equation}
\dd s^2 = 4 \pdv{}{\zeta} \left( \pdv{F}{\zeta^*} \right) \, \dd \zeta \dd \zeta^*.
\end{equation}
The areas are measured with the symplectic $2$-form
\begin{equation}
\omega = 2 i  \frac{\dd \zeta \wedge \dd \zeta^* }{(1+ | \zeta |^2) ^2}.
\end{equation}

\subsection{The complex projective line}

The set $\mathbb{C}^2$ is a complex vector space. A \textit{vector line} $d$ is a uni-dimensional linear subspace of $\mathbb{C}^2$. The \textit{complex projective line} $\mathbb{C}P^1$ is the set of vector lines of $\mathbb{C}^2$. 

A more sophisticated way of saying the same thing consists in defining an equivalence relation between two points $(z_0,z_1)$ and $(w_0,w_1)$, each in $\mathbb{C}^2$, such as
\begin{equation}
(z_0,z_1) \sim (w_0,w_1) \qquad \Leftrightarrow \qquad \exists \lambda \in \mathbb{C}^*, \quad (z_0,z_1) = (\lambda w_0,\lambda w_1).
\end{equation}
$\mathbb{C}P^1$ is then defined as the quotient space:
\begin{equation}
\mathbb{C}P^1 \overset{\text{def}}= (\mathbb{C}^2 \setminus \{0\})/\sim.
\end{equation}
If $(x_0,x_1) \in \mathbb{C}^2 \setminus \{0\}$, the vector line $d$ passing through it is denoted $[x_0:x_1]$, and $(x_0,x_1)$ are called the \textit{homogeneous coordinates} of $d$. The surjective function,
\begin{equation}\label{eq:surjective-projection}
p : \left\{ 
\begin{array}{lll}
\mathbb{C}^2 \setminus \{ 0 \} & \rightarrow & \mathbb{C}P^1 \\
(x_0,x_1) & \mapsto & [x_0:x_1]
\end{array} \right.
\end{equation}
induces a topology on $\mathbb{C}P^1$ ($U \subset \mathbb{C}P^1$ is an open set if, and only if,  $p^{-1}(U)$ is an open set), so that $p$ is continuous. In fact, $\mathbb{C}P^1$ is a complex manifold, with the open cover given by 
\begin{equation}
\begin{split}
U_0 \overset{\text{def}}= \left\{ [x_0:x_1] \mid x_0 \neq 0 \right\} \\
U_1 \overset{\text{def}}= \left\{ [x_0:x_1] \mid x_1 \neq 0 \right\} 
\end{split}
\end{equation}
The \textit{north pole} is $N \overset{\text{def}}= [1:0] \in U_0$ and the \textit{south pole} is $S \overset{\text{def}}= [0:1] \in U_1$. In fact, $U_0 = \mathbb{C}P^1 \setminus \{ S \}$ and $U_1 = \mathbb{C}P^1 \setminus \{ N \}$. The north map is
\begin{equation}
\zeta_0 : \left\{ 
\begin{array}{lll}
U_0 & \rightarrow & \mathbb{C} \\
\left[x_0 : x_1\right] & \mapsto & \frac{x_1}{x_0}.
\end{array} \right.
\end{equation}
The south map $\zeta_1$ is defined similarly. It is well-defined whatever the choice of homogeneous coordinates. $\zeta_0$ and $\zeta_1$ are homeomorphisms and $\zeta_0 \circ \zeta_1^{-1} (z) = \frac 1z $ is holomorphic. Thus $\mathbb{C}P^1$ is diffeomorphic to $\bar{\mathbb{C}}$ (and thus to $S^2$)
\begin{equation}
\mathbb{C}P^1 \cong \bar{\mathbb{C}}  .
\end{equation}

\section{Tautological bundle $\mathcal{O}(-1)$}
\label{sec:tautological}

We now introduce an essential fibre bundle which plays a crucial role in the representation of \SL, as described in chapter \ref{ch:representation-SL2C}. 

The set $\mathcal{O}(-1)$ is the disjoint union of vector lines of $\mathbb{C}^2$:
\begin{equation}
\mathcal{O}(-1) \overset{\text{def}}= \bigsqcup_{d \in \mathbb{C}P^1} d = \bigcup_{d \in \mathbb{C}P^1} \left\{ (d,x) \mid x \in d \right\}.
\end{equation}
It is a subset of $\mathbb{C}P^1 \times \mathbb{C}^2$. Note the difference with the simple union and the set of vector lines:
\begin{equation}
\bigcup_{d \in \mathbb{C}P^1} d = \mathbb{C}^2 \quad \text{and} \quad \bigcup_{d \in \mathbb{C}P^1} \{ d \} = \mathbb{C}P^1.
\end{equation}
Notations are subtle. The specificity of $\mathcal{O}(-1)$ is its structure of holomorphic line bundle:
\begin{equation}
\mathbb{C} \to \mathcal{O}(-1) \to \mathbb{C}P^1,
\end{equation}
with the usual notation "fibre $\to$ bundle $\to$ base". It is called the \textit{tautological bundle} because the fibre over a point $d \in \mathbb{C}P^1$ is $d$ itself. 

\paragraph{Definition.}
The projection over the base space is
\begin{equation}
\pi : \left\{ 
\begin{array}{lll}
\mathcal{O}(-1) & \rightarrow & \mathbb{C}P^1 \\
(d,x) & \mapsto & d.
\end{array} \right.
\end{equation}
The fibre over $d \in \mathbb{C}P^1$ is $\pi^{-1}(d) \cong \mathbb{C}$. $(U_0, U_1)$ forms an open covering of $\mathbb{C}P^1$ so that $(\pi^{-1}(U_0),\pi^{-1}(U_1))$ is an open covering of $\mathcal{O}(-1)$. In fact, $\mathcal{O}(-1)$ is locally trivialised on this covering by
\begin{equation}
\begin{split}
t_0 : \left\{ 
\begin{array}{lll}
 U_0 \times \mathbb{C}& \rightarrow & \pi^{-1}(U_0)  \\
(d, z) & \mapsto & (d,( z , \zeta_0(d) z))
\end{array} \right. \\
t_1 : \left\{ 
\begin{array}{lll}
 U_1 \times \mathbb{C}& \rightarrow & \pi^{-1}(U_1)  \\
(d, z) & \mapsto & (d,( \zeta_1(d) z , z))
\end{array} \right.
\end{split}
\end{equation}
As trivialisations, $t_0$ and $t_1$ are diffeomorphisms satisfying $\pi \circ t (d,z) = d$. They are related by a transition function
\begin{equation}
t_{10} : \left\{ 
\begin{array}{cll}
U_0 \cap U_1 & \rightarrow & \mathbb{C}  \\
d & \mapsto & \zeta_0(d)
\end{array} \right.
\end{equation}
which satisfies
\begin{equation}
t_0(d,x) = t_1(d, t_{10}(d)x).
\end{equation}
All these properties define a line bundle. The manifold $\mathcal{O}(-1)$ and $\mathbb{C}P^1$ are complex, and the trivialisations and the transition functions are holomorphic, so that the $\mathcal{O}(-1)$ is a holomorphic line bundle.

\paragraph{Notation.} 
The transition function $t_{10}$ is nothing but $\zeta_0$ restricted to $U_0 \cap U_1$. In fact, for any $k \in \mathbb{Z}$, one can define a holomorphic line bundle $\mathcal{O}(k)$ over $\mathbb{C}P^1$, for which the transition function $t_{10}$ is the restriction of $\zeta_0^{-k}$. When $k=-1$, we get the tautological bundle $\mathcal{O}(-1)$. When $k=0$, we get the trivial bundle $\mathcal{O}(0) = \mathbb{C}P^1  \times \mathbb{C}$.
It is a theorem that any holomorphic line bundle over $\mathbb{C}P^1$ is one of the $\mathcal{O}(k)$. Over other Riemann surfaces, there are generally many more line bundles, sometimes even a continuous family of them. A hard theorem by Grothendieck states that any holomorphic vector bundle over $\mathbb{C}P^1$ is a direct sum of some $\mathcal{O}(k)$ \cite{hazewinkel1982}.

\paragraph{Section.} A global section of $\mathcal{O}(-1)$ is a continuous map $s : \mathbb{C}P^1 \rightarrow \mathcal{O}(-1)$ satisfying $\pi \circ s = Id_{\mathbb{C}P^1}$. Actually, the only global holomorphic section of $\mathcal{O}(-1)$ is the null section ($s_0 : d \mapsto (d,(0,0))$). So, it is more interesting to look at local sections, defined over open sets of $\mathbb{C}P^1$. The set of local sections has the mathematical structure of a sheaf.

Let's consider sections over $U_0$. Any such section $s : U_0 \rightarrow \mathcal{O}(-1)$ can be written as $s(d) = (d, f(d))$, with $f : U_0 \rightarrow \mathbb{C}^2$ a continuous function satisfying $f(d) \in d$. Then, the map $\sigma \overset{\text{def}}= f \circ \zeta_0^{-1} : \mathbb{C} \rightarrow \mathbb{C}^2$ is a continuous function satisfying $\sigma_1(z) = z \, \sigma_0(z)$. Thus, any section over $U_0$ is uniquely characterised by a continuous function $\sigma_0 : \mathbb{C} \rightarrow \mathbb{C}$, and conversely any such function defines a section.

\paragraph{Integration.}
$\mathcal{O}(-1)$ is a real differentiable manifold of dimension $4$. The image of a section $s(U_0) \subset \mathcal{O}(-1)$ is a $2$-dimensional real submanifold. A $2$-form $\alpha$ defined over $\mathcal{O}(-1)$ can be integrated over $s(U_0)$:
\begin{equation}
\int_{s(U_0)}  \alpha
\end{equation}
Such a $2$-form $\alpha$ is itself a section of the vector bundle $\Lambda^2 T^*\mathcal{O}(-1)$ of base space $\mathcal{O}(-1)$. In this bundle, a basis of sections over $\pi^{-1}(U_0)$ is given by 
\begin{equation*}
dz_0 \wedge dz_1,\quad dz_0 \wedge d\bar{z}_0,\quad dz_0 \wedge d\bar{z}_1,\quad d\bar{z}_0 \wedge dz_1,\quad dz_1 \wedge d\bar{z}_1,\quad d\bar{z}_0 \wedge d\bar{z}_1,
\end{equation*}
denoted $D_i(z_0,z_1)$ with $i \in \{1,...,6\}$. So $\alpha$ can be written in this basis and
\begin{equation}
\int_{s(U_0)}  \alpha  = \int_{\mathbb{C}} \alpha_i(\sigma_0(z),\sigma_1(z)) D_i(\sigma_0(z),\sigma_1(z)).
\end{equation}
A $2$-form $\alpha$ is homogeneous of degree $0$ if it is constant on each fibre, \ie
\begin{equation}
\forall \lambda \in \mathbb{C}, \quad \alpha(z_0,z_1) = \alpha(\lambda z_0, \lambda z_1).
\end{equation}
In such a case, the integral does not depend on the chosen section: 
\begin{equation}
\int_{s(U_0)}  \alpha = \int_{\mathbb{C}} \alpha_i(1,z) D_i(1,z).
\end{equation}
This fact will be used in section \ref{sec:principal series} to define the principal series of representations of \SL.

\section{Hopf fibration}
\label{sec:hopf}

In the preceding section, we have seen the tautological bundle over $\mathbb{C}P^1$. We now describe another useful bundle over $\mathbb{C}P^1$, the \textit{Hopf fibration}:
\begin{equation}
S^1 \rightarrow S^3 \rightarrow S^2.
\end{equation}
To illustrate the central role of the Hopf fibration in physics, we refer the reader to the beautiful review paper \cite{urbantke2003}.

$\mathbb{C}^2$ and \su are respectively the fundamental and the adjoint representations of \SU. The fundamental representation is simply given by the multiplication of a matrix by a vector, and the adjoint representation is given by 
\begin{equation}\label{eq:adjoint_representation}
\forall u \in \SU, \ \forall x \in \su, \quad u \cdot x = u \, x \, u^{-1}. 
\end{equation}
The \textit{Hopf map} 
\begin{equation}
p : \left\{ \begin{matrix}
\mathbb{C}^2 & \rightarrow & \su \\
\pmb z & \mapsto & \sum_{k=1}^3 (\pmb z^\dagger  \sigma_k \pmb z)   \ i \sigma_k
\end{matrix} \right.
\end{equation}
\marginpar{We use the shorthand $\pmb z = (z_0,z_1)$, and $a \, \pmb z$ denotes the matrix multiplication of $a$ with $\mqty(z_0 \\ z_1)$.}is an intertwiner of the representations of \SU. It satisfies notably
\begin{equation}
\pmb z \pmb z^\dagger = \frac{1}{2} \left( \| p (\pmb z)\| \mathbb{1} + p(\pmb z) \cdot \vec \sigma \right).
\end{equation}
Using the standard isomorphism between \su and $\mathbb{R}^3$ (see section \ref{sec:basics-SU(2)}), the Hopf map reads
\begin{equation}
p : \left\{ \begin{matrix}
\mathbb{C}^2 & \rightarrow & \mathbb{R}^3 \\
(z,w) & \mapsto & (zw^*+wz^*,i(zw^*-wz^*),|z|^2 - |w|^2) 
\end{matrix} \right.
\end{equation}
Then using the standard isomorphism between $\mathbb{C}^2$ and $\mathbb{R}^4$, it reads
\begin{equation}
p : \left\{ \begin{matrix}
\mathbb{R}^4 & \rightarrow & \mathbb{R}^3 \\
(x,y,z,w) & \mapsto & (2(xz+yw),2(xw-yz),x^2+y^2-z^2-w^2)
\end{matrix} \right.
\end{equation}
The sphere $S^3$ is defined as
\begin{equation}
S^3 \overset{\text{def}}= \left\{ (x,y,z,t) \in \mathbb{R}^4 \mid x^2 + y^2 + z^2 + t^2 =1 \right\}.
\end{equation}
When we restrict $p$ to $S^3$, we get the \textit{Hopf projection}
\begin{equation}
\mathfrak{p} : S^3 \rightarrow S^2.
\end{equation}
In terms of Euler angles and spherical coordinates, it reads
\begin{equation}
\mathfrak{p} : \left\{ \begin{matrix}
S^3 & \rightarrow & S^2 \\
(\theta,\phi,\psi) & \mapsto & (\theta,\phi)
\end{matrix} \right.
\end{equation}
Using the fact $S^2 \cong \mathbb{C}P^1$ and the following identification 
\begin{equation}
S^3 \cong \left\{ (z_0,z_1) \in \mathbb{C}^2 \mid |z_0|^2 + |z_1|^2 = 1 \right\},
\end{equation}
we can show that the Hopf projection reads
\begin{equation}
\mathfrak{p} : \left\{ \begin{matrix}
S^3 & \rightarrow & \mathbb{C}P^1 \\
(z_0,z_1) & \mapsto & [z_0 : z_1]
\end{matrix} \right.
\end{equation}
So the Hopf projection is a restriction of the surjective map \eqref{eq:surjective-projection} that sends a point of $\mathbb{C}^2$ to the complex line to which it belongs. Then, for any $d \in \mathbb{C}P^1$, we have
\begin{equation}
\mathfrak{p}^{-1}(d) \cong \U.
\end{equation}
$\mathfrak{p}^{-1}(U_0)$ and $\mathfrak{p}^{-1}(U_1)$ form an open cover of $S^3$. In fact, $S^3$ is locally trivialised over each of them through:
\begin{equation}
\begin{split}
\tau_0 : \left\{ 
\begin{array}{lll}
U_0 \times \U & \rightarrow & \mathfrak{p}^{-1}(U_0)  \\
(d, \lambda) & \mapsto & \left( \frac{\lambda}{\sqrt{1 + |\zeta_0(d)|^2}},\frac{ \zeta_0 (d) \lambda}{ \sqrt{1 +|\zeta_0(d)|^2}} \right)
\end{array} \right.
\\
\tau_1 : \left\{ 
\begin{array}{lll}
U_1 \times \U & \rightarrow & \mathfrak{p}^{-1}(U_1)  \\
(d, \lambda) & \mapsto & \left( \frac{\zeta_1(d)\lambda}{\sqrt{1 + |\zeta_1(d)|^2}},\frac{\lambda}{ \sqrt{1 +|\zeta_1(d)|^2}} \right)
\end{array} \right.
\end{split}
\end{equation}
$\tau_0$ and $\tau_1$ are indeed diffeomorphisms satisfying $\mathfrak{p} \circ \tau (d,\lambda) = d$. The transition function is
\begin{equation}
\tau_{10} : \left\{ 
\begin{array}{cll}
U_0 \cap U_1 & \rightarrow & \U  \\
d & \mapsto & e^{i \arg \zeta_0(d)} = \dfrac{\zeta_0(d)}{|\zeta_0(d)|}
\end{array} \right.
\end{equation}
So, it satisfies
\begin{equation}
\tau_0(d,\lambda) = \tau_1(d, \tau_{10}(d) \lambda).
\end{equation}
All these properties, define indeed the structure of a fibre bundle. $S^3$ is actually a \U-principal bundle over $\mathbb{C}P^1$. Using standard isomorphisms, we get the nice stacking of spheres:
\begin{equation}
S^1 \rightarrow S^3 \rightarrow S^2.
\end{equation}
Finally, the Hopf fibration can also be seen as a re-branding of the quotient
\begin{equation}
S^2 \cong  \SU/\U.
\end{equation}
This is the consequence of a more general theorem. If $G$ is a Lie group and $H$ a closed subgroup, then there exists a unique structure of smooth manifold on $G/H$ such that the quotient map $\pi : G \rightarrow G/H$ defines a fibre bundle of fibre $H$.
\chapter{Representation theory of \texorpdfstring{\SU}{SU(2)}}
\label{ch:representation-SU2}

What mathematicians call the \textit{representation theory of \SU}, physicists call it the \textit{quantum theory of the angular momentum}. The latter emerged with the very first steps of quantum mechanics as it is essential to understand how electrons turn around nuclei. The first textbook dealing with the subject seems to have been \textit{The Theory of Atomic Spectra} \cite{condon1959} by Condon and Shortley, first published in 1935. The monograph offers a synthesis of the achievements in the understanding of the line spectra of atoms, the very problem that had lighted up the quantum mechanical revolution. In 1957, Edmonds published \textit{Angular Momentum in Quantum Mechanics} \cite{edmonds1957} which tackles the subject of angular momentum with more details that were discovered meanwhile. In 1960, Yutsis, Levinson and Vanagas\footnote{Their book was originally published in Russian, but an English version was released in 1962. Their name is here translated from Russian, while they are actually Lithuanian, from which language their name is sometimes translated as Jucys, Levinsonas and Vanagas.} published \textit{Mathematical Apparatus of the Theory of Angular Momentum} \cite{yutsis1962}, which completes \cite{edmonds1957} in some aspects.

A lot of the mathematical content in these textbooks was already known by mathematicians before. Indeed, the representation theory of \SU did not wait for the theory of atomic spectra to be discovered. It is an abstract theory independent of the contingency of the physical contexts. Physics sometimes helps to make it more tangible, but it can also hinder some general understanding, for instance when it drags old-fashioned jargon.

In this chapter, we introduce the representations of \SU over finite-dimensional Hilbert spaces. Note that real representations of \SU also exist (see \cite{itzkowitz1991}) but they are ignored by physicists. Due to Peter-Weyl's theorem, a complex finite representation of a compact group can be decomposed into a direct sum of irreps. So we will focus on irreps of \SU, going through several equivalent realisations that co-exist across various communities.

\section{Irreps of \texorpdfstring{\SU}{SU(2)}}
\label{sec:abstract representation}	

To start with, it is important to notice the following one-to-one correspondence between sets of finite-dimensional representations:
\begin{enumerate}
\setlength\itemsep{0 \baselineskip}
\item Holomorphic representations\footnote{Here, "holomorphic" means that the map defined by the representation over the vector space is holomorphic.} of \SL,
\item Representations of \SU,
\item Representations of \su,
\item $\mathbb{C}$-linear representations\footnote{"$\mathbb{C}$-linear" means that we regard \slc as a complex (and not real) vector space. We will care about the $\mathbb{R}$-linear representations of \slc in section \ref{sec:finite irreps}.} of \slc.
\end{enumerate}
It is a particular case of the so-called \textit{Weyl's unitary trick}. Concretely, we go from one set to another through:
\begin{description}
\item [$(\text{1}) \Rightarrow (\text{2})$] Restriction of the action of \SL to its subgroup \SU.
\item [$(\text{2}) \Rightarrow (\text{3})$] Differentiation as shown in equation \eqref{eq:diff}.
\item [$(\text{3}) \Rightarrow (\text{4})$] Using $\slc \cong \su \oplus i \, \su$. 
\item [$(\text{4}) \Rightarrow (\text{1})$] Exponentiation as shown in equation \eqref{eq:exp}.
\end{description}
Importantly, this correspondence preserves invariant subspaces and equivalences of representations. In particular, it means that it is now sufficient to our purpose to find all the $\mathbb{C}$-linear irreps of \slc.

\paragraph{Theorem.}
\textit{For all $n \in \mathbb{N}$, there exists an $(n+1)$-dimensional $\mathbb{C}$-linear irrep of \slc, unique up to equivalence.}
\vspace{\baselineskip}

A proof can be found in \cite{bernard2012}. The $(n+1)$-dimensional irrep is fully characterised by the action of the elements $h,e,f \in \slc$, defined by 
\begin{equation}
\begin{split}
&h \overset{\text{def}}=\sigma_3 = \begin{pmatrix}
1 & 0 \\ 0 & -1
\end{pmatrix} \\
&e \overset{\text{def}}=\frac{\sigma_1+i\sigma_2}{2} = \begin{pmatrix}
0 & 1 \\ 0 & 0
\end{pmatrix} \\
&f \overset{\text{def}}=\frac{\sigma_1-i\sigma_2}{2} = \begin{pmatrix}
0 & 0 \\ 1 & 0
\end{pmatrix},
\end{split}
\end{equation}
which satisfy the commuting relations
\begin{align*}
[h,e]=2e && [h,f]=-2f && [e,f]=h.
\end{align*}
Their action over a basis $(v_i)$, $i \in \{ 0,...,n \}$, is given by
\begin{align}
h \cdot v_k = (n-2k)v_k, && e \cdot v_k = k(n-k+1)v_{k-1}, && f \cdot v_k = v_{k+1}.
\end{align}
The $3$-dimensional complex vector space \slc can also be seen as a $6$-dimensional real vector space, which has \su as its subspace. Thus, by restriction of the action of \slc to \su, the previously found $\mathbb{C}$-linear irreps of \slc, define also irreps of \su. Finally, by exponentiating with \eqref{eq:exponential decomposition}, we find all irreps of \SU over complex vector spaces. 

\section{Angular momentum realisation}
\label{sec:angular momentum realisation}

In physics textbooks, the representations of \slc are indexed by half-integers, called \textit{spins}. To each spin $j\in \mathbb{N}/2$ is associated a Hilbert space $\mathcal{Q}_j$ of dimension $2j+1$. The canonical basis, also called the \textit{magnetic basis}, is composed of the vectors (or "kets" in the Dirac language) denoted 
\begin{equation}
\ket{j,m} \quad \text{with} \quad m \in \{-j,-j+1,...,0,...,j-1,j \}.
\end{equation}
It is made orthonormal by choosing the scalar product that satisfies
\begin{equation}
\braket{j,m}{ j,n} = \delta_{mn}.
\end{equation}
We now define the \textit{angular momentum observables} $J_i \overset{\text{def}}= \frac{1}{2} \sigma_i$, sometimes called simply \textit{generators of \SU} or \textit{generators of rotations}.
\begin{physics}
In some textbooks, the generators are defined as $J_i \overset{\text{def}}= \frac{\hbar}{2} \sigma_i$ where $\hbar$ is the reduced Planck constant, which has the dimension of an action. Indeed, this realisation originally comes from atomic physics, where the $J_i$ represent "observables" of angular momentum. For simplicity, we are working in the units where $\hbar = 1$, keeping in mind the possibility to restore $\hbar$ explicitly at any moment by dimensional analysis. By the way, notice also that since "observables" are required to be hermitian operators, the $J_i$ are elements of $i \mathfrak{su}(2)$ and not of $\mathfrak{su}(2)$. 
\end{physics}
They satisfy
\begin{equation}
\left[ J_i,J_j \right] = i  \epsilon_{ijk} J_k.
\end{equation}
We then define their linear action over $\mathcal{Q}_j$ by
\begin{equation}
\begin{split}
&J_1 \ket{j,m} = \frac 1 2  \sqrt{(j - m)(j+m+1)} \ket{j,m+1} \\
&\hspace{100 pt} + \frac 1 2 \sqrt{(j+m)(j-m+1)} \ket{j,m-1},  \\
&J_2 \ket{j,m} = \frac{1}{2i}  \sqrt{(j - m)(j+m+1)} \ket{j,m+1} \\
&\hspace{100 pt} - \frac{1}{2i} \sqrt{(j+m)(j-m+1)} \ket{j,m-1}, \\
&J_3 \ket{j,m} = m  \ket{j,m}.
\end{split}
\end{equation}
\graffito{Sometimes the action of $J_+$ over $\ket{jm}$ is written with a constant phase $e^{i\delta}$. It defines an equivalent representation, but the choice of $\delta = 0$ (called the \textit{Condon-Shortley convention}, from \cite{condon1959}) is the most widespread.}It is somehow simpler to remember the action of the \textit{ladder operators} $J_\pm  \overset{\text{def}}= J_1 \pm iJ_2 $, \begin{equation}
\begin{aligned}
&J_+ \ket{j,m} = \sqrt{(j - m)(j+m+1)} \ket{j,m+1}, \\
&J_- \ket{j,m} = \sqrt{(j+m)(j-m+1)} \ket{j,m-1}.
\end{aligned}
\end{equation}
The action of the generators $J_i$ over $\mathcal{Q}_j$ defines a $(2j+1)$-dimensional irrep of \SU, called the \textit{spin-$j$ representation}. This is shown by exhibiting the following equivalence with the irreps defined in the previous section:
\begin{align}
J_3 \cong h/2, && J_+ \cong e, && J_- \cong f,
\end{align}
together with the vector correspondence
\begin{equation}
\ket{jm} \cong \sqrt{\frac{(j+m)!}{(2j)!\,(j-m)!}} \, v_{j-m}.
\end{equation}
Finally notice that $\ket{jm}$ is also an eigenvector of the \textit{total angular momentum} $\vec{J}^2 \overset{\text{def}}= J_1^2 + J_2^2 + J_3^2$:
\begin{equation}
\vec{J}^2 \ket{jm} =  j (j+1) \ket{jm}.
\end{equation}
In fact, the $\ket{jm}$ form the unique orthonormal basis that diagonalises simultaneously the commuting operators $J_3$ and $\vec{J}^2$. We say that $J_3$ and $\vec{J}^2$ form a \textit{\ac{CSCO}}. From a mathematical perspective, notice also that $\vec{J}^2$ is not an element of the algebra $i \, \su$, but an element of the \textit{universal enveloping algebra} $\mathcal{U}(i \, \su)$ whose action can be easily computed by successive action of \su. Since $\vec{J}^2$ has the property to be a quadratic element that commutes with all of $\mathcal{U}(i \, \su)$, it is called the \textit{Casimir operator} of $\mathcal{U}(i \, \su)$.

\paragraph{Wigner matrix.}
The exponentiation of the action of the generators of \SU defines a linear action of the group \SU (see  eq. \eqref{eq:exp}). The \textit{Wigner matrix} $D^j(g)$ represents the action of $g \in \SU$ in the $\ket{j,m}$ basis. It is thus a square matrix of size $2j+1$, whose coefficients are the functions
\begin{equation}
D^j_{mn} (g) \overset{\text{def}}= \matrixel{j,m}{ g }{ j,n}.
\end{equation}
\begin{notabene}
One should be aware of a small ambiguity in the notation "$\matrixel{j,m}{ g }{ j,n}$" that arises when $g$ is a matrix that belongs simultaneously to \SU and to \su. Then it should be said explicitly if one considers the group action or the algebra action when computing $\matrixel{j,m}{ g }{ j,n}$, because it gives a different result. This ambiguity comes from the fact that physicists do not usually write explicitly whether they consider the group representation $\rho$, or its differential $D\rho$. Mathematicians would write  $\matrixel{j,m}{\rho(g)}{j,n}$, or $ \matrixel{j,m}{D\rho(g)}{ j,n}$. From equation \eqref{eq:exponential decomposition}, if $ g = e^a \in \SU \cap \su$, with $a \in \su$, then $\rho(g) = e^{D\rho(a)}$, but $\rho(g) \neq {D\rho(e^a)} = D\rho(g)$. In the definition of the Wigner matrix above, it is the group action which is considered.
\end{notabene}
From Schur's lemma, it can be shown that the functions $D^j_{mn}$ form an orthogonal family of $L^2(\SU)$:
\begin{equation}\label{eq:ortho D}
\int_{\SU} dg \ \overline{ D^{j'}_{m'n'}(g)} D^j_{mn}(g) = \frac{1}{2j+1} \delta_{jj'} \delta_{mm'} \delta_{nn'}.
\end{equation}
\begin{proof}
The left hand side (LHS) is the coefficient $\matrixel{j'n'}{A}{jn}$ of the operator $A : \mathcal{Q}_j \to \mathcal{Q}_{j'}$ defined by
\begin{equation}
A \overset{\text{def}}= \int_\SU \dd g \ g^\dagger \dyad{j'm'}{jm} g.
\end{equation}
We first show that $A$ is an intertwiner. If $j \neq j'$, then, by Schur's lemmas (cf. section \ref{sec:intertwiners}), $A = 0$. Otherwise, $j'=j$, and $A$ is bijective, and there exists $\lambda \in \mathbb{C}$ so that $A = \lambda \mathbb{1}$. Taking the trace on both sides, we see that $\lambda = \frac{\delta_{mm'}}{2j+1}$.
\end{proof}
In fact, the Peter-Weyl theorem even asserts that the functions $D^j_{mn}$ form a basis of $L^2(\SU)$, \ie any function $f \in L^2(\SU)$ can be written 
\begin{equation}
f(g) = \sum_{j \in \mathbb{N}/2} \sum_{m = -j}^j \sum_{n = -j}^j f^j_{mn} D^j_{mn}(g),
\end{equation}
with coefficients $f^j_{mn} \in \mathbb{C}$. It implies notably an equivalence between the following Hilbert spaces
\begin{equation}\label{eq:Peter L(SU(2))}
L^2(\SU) \cong \bigoplus_{j \in \mathbb{N}/2} (\mathcal{Q}_j \otimes \mathcal{Q}_j^*).
\end{equation}
The equivalence is not per se a surprise, since all Hilbert spaces of the same dimension are isomorphic, but more interesting is the  specific form of the isomorphism, \ie $D^j_{mn} \cong \ket{j,n} \otimes \bra{j,m}$. We are going now to derive explicit expressions for computing $D^j_{mn}(g)$, but we first need to introduce another realisation of the spin-$j$ irreps. 

\section{Homogeneous realisation}
\label{sec:section homogeneous}

Let $\mathbb{C}_{2j}[z_0,z_1]$ be the vector space of polynomials of two complex variables, homogeneous of degree $2j \in \mathbb{N}$. If $P(z_0,z_1) \in \mathbb{C}_{2j}[z_0,z_1]$, it can be written as
\begin{equation}
P(z_0,z_1) = \sum_{k=0}^{2j} a_k z_0^k z_1^{2j-k},
\end{equation}
with coefficients $a_0,...,a_{2j} \in \mathbb{C}$.The action of \SU given by
\begin{equation}\label{eq:action homogeneous}
g \cdot P(\pmb z) = P(g^T \pmb z)
\end{equation}
defines a $(2j+1)$-dimensional group representation.
\begin{proof}
The action satisfies
\begin{align*}
(g_1g_2) \cdot P = g_1 \cdot (g_2 \cdot P) && \text{and} && e \cdot P = P,
\end{align*}
which defines a group action over the vector space $\mathbb{C}_{2j}[z_0,z_1]$.
\end{proof}
\begin{notabene}
We would get equivalent realisations by defining the action with $P(a^{-1} \pmb z)$, $P(a^\dagger \pmb z)$ or $P(\pmb z a)$. In fact $P(\pmb z a) = P(a^T \pmb z)$. The convention that we have chosen here is the one of Rovelli-Vidotto (\cite{CLQG} p. 173). Moreover our convention is consistent with the choice we have made later for the representations of the principal series (cf. section \ref{sub:principal homogeneous}). In Bernard (\cite{bernard2012} p. 128) the convention $P(a^{-1} \pmb z)$ is used.
\end{notabene}
The group action induces the following action of the generators
\begin{equation}\label{eq:generators homogeneous}
J_+ \cong z_0 \frac{\partial}{\partial z_1} \qquad J_- \cong z_1 \frac{\partial}{\partial z_0} \qquad J_3 \cong \frac{1}{2} \left( z_0 \frac{\partial}{\partial z_0} - z_1 \frac{\partial}{\partial z_1} \right).
\end{equation}
\begin{proof}
Let $g(t) = e^{t M}$. The differential of the representation is given by
\begin{align*}
M \cdot P(z_0,z_1) &= \frac{\dd}{\dd t} \left( g(t) \cdot P(z_0,z_1) \right) \big|_{t=0} \\
&=  \frac{\dd}{\dd t} \left( P(g_{11}(t) z_0 + g_{21}(t)z_1,g_{12}(t) z_0 + g_{22}(t)z_1) \right) \big|_{t=0} \\
&=  (M_{11}z_0 + M_{21} z_1) \frac{\partial P}{\partial z_0} (z_0,z_1) +  (M_{12}z_0 + M_{22} z_1) \frac{\partial P}{\partial z_1} (z_0 , z_1).
\end{align*}
Then it suffices to apply to $J_3, J_+, J_-$.
\end{proof}
This representation is equivalent to the spin-$j$ irrep through the correspondence:
\begin{equation}
\label{eq:equivalence_homogeneous}
\ket{j,m} \cong \left( \frac{(2j)!}{(j+m)!(j-m)!} \right)^{1/2} z_0^{j+m} z_1^{j-m},
\end{equation}
The \ac{RHS} is sometimes denoted with Dirac notations $\braket{z_0z_1}{jm}$.
\begin{proof}
One looks for an intertwiner $\Phi : \mathcal{Q}_j \to \mathbb{C}_{2j}[z_0,z_1]$, such that for all $i$, $J_i \cdot \Phi(\ket{j,m}) = \Phi (J_i \ket{j,m})$. We show easily the necessary condition:
\begin{equation}
\Phi(\ket{j,m}) = c \left( \frac{(2j)!}{(j+m)!(j-m)!} \right)^{1/2} z_0^{j+m} z_1^{j-m},
\end{equation}
We choose $c=1$.
\end{proof}
We have thus found another realisation of the spin-$j$ irrep, called the \textit{homogeneous realisation}. It is very convenient to derive an explicit expression for the Wigner matrix coefficients. 

\paragraph{Wigner matrix formula.}
\begin{multline} \label{eq:Wigner matrix}
D^j_{mn}(g) = \left( \frac{(j+m)!(j-m)!}{(j+n)!(j-n)!} \right)^{1/2} \\
\times \sum_k \binom{j+n}{k} \binom{j-n}{j+m-k} g_{11}^k g_{21}^{j+n-k} g_{12}^{j+m-k} g_{22}^{k-m-n}. 
\end{multline}
The sum is done over the integers $k \in \{\max (0,m+n),...,\min(j+m,j+n)\}$.
\begin{proof} First of all remark that the action \eqref{eq:action homogeneous} is well defined for any $g = \begin{pmatrix}
g_{11} & g_{12} \\ g_{21} & g_{22}
\end{pmatrix} \in \GL$. Explicitly, it acts over the canonical basis of $\mathbb{C}_{2j}[z_0,z_1]$ like
\begin{align*}
g \cdot z_0^k z_1^{2j - k} &= P\left( \begin{pmatrix}
g_{11} & g_{21} \\ g_{12} & g_{22}
\end{pmatrix} \binom{z_0}{z_1} \right) \\
&= (g_{11} z_0 + g_{21} z_1)^k (g_{12} z_0 + g_{22} z_1)^{2j-k} \\
&= \sum_{i=0}^k \sum_{l=0}^{2j-k}  \binom{k}{i} \binom{2j-k}{l} g_{11}^i g_{21}^{k-i} g_{12}^l g_{22}^{2j-k-l} z_0^{i+l} z_1^{2j-i-l} \\
&=  \sum_{n=0}^{2j} \left( \sum_{i= \max (0,n+k-2j)}^{\min (k,n)} \binom{k}{i} \binom{2j-k}{n-i} g_{11}^i g_{21}^{k-i} g_{12}^{n-i} g_{22}^{2j-k-n+i} \right) z_0^{n} z_1^{2j-n}
\end{align*}
This is nothing but mechanical computation. Finally, we recover the canonical basis of equation \eqref{eq:equivalence_homogeneous} with the subtle change of variables:
\begin{equation}
m := n-j \qquad q:=k-j.
\end{equation}
The subtlety is that $j$ is a spin, and we generalise the notation $\sum$ for half-integers bounds with still a step of $1$.
\end{proof}
From this, we show that
\begin{equation}\label{eq:symmetry D conjugate}
\overline{D^j_{mn}(u)} = (-1)^{m-n} D^j_{-m,-n}(u).
\end{equation}

\paragraph{Euler angles expression.}
Wigner proposed also another explicit expression for his matrix, in terms of the Euler angles.
If $u \in \SU$, and $(\alpha, \beta, \gamma) \in \mathbb{R}^3$ are the Euler angles of $u$, such that $u = e^{-\frac{i\alpha}{2} \sigma_3} e^{-\frac{i \beta}{2} \sigma_2} e^{-\frac{i \gamma}{2} \sigma_3}$, then
\begin{align*}
D^j_{m'm}(u)  = e^{-i (\alpha m' + \gamma  m )} d^j_{m'm}(\beta)
\end{align*}
with the \textit{reduced Wigner matrix}
\begin{multline}\label{eq:reduced Wigner}
d^j_{m'm}(\beta) = \left( \frac{(j+m')!(j-m')!}{(j+m)!(j-m)!} \right)^{\frac 12} \\
\times \sum_{k= \max(0,m'+m)}^{\min(j+m',j+m)} (-1)^{m'+j-k}  \binom{j+m}{k} \binom{j-m}{j-k+m'} \\ \times \left(\cos \frac \beta 2 \right)^{2k-m-m'}  \left(\sin \frac \beta 2 \right)^{m+m'+2j-2k}.
\end{multline}
It is already implemented in the Wolfram Language with the command
\begin{equation}
\texttt{WignerD}[\{j,m,n\},\alpha,\beta,\gamma]= e^{i (\alpha m + \gamma  n )} d^j_{mn}(-\beta).
\end{equation}
\begin{proof}
The proof is very easy once given formula \eqref{eq:Wigner matrix}. It only consists in applying it to the matrix 
\begin{align*}
e^{-\frac{i\alpha}{2} \sigma_3} e^{-\frac{i \beta}{2} \sigma_2} e^{-\frac{i \gamma}{2} \sigma_3} =
\begin{pmatrix}
e^{-\frac{i(\alpha + \gamma)}{2}} \cos \beta/2 & - e^{\frac{i ( \gamma - \alpha)}{2}} \sin \beta /2\\
e^{- \frac{i(\gamma - \alpha)}{2}} \sin \beta/2 & e^{\frac{i(\alpha + \gamma)}{2}} \cos \beta /2
\end{pmatrix}.
\end{align*}
This is what Rühl does (\cite{ruhl1970} p. 43), except that in his convention, the Euler angles are defined with no minus sign in front, so $\alpha \mapsto - \alpha$, $\beta \mapsto - \beta$, $\gamma \mapsto - \gamma$, and so $d^j_{m'm}(\beta)|_{\text{Rühl}}  = d^j_{m'm} (-\beta)$, which finally also equals the \ac{RHS} of \eqref{eq:reduced Wigner} provided the coefficient $(-1)^{m'+j-k}$ is changed into $(-1)^{m+j-k}$.

A proof that does not presuppose formula \eqref{eq:Wigner matrix} can be found in Sakurai (\cite{sakurai2011} p. 236--238), which has the same convention as ours for Euler angles. It uses the Schwinger's oscillator model for angular momentum. It obtains the same formula as ours, but written in a slightly different way, changing the index of summation $k \rightarrow j+m-k$.  Finally, Varshalovich (\cite{varshalovich1987} p. 76), which also has the same convention, gives a series of variations in the way of writing the above formula.
\end{proof}

\section{Projective realisation}
\label{sec:projective}

The spin-$j$ irrep can also be realised over $\mathbb{C}_{2j}[z]$, the vector space of complex polynomials of one variable $z$ of degree at most $2j$. This realisation is obtained from the $\mathbb{C}_{2j}[z_0,z_1]$ realisation by the map:
\begin{equation}
\left\{ \begin{array}{ll}
\mathbb{C}_{2j}[z_0,z_1] &\to \mathbb{C}_{2j}[z] \\
P(z_0,z_1) &\mapsto P(z,1)
\end{array} \right.
\end{equation}
This map is constructed from a projection from $\mathbb{C}^2$ to $\mathbb{C}$, hence the name "projective" that we give to this realisation. Sometimes it is also named the "holomorphic" realisation.
From this we deduce the action of \SU
\begin{equation}
a \cdot f(z) = (a_{12} z + a_{22})^{2j} f \left(\frac{a_{11} z + a_{21}}{a_{12} z +a_{22}}\right),
\end{equation}
and of the generators
\begin{equation}
J_+ \cong - z^2 \frac{d}{dz} + 2jz \quad J_3 \cong z\frac{d}{dz} - j \quad J_- \cong \frac{d}{dz}.
\end{equation}
The canonical basis becomes
\begin{equation}
\ket{j,m} \cong \sqrt{\frac{(2j)!}{(j+m)!(j-m)!}} z^{j+m}.
\end{equation}
We can give the following explicit expression for the scalar product that makes the canonical basis orthonormal:
\begin{equation}
\braket{f}{g} \overset{\text{def}}= \frac{i}{2} \frac{2j+1}{\pi} \int_\mathbb{C} \overline{f(z)} g(z) \frac{dzd\overline{z}}{(1+|z|^2)^{2j+2}}.
\end{equation}

	\section{Spinorial realisation}
	\label{sec:spinorial_realisation}

We now introduce a last realisation of the spin-$j$ irreps, which relies on notations developed by Penrose \cite{penrose1984}. It was found useful for \textit{twistor theory} \cite{penrose1986}, and later in quantum gravity for the so-called \textit{twisted geometries} \cite{freidel2010a, langvik2016}. The notions are a kind of gymnastics that needs some time to be learnt, but finally bears fruit in the long run.

\paragraph{Abstract indices.}
We are going to use the clever \textit{conventions of abstract indices} of Penrose (\cite{penrose1984} pp. 68--115). To start with, we need a set of "abstract indices" $\mathcal{L}$, that is to say a countable set of symbols. We use for instance capital letters:
\begin{equation}
\mathcal{L} \overset{\text{def}}= \{ A, B, ..., Z, A_0, ..., Z_0, A_1,... \}.
\end{equation}
Then we denote $\mathfrak{S}^\bullet \overset{\text{def}}= \mathbb{C}^2$, and for any abstract index $A \in \mathcal{L}$, $\mathfrak{S}^A \overset{\text{def}}= \mathfrak{S}^\bullet \times \{ A \}$. Obviously $\mathfrak{S}^A$ is isomorphic to $\mathbb{C}^2$ as a complex vector space. An element of $\mathfrak{S}^A$ will be typically denoted $z^A = (z,A) \in \mathfrak{S}^A$. The abstract index $A$ serves as a marker to "type" the vector $z \in \mathbb{C}^2$ (thus $z^A \neq z^B$). This notation is very efficient to deal with several copies of the same space (here $\mathbb{C}^2$), like in tensor theory. 

The vector space of linear forms from $\mathbb{C}^2$ to $\mathbb{C}$, is called the dual space, and denoted $\mathfrak{S}_\bullet$. Similarly, we denote $\mathfrak{S}_A \overset{\text{def}}= \mathfrak{S}_\bullet \times \{A\}$, which is trivially isomorphic to the dual space of $\mathfrak{S}^A$. Its elements, called covectors, are denoted with an abstract lower capital index, $z_A$. Then the evaluation of a covector $y_A = (y,A)$ on a vector $z^A = (z,A)$ (called a "contraction") is denoted $y_A z^A = y(z) \in \mathbb{C}$ (the order does not matter $y_A z^A = z^A y_A$). 

\paragraph{Spinors.}
Consider the space of \textit{formal} (commutative and associative) finite sums of \textit{formal} (commutative and associative) products of elements, one from each $\mathfrak{S}^{A_1} , ... , \mathfrak{S}^{A_p} , \mathfrak{S}_{B_1} , ... , \mathfrak{S}_{B_q}$. A typical element can be written:
\begin{equation}
t^{A_1 ... A_p}_{\hspace{30 pt} B_1 ... B_q} = \sum_{i=1}^m z_{1,i}^{\hspace{10 pt} A_1} \, ... \,  z_{p,i}^{\hspace{10 pt}  A_p} \ y^{1,i}_{\hspace{10 pt}  B_1} \, ... \, y^{q,i}_{\hspace{10 pt} B_q}.
\end{equation}
Then impose the rules
\begin{enumerate}
\item (Homogeneity) 
\begin{equation}
\forall \alpha \in \mathbb{C}, \quad (\alpha z_1^{\ A_1}) \, z_2^{\ A_2} \, ... \,  z_p^{\ A_p} = z_1^{\ A_1} \, (\alpha z_2^{\ A_2}) \, ... \,  z_p^{\ A_p}
\end{equation}
\item (Distributivity)
\begin{equation}
(z_1^{\ A_1} + z_2^{\ A_2}) \, z_3^{A_3} \, ... \,  z_p^{\ A_p} = z_1^{\ A_1} \, z_3^{\ A_3} \, ... \,  z_p^{\ A_p} + z_2^{\ A_2} \, z_3^{\ A_3} \, ... \,  z_p^{\ A_p}.
\end{equation}
\end{enumerate}
The resulting space is a vector space denoted $\mathfrak{S}^{A_1 ... A_p}_{B_1 ... B_q}$. Its elements are called \textit{spinors} of type $(p,q)$, and its dimension is $2^{p+q}$.
\begin{notabene}
The spinor space $\mathfrak{S}^{A_1 ... A_p}_{B_1 ... B_q}$ is isomorphic, but not equal, to $\mathfrak{S}^{A_1} \otimes ... \otimes \mathfrak{S}^{A_p} \otimes \mathfrak{S}_{B_1} \otimes ... \otimes \mathfrak{S}_{B_q}$. The difference is the commutativity of the product. For instance the formal product of $\mathfrak{S}^{AB}$ is commutative (by assumption) in the sense that, for $z^A \in \mathfrak{S}^A$ and $y^B \in \mathfrak{S}^B$, $z^A y^B = y^B z^A$, whereas the tensor product is not, $z^A \otimes y^B \neq y^B \otimes z^A$, simply because $z^A \otimes y^B \in \mathfrak{S}^A \otimes \mathfrak{S}^B$ and $y^B \otimes z^A \in \mathfrak{S}^B \otimes \mathfrak{S}^A$ do not belong to the same set. The usual tensor product $\otimes$ imposes an arbitrary ordering between the vectors of each space, while the abstract indices notation is a way to allow the commutation of vectors at the price of constantly keeping track of the vector space to which they belong with a label. If you have to correct an exam, either you keep the pile of copies in a rigid and arbitrary defined order, or you ask the students to write their name on their copy, so that it does not really matter if the copies are mixed up while you fall down the stairs. 
\end{notabene}

The spinor space is endowed with a bunch of basic operations defined by a set of rules. It would be utterly non-pedagogical to state these rules in the most general case. On the contrary, they are very intuitive for simple examples and generalise without ambiguity for higher-order spinors.
\begin{enumerate}
\item (Index substitution) If $z^A = (z,A) \in \mathfrak{S}^A$, we denote $z^B = (z,B) \in \mathfrak{S}^B$. Thus $z^A \neq z^B$.
\item (Index permutation) If $t^{AB} = \sum_i z_i^A y_i^B \in \mathfrak{S}^{AB}$, we denote $t^{BA} = \sum_i z_i^B y_i^A  = \sum_i y_i^A z_i^B \in \mathfrak{S}^{AB}$. 
\item (Symmetrisation) $t^{(AB)} \overset{\text{def}}= \frac 12 (t^{AB} + t^{BA} )$ or generally \marginpar{$S_n$ denotes the group of permutation of $\{1,...,n\}$.}
\begin{equation}
z^{(A_1...A_n)} \overset{\text{def}}= \frac{1}{n!} \sum_{\sigma \in S_n}z^{A_{\sigma(1)}...A_{\sigma(n)}}.
\end{equation}
\item (Anti-symmetrisation) $t^{[AB]} \overset{\text{def}}= \frac 12 (t^{AB} - t^{BA} )$ or generally 
\begin{equation}
z^{[A_1...A_n]} \overset{\text{def}}= \frac{1}{n!} \sum_{\sigma \in S_n} \epsilon_\sigma z^{A_{\sigma(1)}...A_{\sigma(n)}},
\end{equation}
with $\epsilon_\sigma$ the signature of the permutation $\sigma$. 
\item (Contraction) If $t^A_{\ B} = \sum_i z_i^A y^i_B \in \mathfrak{S}^A_B$, then $t^A_{\ A} = \sum_i z_i^A y^i_A \in \mathbb{C}$.
\end{enumerate} 

\paragraph{Index dualisation.}
We denote the canonical basis of $\mathbb{C}^2$:
\begin{equation}
e_0 \overset{\text{def}}= \binom{1}{0} \hspace{3 cm} e_1 \overset{\text{def}}= \binom{0}{1}.
\end{equation}
It is easy to show that there exists a unique normalised skew-symmetric spinor of type $(0,2)$. It is denoted $\epsilon_{AB}$, and satisfies by definition:
\begin{equation}
\epsilon_{AB} = - \epsilon_{BA}, \hspace{3 cm} \epsilon_{AB} e_0^A e_1^B = 1.
\end{equation}
$\epsilon_{AB}$ corresponds over $\mathbb{C}^2$ to the unique $2$-form $\epsilon$ normalised by the condition $\epsilon(e_0,e_1)=1$, which is nothing but the determinant over $\mathbb{C}^2$.
For two vectors $z = (z_0,z_1)$ and $y=(y_0,y_1)$, we show easily that:
\begin{equation}\label{eq:determinant spinor}
\epsilon_{AB} z^A y^B = z_0 y_1 - z_1 y_0.
\end{equation}
\marginpar{The sense of contraction of the indices can be remembered with the ace of heart.\begin{center}
\includegraphics[scale=0.25]{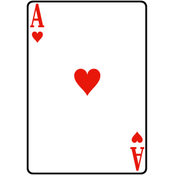}
\end{center}}Interestingly, $\epsilon_{AB}$ defines a canonical mapping between $\mathfrak{S}^A$ and $\mathfrak{S}_A$, given by
\begin{equation}
z^B \mapsto z_B = z^A \epsilon_{AB}.
\end{equation}
It is called \textit{index dualisation}.
\begin{notabene}
The convention $z_A =  \epsilon_{AB} z^B$ is also encountered. It gives the same mapping "up to a sign", since $\epsilon_{AB}$ is skew-symmetric. This other convention is chosen by Rovelli-Vidotto (\cite{CLQG} p. 23). Our convention is the one of Penrose (\cite{penrose1984} p. 104).
\end{notabene}
The covectors of the dual space $\mathfrak{S}_\bullet$ can also be described by a pair of components in the dual basis. The index dualisation can then be expressed in components as $((z_0,z_1),A) \mapsto ((-z_1,z_0),A)$. Similarly to the usual Dirac notation $\ket{z} = (z_0,z_1)$, a notation is sometimes introduced for the dual $[z| = (-z_1,z_0)$. With this choice, the \ac{RHS} of \eqref{eq:determinant spinor} reads $[z|y\rangle$.

\paragraph{Conjugation.}
The conjugate of $z \in \mathbb{C}$ is denoted $\bar z$ or $z^*$. We define the conjugation over $\mathfrak{S}^A$ by $\overline{z^A} = \overline{(z,A)} \overset{\text{def}}=(\overline{z},\dot A) = \bar z^{\dot A} \in \mathfrak{S}^{\dot A}$. Thus we have introduced a new set of abstract indices, the dotted indices:
\begin{equation}
\dot{\mathcal{L}} \overset{\text{def}}= \{ \dot A, \dot B, ..., \dot Z, \dot A_0, ..., \dot Z_0, \dot A_1,... \}.
\end{equation}
We impose moreover that $\overline{\bar z^{\dot A}} = z^A$, \ie $\ddot A = A$, so that the conjugation is an involution.  Importantly, we regard the set $\mathcal{L}$ and $\dot{\mathcal{L}}$ as incompatible classes of abstract indices, meaning that we forbid index substitution between them two. In other words, dotted and undotted indices commute: for any $t^{A \dot B} \in \mathfrak{S}^{A \dot B}$, we have $t^{A \dot B} = t^{\dot B A}$.
\begin{notabene}
One way to formalise this "incompatibility" between dotted and undotted indices would be to define rather $z^A = (z,A,0)$ and $z^{\dot A} = (z,A,1)$. Thus the index substitution $z^A = (z,A,0) \mapsto z^A = (z,B,0) = z^B $ clearly does not enable to translate from a dotted to an undotted index. Only the complex conjugation can through $z^A = (z,A,0) \mapsto (\overline{z},A,1) = z^{\dot A}$.
\end{notabene}

\paragraph{Inner product.} We define the map $J$ by:
\begin{equation}
J \binom{z_0}{z_1} = \binom{-\overline{z_1}}{\overline{z_0}}.
\end{equation}
Using the previously introduced generalised Dirac notation, we read $J \ket{z} = |z]$.
Since $J^2 = - 1$, the map $J$ behaves over $\mathbb{C}^2$ very much as the imaginary number $i$ behaves over $\mathbb{C}$. For this reason the map $J$ is said to define a \textit{complex structure} over $\mathbb{C}^2$. A combination of $\epsilon_{AB}$ and $J$ defines an inner product over $\mathfrak{S}^A$:
\begin{equation}
- \epsilon_{AB} (Jz)^A y^B  = \overline{z_0}y_0 + \overline{z_1} y_1.
\end{equation}
In generalised Dirac notations, we read $- [Jz | y\rangle = \braket{z}{y}$, which is consistent with the usual Dirac notation for the scalar product.
\begin{notabene}
With the matrix action over spinors, defined just below, equation \eqref{eq:action over spinors}, we can see that the inner product is invariant under the action of \SU: $\braket{u \cdot z }{u \cdot y} = \braket{z }{y}$. There is no surprise since it is actually one way of defining \SU. However the inner product is not invariant under \SL (contrary to the determinant). So, to "another choice of \SU", in the sense of a stabilizer of a time direction, would correspond another invariant inner product, and thus another complex structure $J$. For instance, in \cite{livine2011a}, they choose rather $(-J)$ for the complex structure. The choice we have made here is the one of Rovelli-Vidotto (\cite{CLQG} p. 24).
\end{notabene}

\paragraph{Representation.}
The vector space $\mathcal{M}_2(\mathbb{C})$ is isomorphic to $\mathfrak{S}^A_B$, through the isomorphism that associates to any $t \in \mathcal{M}_2(\mathbb{C})$ the unique spinor $t^A_{\ \ B}$ such that:
\begin{equation}\label{eq:action over spinors}
\forall z \in \mathbb{C}^2, \quad (tz)^A = t^A_{\ \ B} \, z^B.
\end{equation} 
Then the groups \SL and \SU can be represented over $\mathfrak{S}^{A_1...A_p}$ such as:
\begin{equation}
u \cdot z^{A_1...A_p} = u^{A_1}_{\ \ B_1} ... u^{A_p}_{\ \ B_p} \,  z^{B_1...B_p}.
\end{equation}
For later purposes, it is important to notice in particular the \SL-invariance of $\epsilon^{AB}$:
\begin{equation}
\label{eq:epsilon_invariance}
\forall u \in \SL, \quad u^A_{\ \ C} u^B_{\ \ D} \epsilon^{CD} = \epsilon^{AB},
\end{equation}
which is a restatement of \SL matrices having unit determinant.

The representation over $\mathfrak{S}^{A_1...A_p}$ is reducible, as it is stable over the subspace of completely symmetric spinors $\mathfrak{S}^{(A_1...A_p)}$.
\begin{proof}
The proof consists in checking the two equalities:
\begin{align*}
u^{(A_1}_{\ \ B_1} ... u^{A_p)}_{\ \ B_p} \,  z^{B_1...B_p} 
= u^{A_1}_{\ \ (B_1} ... u^{A_p}_{\ \ B_p)} \,  z^{B_1...B_p} = u^{A_1}_{\ \ B_1} ... u^{A_p}_{\ \ B_p} \,  z^{(B_1...B_p)},
\end{align*}
so that it is the same to first act with $u$, and then project down to $\mathfrak{S}^{(A_1...A_p)}$, or the other way around.
\end{proof}
Thus \SL and \SU can be represented on the vector space $\mathfrak{S}^{(A_1...A_p)}$ of dimension $p+1$. An orthogonal basis is given by
\begin{equation}
e^{(A_1}_{i_1}... e^{A_p)}_{i_p}  \quad \text{with} \quad i_1,...,i_p \in \{0,1\},
\end{equation}
which can also be denoted as 
\begin{equation}
e^{(A_1}_0...e^{A_s}_0 e^{A_{s+1}}_1 ... e^{A_p)}_1 \quad \text{with} \quad s \in \{0,...,p\}.
\end{equation}
The norm of these vectors is $\sqrt{\frac{s!(p-s)!}{p!}}$.
This representation is irreducible and equivalent to the spin-$(j=\frac{p}{2})$ irrep through the intertwiner:\begin{equation}
e^{(A_1}_0...e^{A_s}_0 e^{A_{s+1}}_1 ... e^{A_p)}_1  \cong z_0^s z_1^{p-s} \cong \sqrt{\frac{s!(p-s)!}{p!}} \ket{\frac{p}{2},s-\frac{p}{2}}.
\end{equation}
\begin{proof}
To see this, it is simpler to write the spinors in the canonical basis of $\mathfrak{S}^{A_1...A_p}$:\begin{equation}
\xi^{A_1...A_p} = \sum_{i_1,i_2,...,i_p=0}^1 c^{i_1,...,i_p} \, e^{A_1}_{i_1}... e^{A_p}_{i_p}
\end{equation}
The total symmetry of $\xi^{A_1...A_p}$ imposes a total symmetry of the coefficients $c^{i_1,...,i_p}$. Then we define the following bijection between $\mathfrak{S}^{(A_1...A_p)}$ and $\mathbb{C}_p[z_0,z_1]$:
\begin{equation}
\xi^{A_1...A_p} \cong \sum_{i_1,...,i_p=0}^1 c^{i_1,...,i_p} z_{i_1} \cdots z_{i_p}
\end{equation}
This defines an intertwiner as can be checked by looking at the action of a group element.
\end{proof}
\chapter{Recoupling theory of \texorpdfstring{\SU}{SU(2)}}

\label{ch:recoupling-SU2}

The \SU irreps provide the fundamental building blocks of quantum spacetime. From a mathematical perspective, irreps are the fundamental bricks from which other representations are built. Indeed any finite representation of \SU is \textit{completely reducible}, \ie it can be written as a direct sum of irreps. In particular, a tensor product of irreps can be \textit{decomposed} into a direct sum of irreps, \ie there exists a bijective intertwiner that maps the tensor product to a direct sum of irreps. Such an intertwiner is sometimes called a "\textit{coupling tensor}" (see Moussouris \cite{moussouris1983} pp. 10--11). This naming comes from quantum physics: when two systems \textit{couple} (\ie interact), the total system is described by states of the tensor product of the Hilbert spaces of the subsystems. Notice that there may exist several coupling tensors between a tensor product and its decomposition into a sum of irreps. It is precisely the goal of "\textit{recoupling theory}" to describe these coupling tensors and to understand how one can translate from one decomposition to another.

		\section{Clebsch-Gordan coefficients}

Given $\mathcal{Q}_{j_1}$ and $\mathcal{Q}_{j_2}$, two irreps of \SU, the tensor representation is defined over $\mathcal{Q}_{j_1} \otimes \mathcal{Q}_{j_2}$. The canonical basis of $\mathcal{Q}_{j_1} \otimes \mathcal{Q}_{j_2}$, also called the \textit{product basis}, is given by the elements
\begin{equation}
\ket{j_1m_1;j_2m_2} \overset{\text{def}}= \ket{j_1,m_1} \otimes \ket{j_2 ,m _2} 
\end{equation}
where $m_1$ and $m_2$ belong to the usual range of magnetic indices. This basis is the unique orthonormal basis that diagonalises simultaneously the following \ac{CSCO}:
\begin{equation}
J_3 \otimes \mathbb{1}, \quad \mathbb{1} \otimes J_3, \quad \vec{J}^2 \otimes \mathbb{1}, \quad \mathbb{1} \otimes \vec{J}^2.
\end{equation}
Another \ac{CSCO} on $\mathcal{Q}_{j_1} \otimes \mathcal{Q}_{j_2}$ is given by
\begin{equation}
J_3, \quad  \vec J^2, \quad \vec{J}^2 \otimes \mathbb{1}, \quad \mathbb{1} \otimes \vec{J}^2.
\end{equation}
Careful, don't be fooled by notations: $J_3$ and $\vec J^2$ act on the tensor space as given by equation \eqref{eq:tensor_representation}, i.e respectively as
\begin{equation}
\begin{split}
&J_3 \otimes \mathbb{1} + \mathbb{1} \otimes J_3, \\
\text{and} \ &\left( J_1 \otimes \mathbb{1} + \mathbb{1} \otimes J_1 \right)^2 + \left( J_2 \otimes \mathbb{1} + \mathbb{1} \otimes J_2 \right)^2 + \left( J_3 \otimes \mathbb{1} + \mathbb{1} \otimes J_3 \right)^2.
\end{split}
\end{equation}
Therefore, there exists an orthonormal basis that diagonalises them simultaneously. It called the \textit{coupled basis}, denoted
\begin{multline}
\ket{j_1j_2;kn} \\
 \text{with} \quad k \in \{|j_1-j_2|,...,j_1+j_2 \} \\
 \text{and} \quad n\in \{-k,...,k\},
\end{multline}
and characterised by the action of the operators:
\begin{equation}
\begin{split}
&J_3 \ket{j_1j_2;kn} = n \ket{j_1j_2;kn}\\
& \vec{J}^2 \ket{j_1j_2;kn} = k(k+1)\ket{j_1j_2;kn}\\
&\vec{J}^2 \otimes \mathbb{1}\ket{j_1j_2;kn} = j_1(j_1+1)\ket{j_1j_2;kn} \\
&\mathbb{1} \otimes \vec{J}^2\ket{j_1j_2;kn} = j_2 (j_2+1) \ket{j_1j_2;kn}.
\end{split}
\end{equation}
A proof can be found in Sakurai (\cite{sakurai2011} pp. 217--231). This result proves that $\mathcal{Q}_{j_1} \otimes \mathcal{Q}_{j_2}$ can be decomposed into a direct sum of irreps, namely we have the following equivalence of representations
\begin{equation}\label{eq:decomposition of 2}
\mathcal{Q}_{j_1} \otimes \mathcal{Q}_{j_2} \cong \bigoplus_{k=|j_1-j_2|}^{j_1+j_2} \mathcal{Q}_k. 
\end{equation}
The equivalence is given by the bijective intertwiner
\begin{equation}\label{eq:intertwiner iota}
\iota \left\{
\begin{array}{l}
\mathcal{Q}_{j_1} \otimes \mathcal{Q}_{j_2} \to \bigoplus_{k=|j_1-j_2|}^{j_1+j_2} \mathcal{Q}_k \\
\ket{j_1,j_2;km} \mapsto \ket{km}.
\end{array} \right.
\end{equation}
The action of $J_3$ on both sides of \eqref{eq:decomposition of 2} implies the existence of an arbitrary phase $\delta \in \mathbb{R}$ such that $\ket{j_1j_1 ; j_2j_2} \cong e^{i \delta} \ket{j_1+j_2,j_1+j_2}$. It is the choice of the Condon-Shortley convention to fix it so that $\ket{j_1+j_2,j_1+j_2} = \ket{j_1j_1 ;j_2,j_2}$. This fixes the intertwiner $\iota$ completely. The matrix coefficients of $\iota$ in the canonical basis are then called the \textit{Clebsch-Gordan coefficients}. In other words,
\begin{equation}
C^{jm}_{j_1m_1j_2m_2} \overset{\text{def}}= \braket{j_1m_1 ; j_2m_2}{jm}
\end{equation}
so that
\begin{equation}\label{eq:CG translation}
\ket{jm} = \sum_{m_1 = -j_1}^{j_1} \sum_{m_2 = -j_2}^{j_2} C^{jm}_{j_1m_1j_2m_2} \ket{j_1m_1j_2m_2}.
\end{equation}

\paragraph{Remarks}
\begin{enumerate}
\item Due to the Condon-Shortley convention for the \SU-action, we have $C^{jm}_{j_1m_1j_2m_2} \in \mathbb{R}$.
\item The previous definition of the coefficients $C^{jm}_{j_1m_1j_2m_2}$ implicitly assumes that the following \textit{Clebsch-Gordan conditions} hold 
\begin{equation}\label{eq:CGcondition}
\begin{split}
j_1+j_2+j \in \mathbb{N} \\
|j_1-j_2|\leq j \leq j_1+j_2.
\end{split}
\end{equation}
When these are not fulfilled, we choose by convention, that $C^{jm}_{j_1m_1j_2m_2} = 0$.
\item The inequality in \eqref{eq:CGcondition} is also known as the \textit{triangle inequality} as it is also the condition for the existence of a triangle with edge lengths $j_1, j_2$ and $j$. It is equivalent to the more symmetric set of inequalities
\begin{equation}
\begin{split}
j \leq j_1+j_2 \\
j_1 \leq j+j_2 \\
j_2 \leq j_1+j.
\end{split}
\end{equation}
\item If $m \neq m_1+m_2$, then $C^{jm}_{j_1m_1j_2m_2} =0$.
\item The previous conditions do not exhaust all the zeros of $C^{jm}_{j_1m_1j_2m_2}$. For instance,$C^{2,0}_{2,0,1,0} = 0$ although the Clebsch-Gordan conditions are fulfilled and $m=m_1+m_2$.
\item Since the $\ket{jm}$ form an orthonormal basis, we have the following "orthogonality relation"
\begin{equation}\label{eq:CG orthogonalite}
\sum_{m_1 = -j_1}^{j_1} \sum_{m_2 = -j_2}^{j_2} C^{jm}_{j_1m_1j_2m_2} C^{j'm'}_{j_1m_1j_2m_2} = \delta_{j,j'} \delta_{m,m'}.
\end{equation}
\item Another consequence is the decomposition of products of Wigner matrices into sums, like  
\begin{multline}\label{eq:exercise 1}
D^{j_1}_{m_1n_1}(g) D^{j_2}_{m_2n_2} (g) \\
= \sum_{j \in \mathbb{N}/2} \sum_{m=-j}^j \sum_{m'=-j}^{j} C^{jm}_{j_1m_1j_2m_2} C^{jm'}_{j_1n_1j_2n_2} D^j_{mm'}(g).
\end{multline}
\begin{proof}
\begin{align*}
D^{j_1}_{m_1n_1}(g) D^{j_2}_{m_2n_2} (g) &= \bra{j_1m_1} g \ket{j_1n_1} \bra{j_2m_2} g \ket{j_2n_2} \\
&= \bra{j_1m_1 ; j_2m_2} g \ket{j_1n_1 ; j_2n_2} \\
&= \sum_{j \in \mathbb{N}/2} \sum_{m=-j}^j \sum_{j' \in \mathbb{N}/2} \sum_{m'=-j'}^{j'} \braket{j_1m_1j_2m_2}{jm} \\ & \hspace{2 cm} \times \bra{jm} g  \ket{j'm'} \braket{j'm'}{j_1n_1j_2n_2} \\
&= \sum_{j \in \mathbb{N}/2} \sum_{m=-j}^j \sum_{m'=-j}^{j} C^{jm}_{j_1m_1j_2m_2} C^{jm'}_{j_1n_1j_2n_2} D^j_{mm'}(g)
\end{align*}
\end{proof}
\end{enumerate} 
\noindent
The Clebsch-Gordan coefficients are numbers, but their definition is quite implicit. They can be generated with the following iterative procedure
\begin{enumerate}
\item The goal is to express $\ket{jm}$ in the product basis $\ket{j_1m_1;j_2m_2}$. The states $\ket{jm}$ can be represented as the nodes of a grid with vertical axis $m$ and horizontal axis $j$.
\item The decomposition of the highest state of the grid is $\ket{j_1+j_2,j_1+j_2} \cong \ket{j_1j_1; j_2j_2}$.
\item The state $\ket{j_1+j_2,j_1+j_2-1}$ is obtained by applying $J_-$ on both sides:
\begin{equation}
\sqrt{j_1+j_2}\ket{j_1+j_2,j_1+j_2-1} = \sqrt{j_1} \ket{j_1,j_1-1; j_2j_2} + \sqrt{j_2} \ket{j_1j_1; j_2,j_2-1}
\end{equation}
\item The state $\ket{j_1+j_2-1,j_1+j_2-1}$ is decomposed as a sum of $\ket{j_1,j_1-1; j_2j_2}$ and $\ket{j_1j_1; j_2,j_2-1}$ whose coefficients are uniquely determined by the fact that it is normalised and orthogonal to $\ket{j_1+j_2,j_1+j_2-1}$.
\item The two last steps can be repeated to reach any other state.
\end{enumerate}
The previous algorithm is good to understand the mechanics but for practical purposes it is simpler to refer directly to tables like the one of the Particle Data Group:
\begin{center}
\url{https://pdg.lbl.gov/2020/reviews/rpp2020-rev-clebsch-gordan-coefs.pdf}.
\end{center}
\marginpar{Other formulae are found in Varshalovich (\cite{varshalovich1987} p. 238), notably in terms of the hypergeometric function ${}_3F_2$.}It can also be convenient to resort to the following explicit formula:
\begin{multline}\label{eq:CG explicit}
C^{jm}_{j_1m_1j_2m_2} = \delta_{m,m_1+m_2} \sqrt{2j+1} \\
\times \sqrt{\frac{(j+m)!(j-m)!(-j+j_1+j_2)!(j-j_1+j_2)!(j+j_1-j_2)!}{(j+j_1+j_2+1)!(j_1+m_1)!(j_1-m_1)!(j_2+m_2)!(j_2-m_2)!}} \\
\times \sum_k \frac{(-1)^{k+j_2+m_2} (j+j_2+m_1-k)!(j_1-m_1+k)!}{(j-j_1+j_2-k)!(j+m-k)!k!(k+j_1-j_2-m)!}
\end{multline}
In the Wolfram Language, they are implemented as 
\begin{equation}
C^{j_3m_3}_{j_1m_1j_2m_2} = \texttt{ClebschGordan}[\{j_1, m_1\}, \{j_2, m_2\}, \{j_3, m_3\}] .
\end{equation}
In SageMath, the function is
\begin{equation}
C^{j_3m_3}_{j_1m_1j_2m_2} = \texttt{clebsch\_gordan}(j_1, j_2, j_3, m_1, m_2, m_3).
\end{equation}

\section{Invariant subspace}

Generally speaking, a tensor product of $n$ irreps can be decomposed into a direct sum
\begin{equation}
\bigotimes_{i=1}^n \mathcal{Q}_{j_i} \cong \bigoplus_{k=0}^{J} \left( \underbrace{\mathcal{Q}_{k} \oplus ... \oplus \mathcal{Q}_k}_{d_k \ \text{times}} \right),
\end{equation}
where $J\overset{\text{def}}=\sum_i j_i$ and $d_k$ is the degeneracy of the irrep $\mathcal{Q}_k$. Here, "decomposing" means "finding a bijective intertwiner between the two spaces". Concretely, such a decomposition is obtained by applying successively the decomposition of a product of only two irreps, as given by equation \eqref{eq:decomposition of 2}. It is usual to denote the operator $(J_i)_k \overset{\text{def}}= \mathbb{1} \otimes ... \otimes J_i \otimes ... \otimes \mathbb{1}$ corresponding to $J_i$ acting on the $k^{th}$ Hilbert space of the product. The three components $(J_1)_k$, $(J_2)_k$, $(J_3)_k$ form the vectorial operator $\vec{J}_k$.
\begin{physics}
In quantum gravity, such tensor spaces appear in the \textit{kinematical Hilbert space} $\mathcal{H}$. The description of the dynamics requires to impose constraints that select subspaces of $\mathcal{H}$. One important constraint is the \textit{Gauss constraint} which reduces $\mathcal{H}$ to its \SU-invariant subspace $\text{Inv}_{\SU} \mathcal{H}$ that we define below. 
\end{physics} 
We define the \SU-invariant subspace as
\begin{equation}
\text{Inv}_{\SU} \left( \bigotimes_{i=1}^n \mathcal{Q}_{j_i} \right) \overset{\text{def}}= \left\{ \psi \in \bigotimes_{i=1}^n \mathcal{Q}_{j_i} \mid \forall g \in \SU, \ g \cdot \psi = \psi \right\}.
\end{equation}
It can also be characterized by the action of the algebra:
\begin{equation}\label{eq:invariant J=0}
\text{Inv}_{\SU} \left( \bigotimes_{i=1}^n \mathcal{Q}_{j_i} \right) = \left\{ \psi \in \bigotimes_{i=1}^n \mathcal{Q}_{j_i} \mid \forall s \in \su, \ s \cdot \psi = 0 \right\}.
\end{equation}
From this, it is easy to see that
\begin{equation}
\text{Inv}_{\SU} \left( \bigotimes_{i=1}^n \mathcal{Q}_{j_i} \right) \cong  \underbrace{\mathcal{Q}_{0} \oplus ... \oplus \mathcal{Q}_0}_{d_0 \ \text{times}},
\end{equation}
where $\mathcal{Q}_0 \cong \mathbb{C}$ is the trivial representation.
Interestingly, we also have the following isomorphism:
\begin{equation}\label{eq:iso intertwiners}
\text{Inv}_{\SU} \left(\bigotimes_{i=1}^n \mathcal{Q}_{j_i} \right) \cong \text{Hom}_{\SU} \left(\bigotimes_{i=1}^n \mathcal{Q}_{j_i} , \mathcal{Q}_{0} \right),
\end{equation}
where the \ac{RHS} is the vector space of \SU-intertwiners between \\ $\bigotimes_{i=1}^n \mathcal{Q}_{j_i}$ and $\mathcal{Q}_{0}$. It is a particular case of equation \eqref{eq:hom-inv}.
\begin{proof}
Let $T:\bigotimes_{k=1}^n \mathcal{Q}_{j_k} \rightarrow \mathcal{Q}_0$ be an intertwiner. Since $T$ is a linear form, there exists $\psi_T \in \bigotimes_{k=1}^n \mathcal{Q}_{j_k}$ such that $T (\phi) = \braket{\phi}{\psi_T}$. Since $T$ is also an intertwiner, we have for all $\phi \in \bigotimes_{k=1}^n \mathcal{Q}_{j_k}$ and $u \in \SU$, $\braket{\phi}{u \cdot \psi_T} = \braket{u^\dagger \cdot \phi}{ \psi_T} = T(u^\dagger \cdot \phi) = u^\dagger \cdot T(\phi) = T(\phi) =  \braket{\phi}{\psi_T}$. So $u \cdot \psi_T = \psi_T$. We can check that the map $T \mapsto \psi_T$ is linear and bijective. QED.
\end{proof}

\paragraph{Orthogonal projector.} By definition, the orthogonal projector
\begin{equation}
P : \bigotimes_{i=1}^n \mathcal{Q}_{j_i} \to \text{Inv}_{\SU}\left(\bigotimes_{i=1}^n \mathcal{Q}_{j_i} \right)
\end{equation}
satisfies 
\begin{equation}
P^2 = P \quad \text{and} \quad P^\dagger = P.
\end{equation}
It is easy to show that 
\begin{equation}\label{eq:projector integral}
P = \int_{\SU} du \, \bigotimes_{k=1}^n D^{j_k}(u).
\end{equation}
If $\ket{\iota}$ is an orthonormal basis of $\text{Inv}_{\SU}\left(\bigotimes_{i=1}^n \mathcal{Q}_{j_i} \right)$, then $P$ can also be written as 
\begin{equation}\label{eq:projector dyad}
P = \sum_{\iota} \dyad{\iota}{\iota}.
\end{equation}

		\section{Wigner's \texorpdfstring{$3jm$}{3jm}-symbol}
		
We can decompose $\mathcal{Q}_{j_1} \otimes \mathcal{Q}_{j_2} \otimes \mathcal{Q}_{j_3}$ into a direct sum by applying equation \eqref{eq:decomposition of 2} twice, and first on the left tensor product: 
\begin{equation}
\mathcal{Q}_{j_1} \otimes \mathcal{Q}_{j_2} \otimes \mathcal{Q}_{j_3} \rightarrow \left( \bigoplus_{j_{12}} \mathcal{Q}_{j_{12}} \right) \otimes \mathcal{Q}_{j_3} \rightarrow  \bigoplus_{j_{12}=|j_1-j_2|}^{j_1+j_2} \bigoplus_{k=|j_{12}-j_3|}^{j_{12}+j_3} \mathcal{Q}_k
\end{equation}
Thus we construct an orthonormal basis of $\mathcal{Q}_{j_1} \otimes \mathcal{Q}_{j_2} \otimes \mathcal{Q}_{j_3}$ given by the states
\begin{multline}\label{eq:3CG translation}
\ket{(j_1j_2)j_3;j_{12}kn} \\
= \sum_{m_1,m_2,m_3,m_{12}} C^{j_{12}m_{12}}_{j_1m_1j_2m_2} C^{kn}_{j_{12}m_{12}j_3m_3} \bigotimes_{i=1}^3 \ket{j_i,m_i}, \\
\text{with} \ j_{12} \in \{|j_1-j_2|,...,j_1+j_2 \} \\
\text{and} \ k \in \{|j_{12}-j_3|,...,j_{12} + j_3 \} \\
\text{and} \ n\in \{-k,...,k\}.
\end{multline}
\begin{proof}
We apply twice equation \eqref{eq:decomposition of 2}, first to the left tensor product. Notice that applying it first on the right would instead build the states:
\begin{equation}
\ket{j_1(j_2j_3);j_{23}kn} = \sum_{m_1,m_2,m_3,m_{23}} C^{j_{23}m_{23}}_{j_2m_2j_3m_3} C^{kn}_{j_1m_1j_{23}m_{23}} \bigotimes_{i=1}^3 \ket{j_i,m_i}.
\end{equation}
\end{proof}
If the Clebsch-Gordan conditions \eqref{eq:CGcondition} are satisfied, then one can be show that
\begin{equation}
\text{Inv}_{\SU} \left( \mathcal{Q}_{j_1} \otimes \mathcal{Q}_{j_2} \otimes \mathcal{Q}_{j_3} \right) = \text{Span} \left\{ \ket{(j_1j_2)j_3;j_300} \right\},
\end{equation}
so that $\text{Inv}_{\SU} ( \mathcal{Q}_{j_1} \otimes \mathcal{Q}_{j_2} \otimes \mathcal{Q}_{j_3})$ is one dimensional. Otherwise
\begin{equation}
\text{Inv}_{\SU} ( \mathcal{Q}_{j_1} \otimes \mathcal{Q}_{j_2} \otimes \mathcal{Q}_{j_3}) = \{ 0 \}.
\end{equation}
\begin{proof} First we show the equivalence:
\begin{align*}
\vec{J}^2 \ket{(j_1j_2)j_3;j_{12}kn} = 0 \quad \Leftrightarrow \quad k=0.
\end{align*} 
We conclude using the characterisation \eqref{eq:invariant J=0}.
\end{proof}
Supposing the conditions are satisfied, an example of a unit vector in $\text{Inv}_{\SU} ( \mathcal{Q}_{j_1} \otimes \mathcal{Q}_{j_2} \otimes \mathcal{Q}_{j_3})$ is given by
\begin{equation}
\ket{0} \overset{\text{def}}=  \sum_{m_1,m_2,m_3} \begin{pmatrix}
j_1 & j_2 & j_3 \\
m_1 & m_2 & m_3
\end{pmatrix} \bigotimes_{k=1}^3 \ket{j_k,m_k},
\end{equation}
with the so-called \textit{Wigner's $3jm$-symbol}
\begin{equation}\label{eq:definition 3jm}
	\begin{pmatrix}
		j_1 & j_2 & j_3 \\
		m_1 & m_2 & m_3
	\end{pmatrix}
	\overset{\text{def}}= \frac{(-1)^{j_1 - j_2 - m_3}}{\sqrt{2 j_3 + 1}} C^{j_3,-m_3}_{j_1m_1j_2m_2}.
\end{equation}
This symbol was introduced by Wigner around 1940, and published in 1965 \cite{wigner1993}, to get real recoupling coefficients with the following symmetry properties:
\begin{equation}\label{eq:sym 3jm def}
\begin{pmatrix}
    j_1 & j_2 & j_3 \\
    m_1 & m_2 & m_3
\end{pmatrix}
=\begin{pmatrix}
    j_3 & j_1 & j_2 \\
    m_3 & m_1 & m_2
\end{pmatrix}
=\begin{pmatrix}
    j_2 & j_3 & j_1 \\
    m_2 & m_3 & m_1
\end{pmatrix},
\end{equation}
and
\begin{equation}\label{eq:sym 3jm}
	\begin{split}
		\begin{pmatrix}
			j_1 & j_2 & j_3 \\
			m_1 & m_2 & m_3
		\end{pmatrix}
		&= (-1)^{j_1+j_2+j_3}
		\begin{pmatrix}
			j_2 & j_1 & j_3 \\
			m_2 & m_1 & m_3
		\end{pmatrix} \\
		&= (-1)^{j_1+j_2+j_3}
		\begin{pmatrix}
			j_1 & j_2 & j_3 \\
			-m_1 & -m_2 & -m_3
		\end{pmatrix}.
	\end{split}
\end{equation}
In the Wolfram Language, they are given by
\begin{equation}
\texttt{ThreeJSymbol}[\{j_1, m_1 \}, \{j_2, m_2 \}, \{j_3, m_3 \}] = \begin{pmatrix}
    j_1 & j_2 & j_3 \\
    m_1 & m_2 & m_3
\end{pmatrix}.
\end{equation}
In SageMath, the function is
\begin{equation}
\texttt{wigner\_3j}(j_1, j_2, j_3, m_1, m_2, m_3) = \begin{pmatrix}
    j_1 & j_2 & j_3 \\
    m_1 & m_2 & m_3
\end{pmatrix}.
\end{equation}
\begin{proof} In $\text{Inv}_\SU \left( \mathcal{Q}_{j_1} \otimes \mathcal{Q}_{j_2} \otimes \mathcal{Q}_{j_3} \right)$, all vectors are proportional to 
\begin{align*}
\ket{(j_1j_2)j_3;j_300} &= \sum_{m_1,m_2,m_3,m} C^{00}_{jmj_3m_3} C^{jm}_{j_1m_1;j_2m_2}  \ket{j_1m_1;j_2m_2;j_3m_3} \\
&=  \sum_{m_1,m_2,m_3,m} \delta_{m,-m_3} \delta_{j,j_3} \frac{(-1)^{j_3 + m_3}}{\sqrt{2j_3 +1}} C^{jm}_{j_1m_1;j_2m_2} \\ &\hspace{4 cm} \times  \ket{j_1m_1;j_2m_2;j_3m_3} \\
&= \sum_{m_1,m_2,m_3} \frac{(-1)^{j_3 + m_3}}{\sqrt{2j_3 +1}} C^{j_3, -m_3}_{j_1m_1;j_2m_2}  \ket{j_1m_1;j_2m_2;j_3m_3}.
\end{align*}
The proportionality factor is chosen to be $(-1)^{j_1 - j_2 + j_3}$ to match the reality and the symmetry requirements. 
\end{proof}

\paragraph{Remarks.}
\begin{enumerate}
\item The orthogonality relation \eqref{eq:CG orthogonalite} becomes
\begin{equation}
\sum_{jm} (2j+1) \begin{pmatrix}
j_1 & j_2 & j \\ m_1 & m_2 & m
\end{pmatrix}
\begin{pmatrix}
j_1 & j_2 & j \\ m_1' & m_2' & m
\end{pmatrix} = \delta_{m_1 m_1'} \delta_{m_2 m_2'}
\end{equation}
\begin{equation}\label{eq:ortho 3jm sum mm}
\sum_{m_1m_2} (2j+1) \begin{pmatrix}
j_1 & j_2 & j \\ m_1 & m_2 & m
\end{pmatrix}
\begin{pmatrix}
j_1 & j_2 & j' \\ m_1 & m_2 & m'
\end{pmatrix} = \delta_{jj'} \delta_{m m'}
\end{equation}
\item Equating \eqref{eq:projector integral} and \eqref{eq:projector dyad} in the magnetic basis, we find
\begin{multline}\label{eq:projector 3D}
\int_{\SU} D^{j_1}_{m_1n_1} (u) D^{j_2}_{m_2n_2}(u) D^{j_3}_{m_3n_3}(u) \dd u \\
= \begin{pmatrix}
j_1 & j_2 & j_3 \\
m_1 & m_2 & m_3
\end{pmatrix}
\begin{pmatrix}
j_1 & j_2 & j_3 \\
n_1 & n_2 & n_3
\end{pmatrix}.
\end{multline}
\end{enumerate}

\section{Wigner's \texorpdfstring{$4jm$}{4jm}-symbol}

Similarly to the previous section, we can decompose $\mathcal{Q}_{j_1} \otimes \mathcal{Q}_{j_2} \otimes \mathcal{Q}_{j_3} \otimes \mathcal{Q}_{j_4}$ into a direct sum by applying equation \eqref{eq:decomposition of 2} successively. We get
\begin{equation}
\mathcal{Q}_{j_1} \otimes \mathcal{Q}_{j_2} \otimes \mathcal{Q}_{j_3} \otimes \mathcal{Q}_{j_4} \cong \bigoplus_{j_{12}=|j_1-j_2|}^{j_1+j_2} \bigoplus_{k=|j_{12}-j_3|}^{j_{12}+j_3} \bigoplus_{l=|k-j_4|}^{k+j_4} \mathcal{Q}_l
\end{equation}
In particular, we can see that 
\begin{equation}
\text{Inv}_{\SU}\left(\bigotimes_{i=1}^4 \mathcal{Q}_{j_i} \right) \cong \underbrace{\mathcal{Q}_0 \oplus ... \oplus \mathcal{Q}_0}_{d_0 \ \text{times}}
\end{equation}
with $d_0 =\min (j_1+j_2,j_3+j_4) - \max (|j_1-j_2|,|j_3-j_4|) + 1$.
An orthonormal basis of $\text{Inv}_{\SU}\left(\bigotimes_{i=1}^4 \mathcal{Q}_{j_i} \right) $ is given by
\begin{multline}\label{eq:intertwiner 4jm}
\ket{j}_{12} = \sqrt{2j+1} \sum_{\substack{ m_1,m_2,\\ m_3,m_4}}
\begin{pmatrix}
j_1 & j_2 & j_3 & j_4 \\
m_1 & m_2 & m_3 & m_4
\end{pmatrix}^{(j)} \bigotimes_{k=1}^4 \ket{j_k,m_k},
\end{multline}
with $j \in \{ \max (|j_1-j_2|,|j_3-j_4|) ,...,\min (j_1+j_2,j_3+j_4) \}$, and
\begin{multline}
\begin{pmatrix}
j_1 & j_2 & j_3 & j_4 \\
m_1 & m_2 & m_3 & m_4
\end{pmatrix}^{(j)} \\
\overset{\text{def}}= 
\sum_{m} (-1)^{j - m}
\begin{pmatrix}
j_1 & j_2 & j \\
m_1 & m_2 & m
\end{pmatrix}
\begin{pmatrix}
j & j_3 & j_4 \\
-m & m_3 & m_4
\end{pmatrix}
\end{multline}
\begin{proof}
First, we construct an orthonormal basis of $\mathcal{Q}_{j_1} \otimes \mathcal{Q}_{j_2} \otimes \mathcal{Q}_{j_3} \otimes \mathcal{Q}_{j_4}$ given by the states 
\begin{multline}
\ket{((j_1j_2)j_3)j_4;jklm} = \sum_{m_1,m_2,m_3,\mu,n,m_4} C^{j\mu}_{j_1m_1j_2m_2} C^{kn}_{j\mu j_3m_3}  C^{lm}_{knj_4m_4} \\ \hfill \times \bigotimes_{i=1}^4 \ket{j_i,m_i}, \\
\hfill \text{with} \ j \in \{|j_1-j_2|,...,j_1+j_2 \} \quad \text{and} \ k \in \{|j-j_3|,...,j + j_3 \} \\
\text{and} \ l \in \{|k-j_4|,...,k + j_4 \} \quad \text{and} \ m\in \{-l,...,l\}.
\end{multline}
$\text{Inv}_\SU \left(\bigotimes_{i=1}^4 \mathcal{Q}_{j_i} \right)$ is spanned by the vectors with $l=0$. Similarly to the case $n=3$, we compute
\begin{multline}
\ket{((j_1j_2)j_3)j_4;jj_400} 
= (-1)^{j_4 + j_1 - j_2  - j_3} \sum_{m_1,m_2,m_3,m_4} \sqrt{2j + 1} \\ \times \begin{pmatrix}
j_1 & j_2 & j_3 & j_4 \\
m_1 & m_2 & m_3 & m_4
\end{pmatrix}^{(j)}  \bigotimes_{i=1}^4 \ket{j_i,m_i}.
\end{multline}
\end{proof}
This basis has the interesting property that it diagonalises $(\vec{J}_1 + \vec{J}_2)^2$:
\begin{equation}
(\vec{J}_1 + \vec{J}_2)^2 \ket{j}_{12} = j(j+1) \ket{j}_{12},
\end{equation}
and this explains the notation "$12$" in index.
The $4jm$-symbol also satisfy orthogonality relations:
\begin{multline}
 \sum_{m_1,m_2,m_3} \Wfour{j_1}{j_2}{j_3}{j_4}{m_1}{m_2}{m_3}{m_4}{j_{12}} \Wfour{j_1}{j_2}{j_3}{l_4}{m_1}{m_2}{m_3}{n_4}{l_{12}} \\
 = 
 \frac{\delta_{j_{12}l_{12}}}{2 j_{12} +1} \frac{\delta_{j_{4}l_{4}} \delta_{m_4n_4}}{2 j_{4} + 1}.\label{eq:orto4j}
\end{multline}
Finally we can show, similarly to equation \eqref{eq:projector 3D}, that
\begin{multline}\label{eq:4 projected D}
\int_{\SU} D^{j_1}_{m_1n_1} (u) D^{j_2}_{m_2n_2}(u) D^{j_3}_{m_3n_3}(u) D^{j_4}_{m_4n_4} (u) du \\
=  \sum_{j}  (2j+1)
\begin{pmatrix}
j_1 & j_2 & j_3 & j_4 \\
m_1 & m_2 & m_3 	& m_4
\end{pmatrix}^{(j)}
\begin{pmatrix}
j_1 & j_2 & j_3 & j_4\\
n_1 & n_2 & n_3 & n_4
\end{pmatrix}^{(j)}.
\end{multline}

		\section{Wigner's \texorpdfstring{6j}{6j}-symbol}
		
In the previous section, we have exhibited an orthonormal basis for $\text{Inv}_{\SU}\left(\bigotimes_{i=1}^4 \mathcal{Q}_{j_i} \right)$. It is built from one possible decomposition of $\bigotimes_{i=1}^4 \mathcal{Q}_{j_i}$ into irreps. Another possible decomposition leads to another basis
\begin{equation}
\ket{j}_{23} = \sum_{\substack{m_1,m_2, \\ m_3,m_4}} \sqrt{2j +1} \begin{pmatrix}
j_4 & j_1 & j_2 & j_3 \\
m_4 & m_1 & m_2 & m_3
\end{pmatrix}^{(j)} \bigotimes_{i=1}^4 \ket{j_i,m_i} ,
\end{equation}
that diagonalises $(\vec{J}_2 + \vec{J}_3)^2$.
\begin{proof} From
\begin{equation}\label{3CG translation2}
\ket{(j_1(j_2j_3))j_4;jklm} = \sum_{m_1,m_2,m_3,\mu,n,m_4} C^{j\mu}_{j_2m_2j_3m_3} C^{kn}_{j_1m_1j\mu}  C^{lm}_{knj_4m_4} \bigotimes_{i=1}^4 \ket{j_i,m_i},
\end{equation}
we show that
\begin{multline}
\ket{(j_1(j_2j_3))j_4;j_{23}j_400} 
= (-1)^{j_1+j_2-j_3+j_4}  \sum_{m_1,m_2,m_3,m_4} \sqrt{2j_{23} +1}  \\ \begin{pmatrix}
j_4 & j_1 & j_2 & j_3 \\
m_4 & m_1 & m_2 & m_3
\end{pmatrix}^{(j_{23})} \bigotimes_{i=1}^4 \ket{j_i,m_i} .
\end{multline}
\end{proof}
The change of basis is given by
\begin{multline}
{}_{12}\braket{j}{k}_{23} \\ = \sqrt{2j+1}  \sqrt{2k +1} (-1)^{j_1 + j_2 + j_3 - j_4 - 2j - 2k} \begin{Bmatrix}
j_1 & j_2 & j \\
j_3 & j_4 & k
\end{Bmatrix}
\end{multline}
where we have defined a new symbol:
\begin{multline}
 \begin{Bmatrix}
    j_1 & j_2 & j_3\\
    j_4 & j_5 & j_6
  \end{Bmatrix} \overset{\text{def}}=
   \sum_{m_1, \dots, m_6} (-1)^{\sum_{i=1}^6 (j_i -m_i)} \\
   \times 
  \begin{pmatrix}
    j_1 & j_2 & j_3\\
    -m_1 & -m_2 & -m_3
  \end{pmatrix} 
  \begin{pmatrix}
    j_1 & j_5 & j_6\\
    m_1 & -m_5 & m_6
  \end{pmatrix} \\
  \times
  \begin{pmatrix}
    j_4 & j_2 & j_6\\
    m_4 & m_2 & -m_6
  \end{pmatrix}
  \begin{pmatrix}
    j_3 & j_4 & j_5\\
    m_3 & -m_4 & m_5
  \end{pmatrix}
\end{multline}
In the Wolfram Language, it is returned by the function 
\begin{equation}
\texttt{SixJSymbol}[\{j_1, j_2, j_3\}, \{j_4, j_5, j_6\}] =  \begin{Bmatrix}
    j_1 & j_2 & j_3\\
    j_4 & j_5 & j_6
  \end{Bmatrix}.
\end{equation}
In SageMath, it is given by
\begin{equation}
\texttt{wigner\_6j}(j_1, j_2, j_3, j_4, j_5, j_6) =  \begin{Bmatrix}
    j_1 & j_2 & j_3\\
    j_4 & j_5 & j_6
  \end{Bmatrix}.
\end{equation}
These symbols are invariant under any permutation of the columns:
\begin{equation}
\label{eq:6j_sym_1}
 \begin{Bmatrix}
    j_1 & j_2 & j_3\\
    j_4 & j_5 & j_6
  \end{Bmatrix}
  = \begin{Bmatrix}
    j_1 & j_3 & j_2\\
    j_4 & j_6 & j_5
  \end{Bmatrix}
  = \begin{Bmatrix}
    j_3 & j_2 & j_1\\
    j_6 & j_5 & j_4
  \end{Bmatrix}
  = \begin{Bmatrix}
    j_2 & j_1 & j_3\\
    j_5 & j_4 & j_6
  \end{Bmatrix}.
\end{equation}
They are also symmetric if the upper and lower arguments are exchanged in any two columns:
\begin{equation}
\label{eq:6j_sym_2}
 \begin{Bmatrix}
    j_1 & j_2 & j_3\\
    j_4 & j_5 & j_6
  \end{Bmatrix}
  = \begin{Bmatrix}
    j_1 & j_5 & j_6\\
    j_4 & j_2 & j_3
  \end{Bmatrix}
  = \begin{Bmatrix}
    j_4 & j_2 & j_6\\
    j_1 & j_5 & j_3
  \end{Bmatrix}
  = \begin{Bmatrix}
    j_1 & j_5 & j_3\\
    j_4 & j_2 & j_6
  \end{Bmatrix}.
\end{equation}
Similarly one can define the symbols 9j and 15j (see next section).
\begin{physics}
The $\{6j\}$-symbol appeared in quantum gravity when Ponzano and Regge realised that the $\{6j\}$-symbol approximates the action of general relativity in the semi-classical limit \cite{ponzano1968}. More precisely, consider the tetrahedron depicted in the diagram of equation \eqref{eq:graph_6j}, where the labels $j_i$ are understood as the edge lengths. Denoting $V$ the volume of such a tetrahedron, one can show the following asymptotic limit when $\lambda \to \infty$:
\begin{equation}
\label{eq:PonzanoRegge}
    \begin{Bmatrix}
    \lambda j_1 & \lambda j_2 & \lambda j_3 \\
    \lambda j_4 & \lambda j_5 & \lambda j_6
    \end{Bmatrix} \sim \frac{1}{4\sqrt{3 \pi \lambda^3 V}} \left( e^{i S} + e^{-i S} \right)
\end{equation}
with the action
\begin{equation}
    S \overset{\text{def}}= \sum_i \left( \lambda j_i + \frac{1}{2} \right) \xi_i + \frac{\pi}{4} 
\end{equation}
with $\xi_i$ the exterior dihedral angle along the edge $i$ \cite{roberts1999}.
Letting aside the constant $\frac{\pi}{4}$, $S$ is the so-called \textit{Regge action} of the tetrahedron, which is a discrete euclidean 3D version of the Einstein-Hilbert action. This result was an important source of inspiration for later development of spin-foams.
\end{physics}

\section{Yutsis diagrams}
\label{sec:graphical calculus}

The recoupling theory of \SU can be nicely implemented graphically. The underlying philosophy of it is to take advantage of the two dimensions offered by our sheets of paper and our blackboards to literally \textit{draw} our calculations, rather than restricting oneself to the usual one-dimensional lines of calculations. If done properly, the method can help to understand the structure of analytical expressions and make computations faster.

Of course, the first principle of graphical calculus is that there should be a one-to-one correspondence between analytical expressions and diagrams. There exist various conventions for this correspondence in the literature, so we have chosen to stick to the original one, introduced by Yutsis in 1960 \cite{yutsis1962}. The same convention is chosen by Varshalovich (\cite{varshalovich1987}, Chap. 11) and it is quite popular in the quantum gravity literature \cite{sarno2018}.

\paragraph{Definitions.}
The basic object of this graphical calculus is the $3$-valent node, that represents the Wigner's $3jm$ symbol:
\begin{equation}\label{eq:graphical 3jm}
 \Wthree{j_1}{j_2}{j_3}{m_1}{m_2}{m_3}  = \begin{array}{c}
  \begin{overpic}[scale = 0.6]{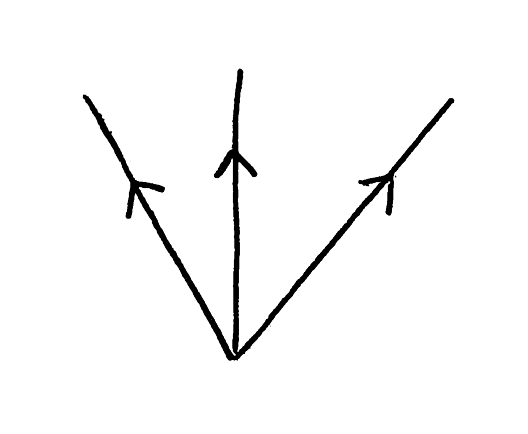}
 \put (-3,67) {$j_1m_1$}
  \put (37,73) {$j_2m_2$}
   \put (81,70) {$j_3m_3$}
   \put (45,5) {$-$}
\end{overpic}
 \end{array}
 = \begin{array}{c}
  \begin{overpic}[scale = 0.6]{figures/3CG-out.png}
 \put (-3,67) {$j_1m_1$}
  \put (37,73) {$j_3m_3$}
   \put (81,70) {$j_2m_2$}
   \put (45,5) {$+$}
\end{overpic}
 \end{array}.
\end{equation}
Remarks:
\begin{enumerate}
\setlength\itemsep{0 \baselineskip}
\item The signs $+/-$ on the nodes indicate the sense of rotation (anticlockwise/clockwise) in which the spins must be read. To alleviate notations we decide not to write them in the following by choosing conventionally that the default sign of the nodes is minus, if not otherwise specified.
\item We implicitly assume that the Clebsch-Gordan conditions are satisfied everywhere.
\item The symmetry properties \eqref{eq:sym 3jm def} are naturally implemented in the diagram. It guarantees the one-to-one correspondence between the analytical expression and the diagram.
\item Only the topology of the diagram matters, which means that all topological deformations are allowed. This property is also called planar isotopy.
\begin{equation}
\begin{array}{c}
  \begin{overpic}[scale = 0.6]{figures/3CG-out.png}
	\put (-3,67) {$j_1m_1$}
  	\put (37,73) {$j_2m_2$}
   	\put (81,70) {$j_3m_3$}
\end{overpic}
 \end{array} 
 = \begin{array}{c}
  \begin{overpic}[scale = 0.6]{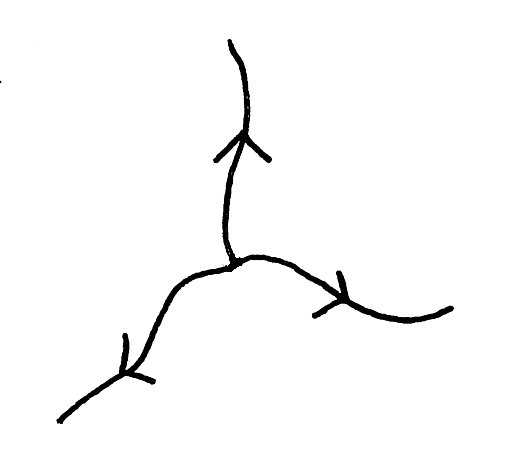}
 \put (0,-5) {$j_1m_1$}
  \put (35,85) {$j_2m_2$}
   \put (90,25) {$j_3m_3$}
\end{overpic}
 \end{array} 
 = \begin{array}{c}
  \begin{overpic}[scale = 0.6]{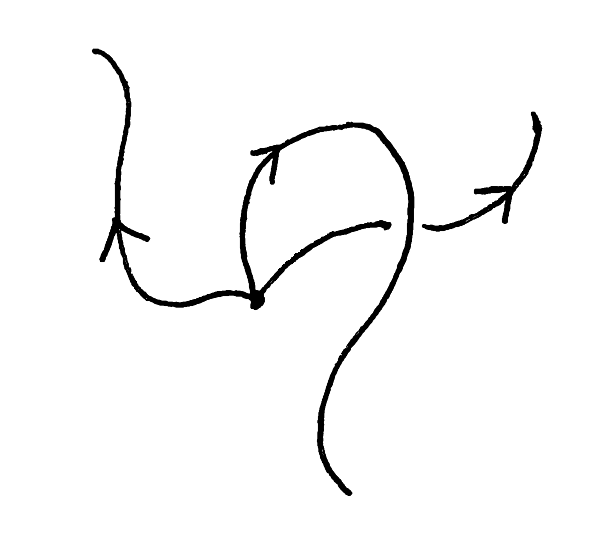}
 \put (5,85) {$j_1m_1$}
  \put (55,-5) {$j_2m_2$}
   \put (79,77) {$j_3m_3$}
\end{overpic}
 \end{array}
\end{equation}
This principle of planar isotopy is a strong principle of graphical calculus, that will hold for any other diagram constructed later.
\end{enumerate}
Then we can define graphically the two basic operations of algebra: multiplication and summation. Multiplication is implemented simply by juxtaposition of diagrams:
\begin{multline}
\begin{array}{c}
  \begin{overpic}[scale = 0.6]{figures/3CG-out.png}
	\put (-3,67) {$j_1m_1$}
  	\put (37,73) {$j_2m_2$}
   	\put (80,70) {$j_3m_3$}
\end{overpic}
 \end{array} \quad \begin{array}{c}
  \begin{overpic}[scale = 0.6]{figures/3CG-out.png}
	\put (-3,67) {$j_4m_4$}
  	\put (37,73) {$j_5m_5$}
   	\put (80,70) {$j_6m_6$}
\end{overpic}
 \end{array} \\
 = \Wthree{j_1}{j_2}{j_3}{m_1}{m_2}{m_3} \Wthree{j_4}{j_5}{j_6}{m_4}{m_5}{m_6} 
\end{multline}
To define summation, we shall first reveal the meaning of the arrows on the wires. So far they were all outgoing, but we now declare that an \textit{ingoing} orientation corresponds to taking \textit{minus} the magnetic index. For instance
\begin{equation}
\begin{array}{c}
  \begin{overpic}[scale = 0.6]{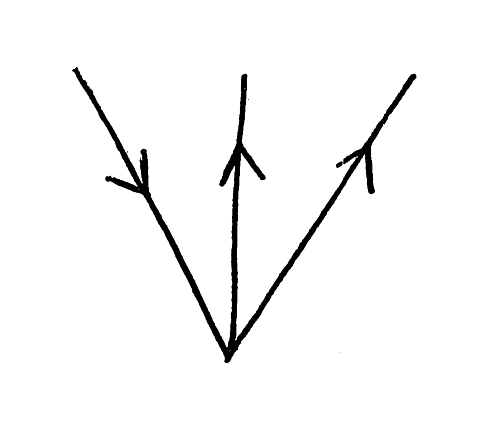}
	\put (-3,72) {$j_1m_1$}
  	\put (37,74) {$j_2m_2$}
   	\put (80,74) {$j_3m_3$}
\end{overpic}
 \end{array}= \Wthree{j_1}{j_2}{j_3}{-m_1}{m_2}{m_3}\,.
\end{equation}
Now we define the gluing of two external wires with the same label $jm$, but opposite directions, as
the sum over $m$ (from $-j$ to $j$), with the additional factor $(-1)^{j-m}$ in the summand, like:
\begin{equation}\label{eq:def summation}
\begin{array}{c}
  \begin{overpic}[scale = 0.6]{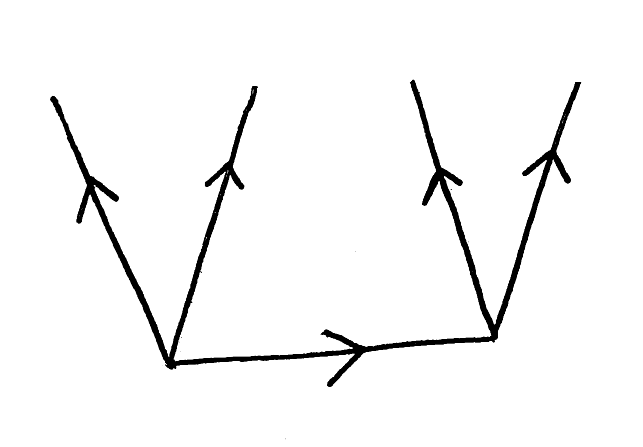}
 \put (-7,59) {$j_1m_1$}
  \put (25,59) {$j_2m_2$}
   \put (59,60) {$j_3m_3$}
   \put (91,60) {$j_4m_4$}
   \put (60,5) {$j$}
\end{overpic}
 \end{array}
=
\sum_{m=-j}^j (-1)^{j-m}
\begin{array}{c}
  \begin{overpic}[scale = 0.6]{figures/3CG-out.png}
 \put (-3,70) {$j_1m_1$}
  \put (35,75) {$j_2m_2$}
   \put (85,68) {$jm$}
\end{overpic}
 \end{array}
 \begin{array}{c}
  \begin{overpic}[scale = 0.6]{figures/3CG-outin.png}
 \put (5,75) {$jm$}
  \put (37,73) {$j_3m_3$}
   \put (84,74) {$j_4m_4$}
\end{overpic}
 \end{array}
\end{equation}
On the \ac{RHS}, we recognise the definition of the $4jm$-symbol:
\begin{equation}
\begin{array}{c}
  \begin{overpic}[scale = 0.6]{figures/4CG.png}
	\put (-7,59) {$j_1m_1$}
  	\put (25,59) {$j_2m_2$}
   	\put (59,60) {$j_3m_3$}
   	\put (91,60) {$j_4m_4$}
   	\put (60,5) {$j$}
\end{overpic}
 \end{array} = 
\Wfour{j_1}{j_2}{j_3}{j_4}{m_1}{m_2}{m_3}{m_4}{j} 
\label{eq:CG4}
\end{equation}
The wire between the two nodes, whose magnetic index is summed over, is called an \textit{internal} wire. Reversing the arrow of an internal wire gives an overall phase:
\begin{equation}
\begin{array}{c}
  \begin{overpic}[scale = 0.6]{figures/4CG.png}
	\put (-10,59) {$j_1m_1$}
  	\put (25,59) {$j_2m_2$}
   	\put (59,60) {$j_3m_3$}
   	\put (91,60) {$j_4m_4$}
   	\put (60,5) {$j$}
\end{overpic}
 \end{array}
 =(-1)^{2j}
\begin{array}{c}
  \begin{overpic}[scale = 0.6]{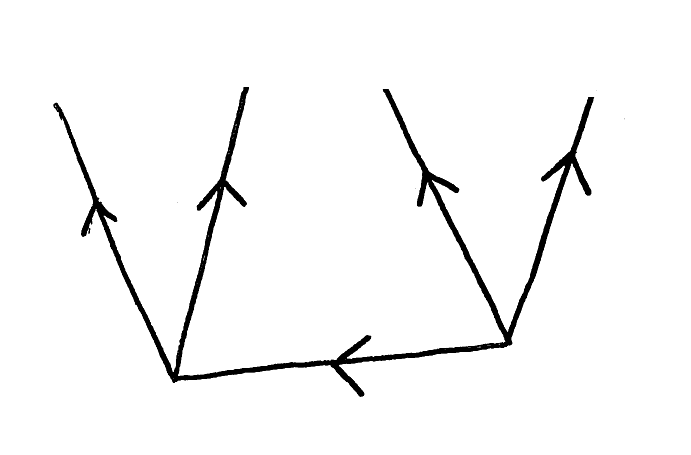}
	\put (-12,59) {$j_1m_1$}
  	\put (20,59) {$j_2m_2$}
   	\put (54,60) {$j_3m_3$}
   	\put (86,60) {$j_4m_4$}
  	\put (60,5) {$j$}
\end{overpic}
 \end{array}.
\nonumber
\end{equation} 
Now we come to an even simpler objet, the single wire:
\begin{equation}
\begin{array}{c}
  \begin{overpic}[scale = 0.6]{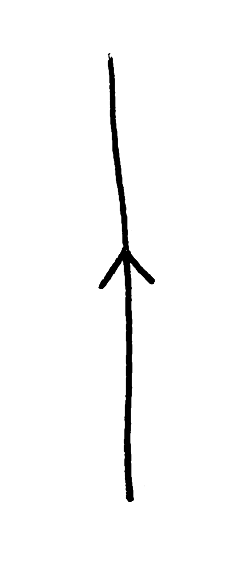}
 \put (15,95) {$j_1m_1$}
   \put (15,2) {$j_2m_2$}
\end{overpic}
 \end{array}= (-1)^{j_2-m_2} \delta_{j_1j_2} \delta_{m_1m_2} \qquad \text{or} \qquad
\begin{array}{c}
  \begin{overpic}[scale = 0.6]{figures/up.png}
 \put (10,60) {$j$}
  \put (15,95) {$m$}
   \put (20,5) {$n$}
\end{overpic}
 \end{array}= (-1)^{j-n} \delta_{mn} \,.
\end{equation}
We can apply the rule of summation to compute its trace:
\begin{equation}
 \begin{array}{c}
  \begin{overpic}[scale = 0.6]{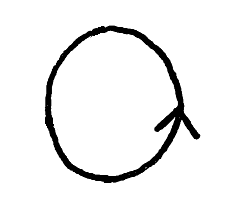}
 \put (5,60) {$j$}
\end{overpic}
 \end{array} = 2j+1.
\end{equation}
This kind of diagram with no external wires encodes numbers. All magnetic indices are summed over, so it is only a function of the internal spins $j_i$ and we call it an \textit{invariant function}. On the contrary, diagrams with external wires encode tensors with one magnetic index on each free end.

\paragraph{Lemmas.}
Most of the equations that were written previously in this chapter can now be drawn. As examples, the following lemmas can be checked:
\begin{enumerate}
\item The symmetry \eqref{eq:sym 3jm} shows the effect of changing the sign of the node
\begin{equation}
\begin{array}{c}
  \begin{overpic}[scale = 0.6]{figures/3CG-out.png}
	\put (-3,67) {$j_1m_1$}
  	\put (37,73) {$j_2m_2$}
   	\put (80,70) {$j_3m_3$}
   	\put (45,5) {$-$}
\end{overpic}
 \end{array} 
 = (-1)^{j_1+j_2+j_3} \begin{array}{c}
  \begin{overpic}[scale = 0.6]{figures/3CG-out.png}
	\put (-3,67) {$j_1m_1$}
  	\put (37,73) {$j_2m_2$}
   	\put (80,70) {$j_3m_3$}
  	\put (45,5) {$+$}
\end{overpic}
 \end{array} 
\end{equation}
and reversing the arrows
\begin{equation}
\begin{array}{c}
  \begin{overpic}[scale = 0.6]{figures/3CG-out.png}
	\put (-3,67) {$j_1m_1$}
  	\put (37,73) {$j_2m_2$}
   	\put (80,70) {$j_3m_3$}
\end{overpic}
 \end{array}
=(-1)^{j_1+j_2+j_3} \begin{array}{c}
  \begin{overpic}[scale = 0.6]{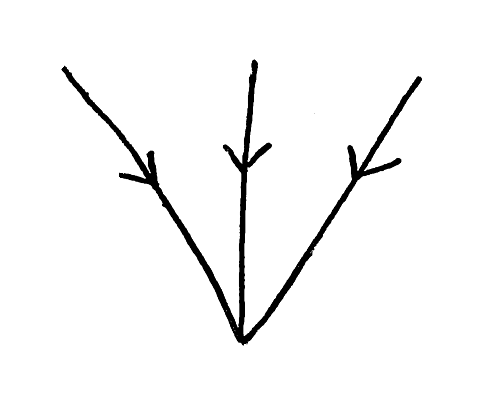}
	\put (-3,70) {$j_1m_1$}
  	\put (37,73) {$j_2m_2$}
   	\put (80,70) {$j_3m_3$}
\end{overpic}
 \end{array} \,.
\label{eq:keep}
\end{equation}
\item The orthogonality relation \eqref{eq:ortho 3jm sum mm} becomes
\begin{equation} 
\begin{array}{c}
  \begin{overpic}[scale = 0.6]{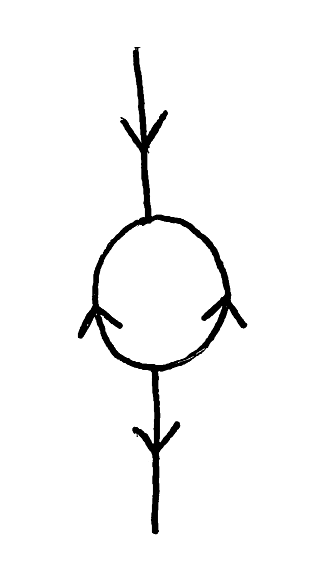}
 \put (15,100) {$j_1m_1$} 
   \put (15,-5) {$j_2m_2$}
  \put (0,50) {$j_3$}
    \put (45,50) {$j_4$}
\end{overpic}
 \end{array}
 \begin{array}{c}
  \begin{overpic}[scale = 0.6]{figures/circle.png}
 \put (75,70) {$j_1$}
\end{overpic}
 \end{array}
 = (-1)^{j_1+j_3+j_4} \begin{array}{c}
  \begin{overpic}[scale = 0.6]{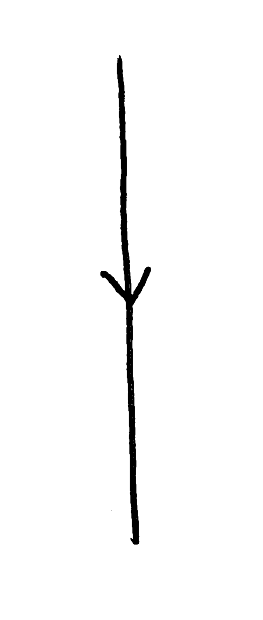}
  \put (5,97) {$j_1m_1$}
   \put (9,-1) {$j_2m_2$}
\end{overpic}
 \end{array}
\label{eq:Rule}
\end{equation}
It gives a way to remove internal loops from diagrams.
\item Taking the trace in the previous relation leads to the \textit{$\Theta$-graph}: 
\begin{equation}
\begin{array}{c}
  \begin{overpic}[scale = 0.6]{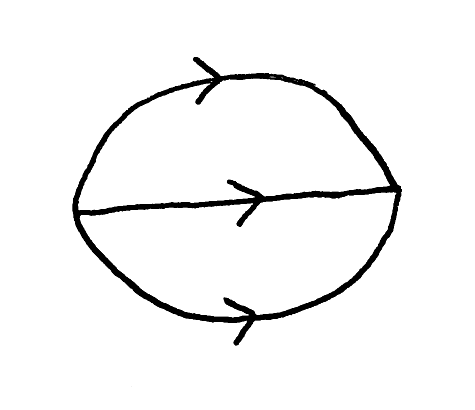}
 \put (20,70) {$j_1$}
  \put (55,50) {$j_2$}
   \put (75,10) {$j_3$}
\end{overpic}
 \end{array}=(-1)^{j_1+j_2+j_3}.
\end{equation}
\item Equation \eqref{eq:orto4j} implies
\begin{equation}
\begin{array}{c}
  \begin{overpic}[scale = 0.6]{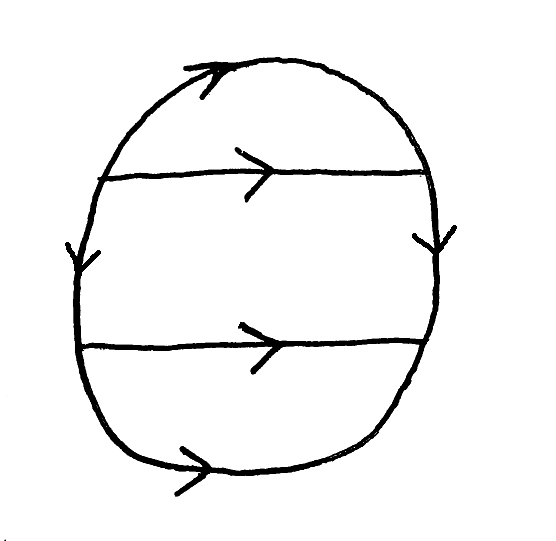}
 \put (60,90) {$j_1$}
  \put (30,55) {$j_2$}
   \put (55,40) {$j_3$}
   \put (70,10) {$j_4$}
   \put (5,40) {$i$}
   \put (85,40) {$k$}
\end{overpic}
 \end{array} =\frac{\delta_{ik}}{2i+1}.
\label{eq:thetaeq}
\end{equation} 
\end{enumerate}

\paragraph{Invariant functions.}
One nice thing about this graphical calculus is that it makes it easy to represent and to remember the Wigner 6j-symbol:
\begin{equation}
\label{eq:graph_6j}
\begin{Bmatrix}
    j_1 & j_2 & j_3\\
    j_4 & j_5 & j_6
  \end{Bmatrix} 
  = \begin{array}{c}
  \begin{overpic}[scale = 0.7]{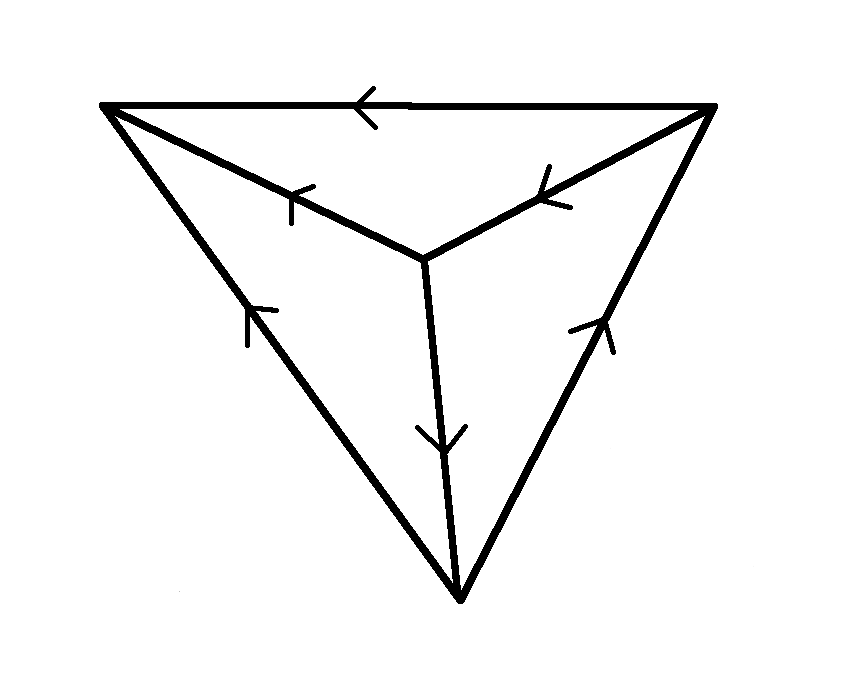}
 \put (50,75) {$j_1$}
  \put (35,50) {$j_2$}
   \put (25,30) {$j_3$}
   \put (54,40) {$j_4$}
   \put (70,30) {$j_5$}
   \put (68,54) {$j_6$}
\end{overpic}
 \end{array}
\end{equation}
The symmetries \eqref{eq:6j_sym_1} and \eqref{eq:6j_sym_2} are built-in! We can define other invariant functions in the same spirit, like the 9j-symbol:
\begin{equation}\label{eq:9j}
\begin{Bmatrix}
    j_1 & j_2 & j_3\\
    l_1 & l_2 & l_3\\
    k_1 & k_2 & k_3
  \end{Bmatrix} 
  = \begin{array}{c}
  \begin{overpic}[scale = 0.6]{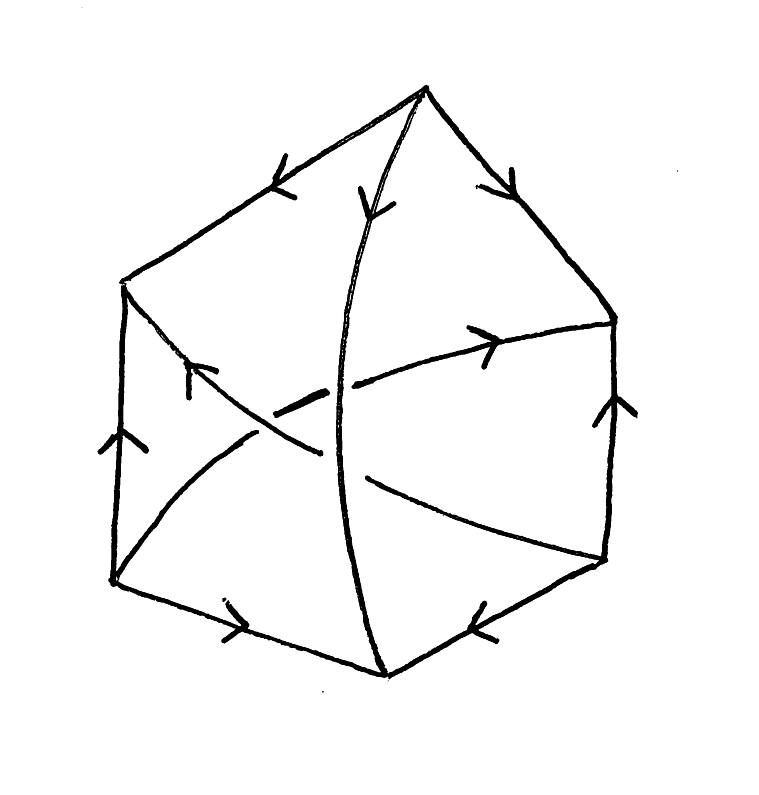}
 \put (25,80) {$j_1$}
  \put (47,65) {$j_3$}
   \put (73,73) {$j_2$}
   \put (60,38) {$k_1$}
   \put (80,40) {$k_2$}
   \put (65,15) {$k_3$}
   \put (25,13) {$l_3$}
   \put (25,33) {$l_2$}
   \put (5,45) {$l_1$}
   \put (7,65) {$+$}
   \put (80,60) {$+$}
   \put (45,7) {$+$}
\end{overpic}
 \end{array}
\end{equation}
This definition is the one given by Yutsis \cite{yutsis1962}. Notice that we could have also defined the 9j-symbol with another underlying graph topology, like:
\begin{equation}
\begin{array}{c}
  \begin{overpic}[scale = 0.6]{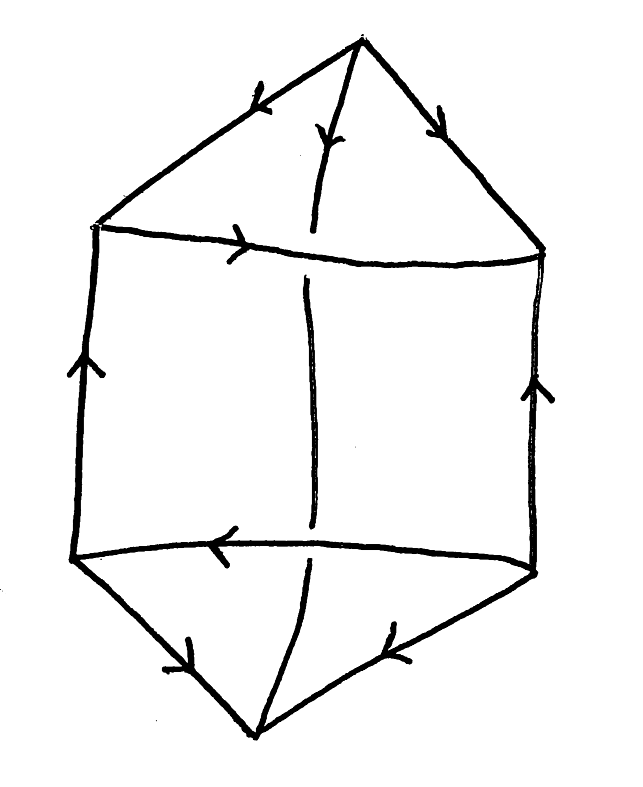}
 \put (25,90) {$j_1$}
  \put (45,75) {$j_2$}
   \put (63,83) {$j_3$}
   \put (50,55) {$k_1$}
   \put (75,40) {$k_2$}
   \put (55,5) {$k_3$}
   \put (10,15) {$l_1$}
   \put (25,35) {$l_2$}
   \put (0,45) {$l_3$}
\end{overpic}
 \end{array}
\end{equation}
but this one can be actually rewritten as a product of two 6j-symbols. Such a decomposition cannot be done with the 9j-symbol of equation \eqref{eq:9j}, so that it is said "irreducible".
We also have the 15j-symbol:
\begin{equation}
\begin{Bmatrix}
    j_1 & j_2 & j_{11} \\
    j_4 & j_5 & j_{15} \\
    j_7 & j_3 & j_{14} \\
    j_9 & j_6 & j_{13} \\
    j_8 & j_{10} & j_{12}
  \end{Bmatrix} 
  = \begin{array}{c}
  \begin{overpic}[scale = 0.6]{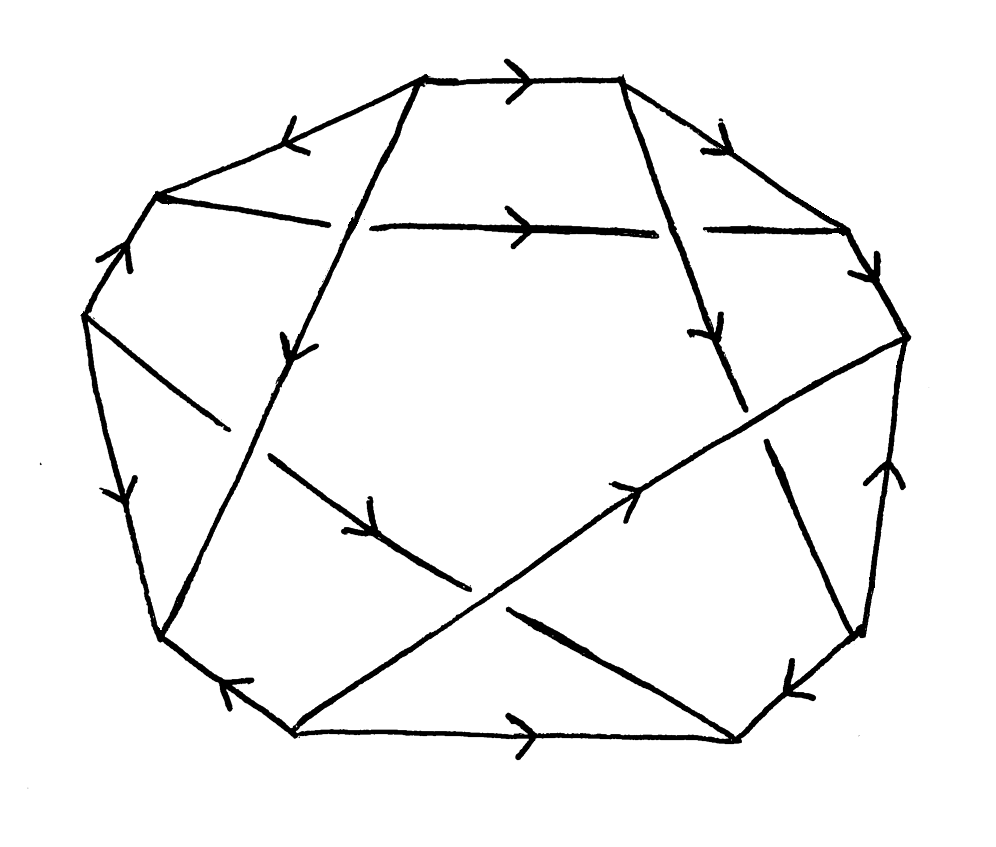}
 \put (25,75) {$j_1$}
  \put (33,47) {$j_2$}
   \put (60,50) {$j_3$}
   \put (80,70) {$j_4$}
   \put (50,55) {$j_5$}
   \put (55,35) {$j_6$}
    \put (95,35) {$j_7$}
  \put (37,37) {$j_8$}
   \put (55,5) {$j_9$}
   \put (0,35) {$j_{10}$}
   \put (45,81) {$j_{11}$}
   \put (0,60) {$j_{12}$}
      \put (15,5) {$j_{13}$}
   \put (80,10) {$j_{14}$}
   \put (93,60) {$j_{15}$}
\end{overpic}
 \end{array}
\end{equation}
which is the definition used by \cite{sarno2018}. It is different from the convention chosen in \cite{ooguri1992}, which is
\begin{equation}
\begin{Bmatrix}
    l_1 & l_2 & l_3 & l_4 & l_5 \\
    j_1 & j_2 & j_3 & j_4 & j_5 \\
    l_{10} & l_9 & l_8 & l_7 & l_6
  \end{Bmatrix} 
  = \begin{array}{c}
  \begin{overpic}[scale = 0.6]{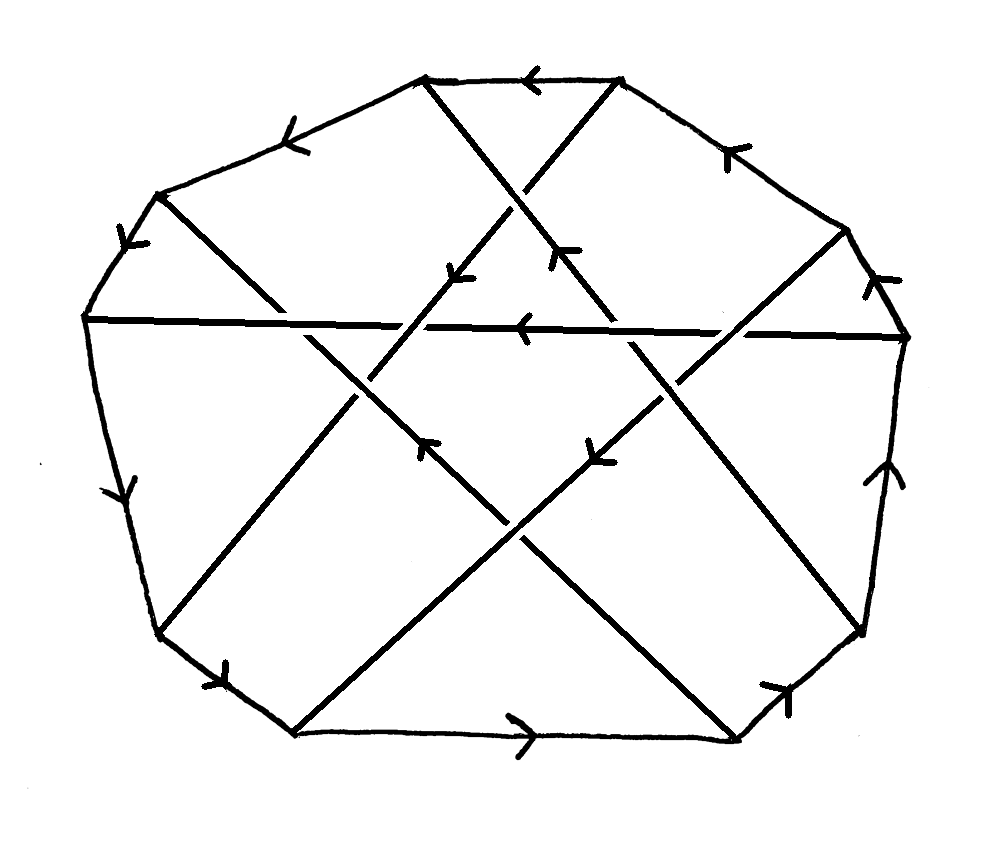}
 \put (25,75) {$l_2$}
  \put (35,60) {$j_4$}
   \put (50,45) {$j_3$}
   \put (80,70) {$l_{10}$}
   \put (60,57) {$j_1$}
   \put (60,30) {$j_5$}
    \put (95,35) {$l_8$}
  \put (37,30) {$j_2$}
   \put (55,5) {$l_6$}
   \put (0,35) {$l_4$}
   \put (45,81) {$l_1$}
   \put (0,60) {$l_3$}
      \put (15,5) {$l_5$}
   \put (80,10) {$l_7$}
   \put (93,60) {$l_9$}
\end{overpic}
 \end{array}
\end{equation}
Contrary to the 6j-symbol, there is no consensus about which convention should be used to define the 15j-symbol, but in all cases it corresponds to an invariant function associated to a 3-valent graph with 15 links. Actually, we can build 5 topologically distinct 15j-symbols\marginpar{For more details on the 9j and 15j-symbol, see Yutsis \cite{yutsis1962}.}. Here we see the power of Yutsis diagrams: it makes huge expressions much more tractable.
\begin{physics}
In the spirit of the result of Ponzano and Regge \eqref{eq:PonzanoRegge}, Ooguri used the 15j-symbol to provide a model of quantum gravity \cite{ooguri1992}. It still plays a major role in the \acs{EPRL} model \cite{speziale2017} (see chapter \ref{ch:loops}).
\end{physics}

\section{Penrose binor calculus}

In section \ref{sec:spinorial_realisation}, we have seen that the spin-$j$ irrep can be built from the symmetrisation of $2j$ copies of $\mathbb{C}^2$. This suggests another graphical calculus that goes under the name of \textit{Penrose binor calculus}. It was introduced in \cite{penrose1971a} as a particular case of a more general abstract tensor system. A more recent introduction can be found in \cite{major1999}.

In this calculus, the identity over $\mathbb{C}^2$ is a single strand
\begin{equation}
	\begin{array}{c}
		\begin{overpic}[height = 1 cm]{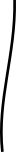}
		\end{overpic}
	\end{array} \cong \dyad{0} + \dyad{1}.
\end{equation}
The free legs carry implicit labels of copies of $\mathbb{C}^2$. There is also a duality between up and down. The cap stands for 
\begin{equation}
	\begin{array}{c}
		\begin{overpic}[height = 1 cm]{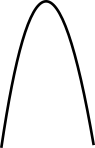}
		\end{overpic} 
	\end{array} \cong i \bra{01} - i \bra{10} \cong i \epsilon_{AB}.
\end{equation}
The cup is
\begin{equation}
	\begin{array}{c}
		\begin{overpic}[height = 1 cm]{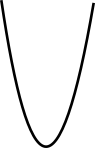}
		\end{overpic} 
	\end{array} \cong i \ket{01} - i \ket{10} .
\end{equation}
With these definitions, the diagrams are then well-behaved under deformation: 
\begin{equation}
	\begin{array}{c}
		\begin{overpic}[height = 1 cm]{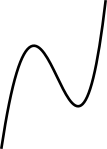}
		\end{overpic} 
	\end{array} = \begin{array}{c}
		\begin{overpic}[height = 1 cm]{figures/binor/straight}
		\end{overpic}
	\end{array}
\end{equation}
Finally, the crossing is the regular swap, but with a global minus sign:
\begin{equation}
	\label{eq:cross_minus}
	\begin{array}{c}
		\begin{overpic}[height = 1 cm]{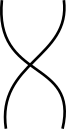}
		\end{overpic} 
	\end{array} = - \dyad{00} - \dyad{10}{01} - \dyad{01}{10} - \dyad{11}.
\end{equation}
These rules guarantee planar isotopy, \ie diagrams can be continuously deformed while preserving their interpretation as linear maps.

\paragraph{Fundamental equations.}

Binor calculus has two core equations that can be deduced from the definitions above.
\begin{equation}
	\label{eq:binor_circle}
	\begin{array}{c}
		\begin{overpic}[height = 1 cm]{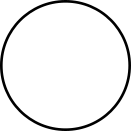}
		\end{overpic} 
	\end{array}
	= -2
\end{equation}
\begin{equation}
	\label{eq:skein_relation}
	\begin{array}{c}
		\begin{overpic}[height = 1 cm]{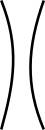}
		\end{overpic} 
	\end{array} + 
	\begin{array}{c}
		\begin{overpic}[height = 1 cm]{figures/binor/cross}
		\end{overpic} 
	\end{array} + 
	\begin{array}{c}
		\begin{overpic}[height = 1 cm]{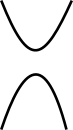}
		\end{overpic} 
	\end{array} = 0
\end{equation}
The value of the loop in the first equation secretely gives the "dimension" of the tensor calculus. Here it is $-2$: the "2" explains the name "binor" and the minus sign explains the title of the original article "negative dimensional tensors" by Penrose \cite{penrose1971a}. The second equation is known under the name of "skein relation" or "binor identity". 

\paragraph{Symmetriser.}
To connect to Yutsis graphical calculus, we must build the spin-$j$ irrep. This is done by symmetrising the strands:
\begin{equation}\label{eq:Penrose-symmetrisation}
	\begin{tikzpicture}
	\begin{pgfonlayer}{nodelayer}
		\node [style=none] (2) at (-1.5, 1.5) {};
		\node [style=none] (3) at (-1.5, -1.5) {};
		\node [style=none] (4) at (-2, 0) {};
		\node [style=none] (5) at (-1, 0) {};
		\node [style=none] (6) at (-2.5, 0) {$j$};
		\node [style=none] (7) at (0, 0) {$\overset{\text{def}}=$};
		\node [style=none] (8) at (1.25, 1.5) {};
		\node [style=none] (9) at (1.25, -1.5) {};
		\node [style=none] (10) at (1, 0) {};
		\node [style=none] (11) at (3, 0) {};
		\node [style=none] (12) at (2.75, 1.5) {};
		\node [style=none] (13) at (2.75, -1.5) {};
		\node [style=none] (14) at (2, 0.5) {$\dots$};
		\node [style=none] (15) at (2, 1) {$2j$};
	\end{pgfonlayer}
	\begin{pgfonlayer}{edgelayer}
		\draw (2.center) to (3.center);
		\draw (4.center) to (5.center);
		\draw (8.center) to (9.center);
		\draw (10.center) to (11.center);
		\draw (12.center) to (13.center);
	\end{pgfonlayer}
\end{tikzpicture} 
	\overset{\text{def}}= \frac{1}{(2j)!}
	\sum_{\sigma \in \mathfrak{S}_{2j}} (-1)^{|\sigma|} \,  \begin{tikzpicture}
	\begin{pgfonlayer}{nodelayer}
		\node [style=none] (0) at (-0.75, 1.5) {};
		\node [style=none] (1) at (-0.75, -1.5) {};
		\node [style=none] (4) at (0.75, 1.5) {};
		\node [style=none] (5) at (0.75, -1.5) {};
		\node [style=none] (6) at (0, 1) {$\dots$};
		\node [style=wide box] (8) at (0, 0) {$\sigma$};
	\end{pgfonlayer}
	\begin{pgfonlayer}{edgelayer}
		\draw (0.center) to (1.center);
		\draw (4.center) to (5.center);
	\end{pgfonlayer}
\end{tikzpicture} 
\end{equation}
where $|\sigma|$ is the parity of $\sigma$ and the $\sigma$-labelled box represents the corresponding permutation of the $2j$ strands from one side to the other. Although this looks like an anti-symmetrisation (since we have the $(-1)^{|\sigma|}$ term), the operator is actually a projector from $\mathbb{C}^2 \otimes \cdots \otimes \mathbb{C}^2$ to $\mathcal{S}_{2j}(\mathbb{C}^2 \otimes \cdots \otimes \mathbb{C}^2 )$, because of the minus sign in~\eqref{eq:cross_minus}. So we have
\begin{equation}\label{eq:binor-jm-vector}
	\begin{tikzpicture}
	\begin{pgfonlayer}{nodelayer}
		\node [style=none] (0) at (-0.75, 1.5) {};
		\node [style=none] (1) at (-0.75, -1.5) {};
		\node [style=none] (2) at (0.75, 1.5) {};
		\node [style=none] (3) at (0.75, -1.5) {};
		\node [style=none] (4) at (0, 0.5) {$\dots$};
		\node [style=none] (5) at (0, 1.25) {$2j$};
		\node [style=none] (6) at (-1.25, 0) {};
		\node [style=none] (7) at (1.25, 0) {};
		\node [style=none] (8) at (0, -2.5) {$\underbrace{0 \dots 1}_{j+m \text{ times } 0}$};
	\end{pgfonlayer}
	\begin{pgfonlayer}{edgelayer}
		\draw (0.center) to (1.center);
		\draw (2.center) to (3.center);
		\draw (6.center) to (7.center);
	\end{pgfonlayer}
\end{tikzpicture} = 
	\mathcal{S}_{2j} (\underbrace{\ket{0}...\ket{1}}_{j+m \text{ times } 0}) = \sqrt{\frac{(j-m)!(j+m)!}{(2j)!}} \ket{j;m}.
\end{equation}

Taking the trace of the symmetrised space (\ie making a loop) gives the value
\begin{equation}
	\begin{tikzpicture}
	\begin{pgfonlayer}{nodelayer}
		\node [style=none] (0) at (-1, 0) {};
		\node [style=none] (1) at (1, 0) {};
		\node [style=none] (2) at (-1, 0) {};
		\node [style=none] (3) at (1, 0) {};
		\node [style=none] (4) at (-1.5, 0) {};
		\node [style=none] (5) at (-0.5, 0) {};
		\node [style=none] (6) at (-2, 0) {$j$};
	\end{pgfonlayer}
	\begin{pgfonlayer}{edgelayer}
		\draw [bend left=90, looseness=1.75] (0.center) to (1.center);
		\draw [bend right=90, looseness=1.75] (2.center) to (3.center);
		\draw (4.center) to (5.center);
	\end{pgfonlayer}
\end{tikzpicture} = 
	\  (-1)^{2j} (2j+1),
\end{equation}
which is the dimension of the space up to a phase.

\paragraph{Invariant tensors.}
Now, let us draw in binor calculus, the analogue of the 3-valent vertex, \ie the $3jm$-symbol. The essential property to be noticed is the invariance of the cup under the action of \SU, \ie for any $u \in \SU$:
\begin{equation}
	\begin{tikzpicture}
	\begin{pgfonlayer}{nodelayer}
		\node [style=none] (0) at (-9, 2) {};
		\node [style=none] (1) at (-7, 2) {};
		\node [style=swirl] (2) at (-7, 1) {$u$};
		\node [style=swirl] (3) at (-9, 1) {$u$};
		\node [style=none] (7) at (0, 0) {$\cong i \, u \cdot (\ket{01} - \ket{10}) = i \,  (\ket{01} - \ket{10}) \cong $};
		\node [style=none] (12) at (7, 2) {};
		\node [style=none] (13) at (9, 2) {};
	\end{pgfonlayer}
	\begin{pgfonlayer}{edgelayer}
		\draw [bend right=90, looseness=5.75] (0.center) to (1.center);
		\draw [bend right=90, looseness=5.75] (12.center) to (13.center);
	\end{pgfonlayer}
\end{tikzpicture}
\end{equation}
Then the diagram
\begin{equation}
	\begin{tikzpicture}
	\begin{pgfonlayer}{nodelayer}
		\node [style=none] (0) at (-4, 2) {};
		\node [style=none] (1) at (-1, 2) {};
		\node [style=none] (4) at (-6, 2) {};
		\node [style=none] (5) at (6, 2) {};
		\node [style=none] (6) at (-5.75, 2) {};
		\node [style=none] (7) at (5.75, 2) {};
		\node [style=none] (8) at (-4.25, 2) {};
		\node [style=none] (9) at (-0.75, 2) {};
		\node [style=none] (10) at (1, 2) {};
		\node [style=none] (11) at (4, 2) {};
		\node [style=none] (12) at (0.75, 2) {};
		\node [style=none] (13) at (4.25, 2) {};
		\node [style=none] (14) at (-6.5, 0) {};
		\node [style=none] (15) at (-3.5, 0) {};
		\node [style=none] (16) at (-1.5, 0) {};
		\node [style=none] (17) at (1.5, 0) {};
		\node [style=none] (18) at (3.5, 0) {};
		\node [style=none] (19) at (6.5, 0) {};
		\node [style=none] (20) at (-5, 0.5) {$\dots$};
		\node [style=none] (21) at (0, 0.5) {$\dots$};
		\node [style=none] (22) at (5, 0.5) {$\dots$};
		\node [style=none] (23) at (-5, 1.25) {$2j_1$};
		\node [style=none] (24) at (0, 1.25) {$2j_2$};
		\node [style=none] (25) at (5, 1.25) {$2j_3$};
	\end{pgfonlayer}
	\begin{pgfonlayer}{edgelayer}
		\draw [bend right=90, looseness=5.00] (0.center) to (1.center);
		\draw [bend right=90, looseness=2.00] (4.center) to (5.center);
		\draw [bend right=90, looseness=2.00] (6.center) to (7.center);
		\draw [bend right=90, looseness=4.50] (8.center) to (9.center);
		\draw [bend right=90, looseness=5.00] (10.center) to (11.center);
		\draw [bend right=90, looseness=4.50] (12.center) to (13.center);
		\draw (14.center) to (15.center);
		\draw (16.center) to (17.center);
		\draw (18.center) to (19.center);
	\end{pgfonlayer}
\end{tikzpicture}
\end{equation}
depicts a vector that belongs to $\text{Inv}_{\SU} ( \mathcal{H}_{j_1} \otimes \mathcal{H}_{j_2} \otimes \mathcal{H}_{j_3})$, so it is proportional to $\ket{j_1,j_2,j_3}$. This should be interpreted as a kind of ``railroad switch'', where the fundamental wires within the three symmetrised bundles redistribute between themselves. Because we are dealing with symmetrised spaces we only care about how many wires go from each bundle to the other bundle. It turns out that there is then actually only one way in which to connect the wires, when it is not impossible. The Clebsch-Gordan conditions~\eqref{eq:CGcondition} precisely state when such a recoupling is possible.

Plugging the vectors of eq. \eqref{eq:binor-jm-vector}, one gets
\begin{multline}
	\label{eq:penrose-3jm}
	\begin{tikzpicture}
	\begin{pgfonlayer}{nodelayer}
		\node [style=none] (0) at (-4, 4.75) {};
		\node [style=none] (1) at (-1, 4.75) {};
		\node [style=none] (2) at (-6, 4.75) {};
		\node [style=none] (3) at (6, 4.75) {};
		\node [style=none] (4) at (-5.75, 4.75) {};
		\node [style=none] (5) at (5.75, 4.75) {};
		\node [style=none] (6) at (-4.25, 4.75) {};
		\node [style=none] (7) at (-0.75, 4.75) {};
		\node [style=none] (8) at (1, 4.75) {};
		\node [style=none] (9) at (4, 4.75) {};
		\node [style=none] (10) at (0.75, 4.75) {};
		\node [style=none] (11) at (4.25, 4.75) {};
		\node [style=none] (12) at (-6.5, 2.75) {};
		\node [style=none] (13) at (-3.5, 2.75) {};
		\node [style=none] (14) at (-1.5, 2.75) {};
		\node [style=none] (15) at (1.5, 2.75) {};
		\node [style=none] (16) at (3.5, 2.75) {};
		\node [style=none] (17) at (6.5, 2.75) {};
		\node [style=none] (18) at (-5, 3.25) {$\dots$};
		\node [style=none] (19) at (0, 3.25) {$\dots$};
		\node [style=none] (20) at (5, 3.25) {$\dots$};
		\node [style=none] (21) at (-5, 4) {$2j_1$};
		\node [style=none] (22) at (0, 4) {$2j_2$};
		\node [style=none] (23) at (5, 4) {$2j_3$};
		\node [style=none] (24) at (-5, 5.75) {$\overbrace{0 0 \dots 1 1 }^{j_1+m_1 \text{ times } 0}$};
		\node [style=none] (25) at (0, 5.75) {$\overbrace{0 0 \dots 1 1 }^{j_2+m_2 \text{ times } 0}$};
		\node [style=none] (26) at (5, 5.75) {$\overbrace{0 0 \dots 1 1 }^{j_3+m_3 \text{ times } 0}$};
	\end{pgfonlayer}
	\begin{pgfonlayer}{edgelayer}
		\draw [bend right=90, looseness=5.00] (0.center) to (1.center);
		\draw [bend right=90, looseness=2.00] (2.center) to (3.center);
		\draw [bend right=90, looseness=2.00] (4.center) to (5.center);
		\draw [bend right=90, looseness=4.50] (6.center) to (7.center);
		\draw [bend right=90, looseness=5.00] (8.center) to (9.center);
		\draw [bend right=90, looseness=4.50] (10.center) to (11.center);
		\draw (12.center) to (13.center);
		\draw (14.center) to (15.center);
		\draw (16.center) to (17.center);
	\end{pgfonlayer}
\end{tikzpicture}  \\ = N(j_1,j_2,j_3) e^{i \phi(j_1,j_2,j_3)} \begin{pmatrix}
		j_1 & j_2 & j_3 \\
		m_1 & m_2 & m_3
	\end{pmatrix} \\ \times \sqrt{\frac{(j_1-m_1)!(j_1+m_1)!(j_2-m_2)!(j_2+m_2)!(j_3-m_3)!(j_3+m_3)!}{(2j_1)!(2j_2)!(2j_3)!}} .
\end{multline}
The functions $N$ and $e^{i\phi}$ can be determined by computing a special case for an easy choice of $m_i$, like $m_1 = j_1, m_2 = j_3 - j_1$ and $m_3 = -j_3$. Then, the diagram is
\begin{equation}
	\begin{tikzpicture}
	\begin{pgfonlayer}{nodelayer}
		\node [style=none] (0) at (-4, 2) {};
		\node [style=none] (1) at (-1, 2) {};
		\node [style=none] (2) at (-6, 2) {};
		\node [style=none] (3) at (6, 2) {};
		\node [style=none] (4) at (-5.75, 2) {};
		\node [style=none] (5) at (5.75, 2) {};
		\node [style=none] (6) at (-4.25, 2) {};
		\node [style=none] (7) at (-0.75, 2) {};
		\node [style=none] (8) at (1, 2) {};
		\node [style=none] (9) at (4, 2) {};
		\node [style=none] (10) at (0.75, 2) {};
		\node [style=none] (11) at (4.25, 2) {};
		\node [style=none] (12) at (-6.5, 0) {};
		\node [style=none] (13) at (-3.5, 0) {};
		\node [style=none] (14) at (-1.5, 0) {};
		\node [style=none] (15) at (1.5, 0) {};
		\node [style=none] (16) at (3.5, 0) {};
		\node [style=none] (17) at (6.5, 0) {};
		\node [style=none] (18) at (-5, 0.5) {$\dots$};
		\node [style=none] (19) at (0, 0.5) {$\dots$};
		\node [style=none] (20) at (5, 0.5) {$\dots$};
		\node [style=none] (21) at (-5, 1.25) {$2j_1$};
		\node [style=none] (22) at (0, 1.25) {$2j_2$};
		\node [style=none] (23) at (5, 1.25) {$2j_3$};
		\node [style=none] (24) at (-5, 2.5) {$0 0 \dots 0 0 $};
		\node [style=none] (25) at (0, 3) {$\overbrace{1 1 \dots 0 0}^{-j_1 + j_2 + j_3  \text{ times } 0}$};
		\node [style=none] (26) at (5, 2.5) {$1 1 \dots 1 1$};
	\end{pgfonlayer}
	\begin{pgfonlayer}{edgelayer}
		\draw [bend right=90, looseness=5.00] (0.center) to (1.center);
		\draw [bend right=90, looseness=2.00] (2.center) to (3.center);
		\draw [bend right=90, looseness=2.00] (4.center) to (5.center);
		\draw [bend right=90, looseness=4.50] (6.center) to (7.center);
		\draw [bend right=90, looseness=5.00] (8.center) to (9.center);
		\draw [bend right=90, looseness=4.50] (10.center) to (11.center);
		\draw (12.center) to (13.center);
		\draw (14.center) to (15.center);
		\draw (16.center) to (17.center);
	\end{pgfonlayer}
\end{tikzpicture} = i^{j_1+j_2+j_3} \frac{(-j_1+j_2+j_3)!(j_1+j_2-j_3)!}{(2j_2)!}.
\end{equation}
while using eq. \eqref{eq:CG explicit}, the RHS gives
\begin{multline}
	N(j_1,j_2,j_3) e^{i \phi(j_1,j_2,j_3)} \sqrt{\frac{(-j_1+j_2+j_3)!(j_1+j_2-j_3)!}{(2j_2)!}}  \\ \times (-1)^{j_1-j_2+j_3} \sqrt{\frac{(2j_1)!(2j_3)!}{(j_1+j_2+j_3+1)!(j_1-j_2+j_3)!}}
\end{multline}
Thus,
\begin{equation}\label{eq:N-factor}
	N(j_1,j_2,j_3) = \sqrt{\frac{(j_1+j_2+j_3+1)!(-j_1+j_2+j_3)!(j_1-j_2+j_3)!(j_1+j_2-j_3)!}{(2j_1)!(2j_2)!(2j_3)!}}.
\end{equation}
and 
\begin{equation}
	e^{i \phi(j_1,j_2,j_3)} = i^{j_1+j_2+j_3} (-1)^{j_1-j_2+j_3}.
\end{equation}

All the invariant functions that were introduced in the previous section can now be translated in terms of binor diagrams. The strength of this calculus lies in the two core equations \eqref{eq:binor_circle} and \eqref{eq:skein_relation}, which are sufficient to simplify the diagrams and compute their actual value. This correspondence between Yutsis and Penrose calculus has been used in \cite{east2021} to translate spin-networks in the language of ZX-calculus.

\chapter{Harmonic analysis over \texorpdfstring{\SU}{SU(2)}}
\label{ch:harmonic}

Harmonic analysis is a subfield of mathematics which studies how functions can be decomposed as a sum of \textit{harmonics}. The word 'harmonic' refers initially to beautiful sound waves, described by the theory of \textit{Musica Universalis} of Pythagoras' school, 6th century BCE. In 1619, Kepler published \textit{Harmonices Mundi}, hypothesising that musical intervals describe the motion of the planets. At the beginning of the 19th century, Joseph Fourier developed a mathematical theory of heat diffusion, nowadays known as \textit{Fourier analysis}. It has later been extended in more abstract contexts under the name of \textit{harmonic analysis}. These mathematics involve periodic functions $f$, which can be decomposed as
\begin{equation}
f(\theta) = \sum_{n \in \mathbb{Z}} c_n e^{  i \frac{2 \pi}{T} n \theta}
\end{equation}
where $T$ is the period of $f$, and $c_n \in \mathbb{C}$. Such functions can be seen as functions over the circle \U. In this short chapter, we first develop the harmonic analysis over \U and then generalise it to \SU. A good and simple reference for this chapter is the thesis \cite{leaser2012}.
 
\section{Harmonic analysis over \U}

What are the unitary irreps of \U? Since \U is abelian, the irreps are one-dimensional. So an irrep of \U acts as a linear function over $\mathbb{C}$, which is just scalar multiplication. Then, one can show that any irrep of \U takes the form
\begin{equation}
\begin{array}{l l}
\chi_m : & \U \to \mathbb{C} \\
& z \mapsto z^m
\end{array} 
\end{equation} 
with $m \in \mathbb{Z}$.

Consider a function $f \in L^2(\U)$. It can be \textit{decomposed into Fourier series} as 
\begin{equation}
f(z) = \sum_{m=-\infty}^\infty f_m \chi_m(z)
\end{equation}
with
\begin{equation}
f_m = \int_{\U} f(z) \chi_m(z^{-1}) \dd \mu(z) 
\end{equation}
with $d\mu$ the Haar measure over \U. Concretely this is
\begin{equation}
f_m = \frac{1}{2 \pi} \int_{-\pi}^{\pi} f(e^{i\theta}) e^{-im \theta} \dd \theta. 
\end{equation}
This motivates for instance to write the Dirac generalised function as
\begin{equation}
\delta(\theta) = \frac{1}{2 \pi} \sum_{m = - \infty}^\infty e^{i m \theta}.
\end{equation}

\section{Harmonic analysis over \SU}

We can mimic the previous definitions for \SU.
Consider a function $f \in L^2(\SU)$. It is said to be \textit{central} if 
\begin{equation}
\forall h,g \in SU(2), \quad f(hgh^{-1}) = f(g).
\end{equation}
Then, any central function $f$ can be \textit{decomposed into Fourier series} as 
\begin{equation}
f(g) = \sum_{j \in \mathbb{N}/2} f_j \, \chi_j(g)
\end{equation}
with the \textit{character} 
\begin{equation}
\chi_j (g) \overset{\text{def}}=  \Tr D^j(g)
\end{equation}
and
\begin{equation}
f_j = \int_{\SU} f(g) \chi_j(g^{-1}) d\mu(g) .
\end{equation}
This motivates to write the Dirac generalised function as
\begin{equation}
\delta(g) =  \sum_{j \in \mathbb{N}/2} (2j+1) \Tr D^j(g).
\end{equation}
\chapter{Representation theory of \texorpdfstring{\SL}{SL(2,C)}}
\label{ch:representation-SL2C}

To put it in a nutshell, the kinematics of \ac{LQG} deal with representations of \SU, while the dynamics, in its spin-foam formulation, lie in the representation theory of \SL. The current models of spin-foams, like the \acs{EPRL} one, extensively use the principal series of \SL.

Whether or not all the representations of \SL, including non-reducible ones, have been classified, is unknown to us, but fortunately all the irreps of \SL are known. In section \ref{sec:finite irreps}, we present the finite-dimensional irreps. In section \ref{sec:infinite irreps}, we summarize the infinite-dimensional ones. Finally, in section \ref{sec:principal series}, we focus on the principal series, which is of main interest for quantum gravity.

\section{Finite irreps}
\label{sec:finite irreps}

\sloppy
The finite irreps of \SL are well-known. They can be obtained from the finite irreps of its $3$-dimensional (complex) Lie algebra \slc. In section \ref{sec:abstract representation}, we have already seen them: they are indexed by a spin $j \in \mathbb{N}/2$. It is also possible to see \slc as a real Lie algebra of dimension $6$, in which case, we will rather denote it $\slc_\mathbb{R}$. In this section, we will describe the (real) linear representations of $\slc_\mathbb{R}$.
We have the following isomorphism between real vector spaces:
\begin{equation}\label{eq:iso sl su}
\slc_{\mathbb{R}} \cong \su \oplus i \, \su .
\end{equation}
As real Lie algebras, $\slc_{\mathbb{R}}$ and $\su \oplus \su$ are actually different: the former is non-compact, while the latter is compact, being the Lie algebra of $\SU \times \SU$. They nevertheless share the same complexification $\slc \oplus \slc$, and hence the same finite-dimensional representation theory. Now, a consequence of Peter-Weyl's theorem is that the irreps of a cartesian product are tensor products of the irreps of the factors. Thus the finite-dimensional irreps of $\slc_{\mathbb{R}} $ are given by the usual tensor representation over $\mathcal{Q}_{j_1} \otimes \mathcal{Q}_{j_2}$, abbreviated by $(j_1,j_2)$. The action is given by:
\begin{multline}
a \cdot (\ket{j_1,m_1} \otimes \ket{j_2,m_2}) \\
\overset{\text{def}}= (a \ket{j_1,m_1}) \otimes \ket{j_2,m_2} + \ket{j_1,m_1} \otimes ( a \ket{j_2,m_2}). 
\end{multline}

The isomorphism \eqref{eq:iso sl su} provides naturally a basis of $\slc_\mathbb{R}$, given by the three Pauli matrices $\sigma_i \in i \, \su$ and the three matrices $i \sigma_i \in \su$. To match the earlier notations introduced in section \ref{sec:angular momentum realisation}, we often denote the \textit{rotation generators} $J_i \overset{\text{def}} = \frac{1}{2} \sigma_i$ and the \textit{boost generators} $K_i \overset{\text{def}}= \frac{i}{2} \sigma_i$. These generators satisfy the commutation relations:
\begin{equation}
\begin{split}
&\left[ J_i,J_j \right]= i \varepsilon_{ijk} J_k \\
&\left[ J_i,K_j \right]= i \varepsilon_{ijk} K_k \\
&\left[ K_i,K_j \right]= - i \varepsilon_{ijk} J_k.
\end{split}
\end{equation}
\begin{notabene}
We can also define the scale operators. Posing $K_\pm \overset{\text{def}}=K_1 \pm i K_2 $ and $J_\pm \overset{\text{def}}= J_1 \pm i J_2$, the scale operators satisfy:
\begin{align*}
&\left[J_3,J_\pm\right] = \pm J_\pm && \left[J_+,J_-\right] = 2 J_3 \\
&\left[K_3,K_\pm\right]=\mp J_\pm && \left[K_+,K_-\right]=-2J_3 \\
&\left[J_+,K_+\right]=\left[J_-,K_-\right]=\left[J_3,K_3\right]=0 && \left[J_\pm,K_\mp\right]= \pm 2 K_3\\
&\left[K_3,J_\pm\right]=\pm K_\pm && \left[J_3,K_\pm\right]=\pm K_\pm
\end{align*}
\end{notabene}
Another basis is given by the \textit{complexified generators}. Posing $A_i = \frac 12 (J_i + i K_i)$ and $B_i = \frac 12 (J_i-i K_i)$, the commutation relations become:
\begin{equation}
\begin{split}
&\left[A_i ,A_j\right] = i\varepsilon_{ijk}A_k \\
&\left[B_i ,B_j\right] = i\varepsilon_{ijk}B_k \\
&\left[A_i ,B_j\right] = 0.
\end{split}
\end{equation}
Then the three realisations, which were described in chapter \ref{ch:representation-SU2} for the action of \su, can be adapted to $\slc_\mathbb{R}$:
\begin{enumerate}
\item (\textit{Homogeneous}) For $m,n \geq 0$, let $\mathbb{C}_{(m,n)}[z_0,z_1;\overline{z_0},\overline{z_1}]$ be the vector space of homogeneous polynomials of degree $m$ in $(z_0,z_1)$ and homogeneous of degree $n$ in $(\overline{z_0},\overline{z_1})$. The action of \SL is given by 
\begin{equation}
g \cdot P(\pmb z ) = P \left(g^T \pmb z  \right).
\end{equation}
The associated action of the algebra $\slc_\mathbb{R}$ is given by
\begin{equation}
\begin{split}
&J_+ \cong z_0 \frac{\partial}{\partial z_1} + \overline{z_0} \frac{\partial}{\partial \overline{z_1}} \\ &J_- \cong z_1 \frac{\partial}{\partial z_0} + \overline{z_1} \frac{\partial}{\partial \overline{z_0}} \\ &J_3 \cong \frac{1}{2} \left( z_0 \frac{\partial}{\partial z_0} - z_1 \frac{\partial}{\partial z_1} + \overline{z_0} \frac{\partial}{\partial \overline{z_0}} - \overline{z_1} \frac{\partial}{\partial \overline{z_1}} \right).
\end{split}
\end{equation}
\item (\textit{Projective}) Let $\mathbb{C}_{(m,n)}[z;\overline{z}]$ be the space of polynomials of degree at most $m$ in $z$ and at most $n$ in $\overline{z}$. The action is given by
\begin{equation}
g \cdot \phi(\xi) = (g_{12} \xi + g_{22})^m \overline{(g_{12} \xi + g_{22})}^n \phi \left( \frac{g_{11} \xi + g_{21}}{g_{12}\xi+g_{22}} \right).
\end{equation} 
\item (\textit{Spinorial}) Over the space of totally symmetric spinors $\mathfrak{S}^{(A_1...A_m)(\dot A_1... \dot A_n)}$, the action is
\begin{equation}
u \cdot z^{A_1...A_m \dot A_1... \dot A_n} = u^{A_1}_{\ \ B_1} ... u^{A_m}_{\ \ B_m} \,  \overline{u}^{\dot A_1}_{\ \ \dot B_1} ... \overline{u}^{\dot A_n}_{\ \ \dot B_n} \,  z^{B_1...B_m \dot B_1... \dot B_n}.
\end{equation}
See Penrose \cite{penrose1984} (p. 142) for details.
\end{enumerate}

The finite representations of \SL cannot be unitary (except the trivial one), because it is a non-compact semisimple Lie group. If we want unitary representations, we shall turn to infinite ones.

\section{Infinite irreps}
\label{sec:infinite irreps}

In this section, we describe all the infinite-dimensional irreps of \SL.  

\begin{notabene}
All the unitary irreps of the Lorentz group have been found simultaneously in 1946 by Gel'fand and Naimark \cite{gelfand1947}, by Harish-Chandra \cite{harish-chandra1947} and by Bargmann \cite{bargmann1947}. It seems nevertheless that Gel'fand and Naimark were the first to publish (unfortunately their article is only in Russian). The question remained to find all the irreps, unitary or not, and this was solved also by Naimark in 1954 \cite{naimark1954}. In 1963, \ac{GMS} published the first book (with English translation) that reviews all these results \cite{gelfand1963}. In 1964, Naimark wrote a more detailed and well-written book that wraps up the subject for mathematically-orientated physicists \cite{naimark1964}.
\end{notabene}
\noindent
The infinite irreps of \SL are parametrised by $(m,\rho) \in \mathbb{Z} \times \mathbb{C}$ with $\Im \rho \geq 0$ and $\rho^2 \neq -(|m| +2n)^2$, with $n \in \mathbb{N}^*$. A realisation is given over the Hilbert space $L^2(\mathbb{C})$ endowed with the scalar product
\begin{equation}
\braket{\varphi }{ \phi} = \frac{i}{2} \int_{\mathbb{C}} \overline{\varphi(\xi)} \phi(\xi) (1+|\xi|^2)^{- \Im \rho} d\xi d\overline{\xi},
\end{equation}
and the action
\begin{equation}\label{eq:action Naimark}
a \cdot f(z) = (a_{12}z+a_{22})^{\frac {m}{2} + \frac{i\rho}{2} -1} \overline{(a_{12} z + a_{22})}^{-\frac {m}{2} + \frac{i \rho}{2}-1} f\left(\frac{a_{11}z+a_{21}}{a_{12}z+a_{22}}\right).
\end{equation}

\paragraph{Remarks.}
\begin{enumerate}
\item If $\rho^2 = -(|m| +2n)^2$ with $n \in \mathbb{N}^*$, the Hilbert space and the action still defines a representation, but a reducible one. Then, if one restricts the action to the subspace of polynomials of degree at most $p= \frac m2 + i \frac \rho 2-1$ in $z$ and $q= -\frac m2 + i \frac \rho 2 -1$ in $\overline{z}$, the representation is irreducible and equivalent to the finite-dimensional representation $(p,q)$.
\item Not all the representations $(m,\rho)$ are unitary. They are unitary in only two cases: when $\rho \in \mathbb{R}$ (principal series); when $m = 0$ and $i \rho \in ]-2,0[$ (complementary series), provided another scalar product is chosen in the latter case (see below).
\item Among these infinite irreps, only the principal representations $(\rho,m)$ and $(-\rho,-m)$ are equivalent.
\item A proof of the result above can be found in Naimark (\cite{naimark1964} pp. 294--295). A sketch of it in the case of the principal series can be found in section \ref{sec:principal series}.
\end{enumerate}

\paragraph{Principal series.} When $\rho \in \mathbb{R}$, the scalar product over $L^2(\mathbb{C})$ becomes the usual one
\begin{equation}
(f_1,f_2) \overset{\text{def}}= \frac i 2\int_\mathbb{C} \overline{f_1(z)} f_2(z)  \dd z \dd \overline{z},
\end{equation}
and the representation $(\rho,m) \in \mathbb{R} \times \mathbb{Z}$ is unitary. The representations $(\rho,m)$ and $(-\rho,- m)$ are unitarily equivalent. They form the so-called principal series, parametrised by $(\rho,m) \in \mathbb{R} \times \mathbb{Z}$. This choice of parametrisation is not universal. Here is a table to translate between different authors:

\begin{tabular}{ll}
Naimark \cite{naimark1964}, \acs{GGV} \cite{gelfand1966} & $(\rho,m)$ \\
Rühl \cite{ruhl1970} & $(\rho_R,m_R) = (\rho, - m)$ \\
\acs{GMS} \cite{gelfand1963} & $(\rho_G, m_G) = (\rho/2,m)$ \\
Rovelli-Vidotto \cite{CLQG}, Barrett \cite{barrett2010} & $(p,k)=(\rho/2, m/2)$
\end{tabular}

\paragraph{Complementary series.} When $m=0$ and $i \rho \in ]-2,0[$, the action becomes
\begin{equation}
a \cdot \phi(\xi) = |a_{12} \xi + a_{22}|^{i\rho-2}  \phi\left( \frac{a_{11} \xi + a_{21}}{a_{12} \xi + a_{22}} \right).
\end{equation}
It also defines a unitary representation for the scalar product
\begin{equation}
\braket{\varphi }{\phi} = \left( \frac{i}{2} \right)^2 \int_{\mathbb{C}^2} \frac{\overline{\varphi(\xi)} \phi(\eta)}{|\xi - \eta|^{2+ i \rho}} d\xi d\overline{\xi} d\eta d\overline{\eta}.
\end{equation}

\section{Principal unitary series}
\label{sec:principal series}

In this section, we review several ways to build the principal series and we expand on its properties.

\subsection{Induced representation}
\label{sub:principal-induced}

The construction of the principal series by Gel'fand and Naimark is based on the induced representation method, which was introduced in section \ref{sec:induced} (see \cite{naimark1964} for details). The principal series is found as the unitary representations of \SL induced by the uni-dimensional representations of the upper-triangular subgroup $K_+$. 

To prove it, the first step is to notice the following diffeomorphism between differentiable manifolds:
\begin{equation}
\SL / K_+ \cong \bar{\mathbb{C}}.
\end{equation}
Then, we can induce the expression of the linear action of \SL over $\bar{\mathbb{C}}$:
\begin{equation}
a \cdot z = \frac{a_{11} z + a_{21}}{a_{12} z + a_{22}}.
\end{equation}
It is nothing but the so-called Möbius transformation. Consider the Hilbert space of square integrable complex functions $L^2(\mathbb{C})$ with the scalar product:
\begin{equation}
(f_1,f_2) \overset{\text{def}}= \frac i2 \int_\mathbb{C} \overline{f_1}(z) f_2(z) \, dz \wedge d\overline{z}.
\end{equation}
Then, we look for a unitary representation over $L^2(\mathbb{C})$ of the form:
\begin{equation}
a \cdot f(z) = \alpha(z,a) f(a \cdot z).
\end{equation}
After lines of computation, we find that for all $(\rho,m) \in \mathbb{R} \times \mathbb{Z}$, there exists a unitary representation of \SL over $L^2(\mathbb{C})$ given by
\begin{multline}
a \cdot f(z) = (a_{12}z+a_{22})^{\frac {m}{2} + \frac{i\rho}{2} -1} \overline{(a_{12} z + a_{22})}^{-\frac {m}{2} + \frac{i \rho}{2}-1} \\
\times f\left(\frac{a_{11}z+a_{21}}{a_{12}z+a_{22}}\right).
\end{multline}
These are called the principal series and we can finally show that they are irreducible!

	\subsection{Homogeneous realisation}
	\label{sub:principal homogeneous}
	
Though rigorous from the mathematical point of view, it is not very intuitive, especially for physicists. In 1962, Gel'fand, Graev and Vilenkin (\acs{GGV}) published a book where they build the principal series from a space of homogeneous functions, which may seem more natural (\cite{gelfand1966} pp. 139--201). A beautiful and concise exposition can be found in the article of Dao and Nguyen (\cite{dao1967} pp. 18--21). We present it here.

Consider $\mathcal{F}(\mathbb{C}^2)$, the vector space of the complex functions over $\mathbb{C}^2$. A function $F \in \mathcal{F}(\mathbb{C}^2)$ is said to be \textit{homogeneous of degree $(\lambda,\mu) \in \mathbb{C}^2$} if it satisfies for all $\alpha \in \mathbb{C}$ :
\begin{equation}
F(\alpha \pmb z) = \alpha^\lambda \overline{\alpha}^\mu F(\pmb z).
\end{equation}
To be consistently defined when $\alpha = e^{2i\pi n}$, the degree should satisfy the condition :
\begin{equation}
\mu - \lambda \in \mathbb{Z}.
\end{equation}
Instead of $(\lambda, \mu)$, we will instead use in the following, the parameters $(p = \frac{\mu + \lambda + 2}{2i}, k =\frac{\lambda - \mu }{2})$ (same choice of parameters as Rovelli-Vidotto \cite{CLQG} p. 182). 
Define $\mathcal{D}^{(p,k)}[z_0,z_1]$ as the subspace of homogeneous functions of degree $(\lambda,\mu)$ infinitely differentiable over $\mathbb{C}^2 \setminus \{ 0 \}$ in the variables $z_0,z_1,\bar{z}_0$ and $\bar{z}_1$ with a certain topology\footnote{The topology is defined by the following property of convergence: a sequence $F_n(z_0,z_1)$ is said to converge to $0$ if it converges to zero uniformly together with all its derivatives on any compact set in the $(z_0,z_1)$-plane which does not contain $(0,0)$ (see \ac{GGV} \cite{gelfand1966} p. 142).}. We define a continuous representation \SL over $\mathcal{D}^{(p,k)}[z_0,z_1]$ by
\begin{equation}
a \cdot F( \pmb z) \overset{\text{def}}= F(a^T \pmb z).
\end{equation}
\begin{notabene}
We could also have defined the action by $F(a^{-1} \pmb z)$, $F(a^\dagger \pmb z)$ or $F(\pmb z a)$. In fact $F(\pmb z a) = F(a^T \pmb z)$, defines the same action. The action with $a^\dagger$ is obtained by the transformation $a_{ij} \mapsto \overline{a_{ij}}$. The action with $a^{-1}$ is obtained by the transformation $a_{ij} \mapsto (2 \delta_{ij} -1 ) \sum_{kl} (1 - \delta_{ik})(1- \delta_{jl}) a_{kl}$. The convention that we have chosen here is the one of \ac{GGV} (\cite{gelfand1966} p. 145), Dao and Nguyen (\cite{dao1967} p. 18), Rühl (\cite{ruhl1970} p. 53), Rovelli-Vidotto (\cite{CLQG} p. 182) and Barrett \cite{barrett2010}. Knapp (\cite{knapp1986} p. 28) is using the convention $F(a^{-1} \pmb z)$.
\end{notabene}
Now define the following $2$-form over $\mathbb{C}^2$:
\begin{align*}
\Omega (z_0,z_1) = \frac i2 (z_0dz_1 - z_1 dz_0) \wedge (\overline{z_0}d\overline{z_1} - \overline{z_1} d\overline{z_0}).
\end{align*} 
Interestingly, it is invariant for the action of \SL: $\Omega(a \pmb z)=\Omega (\pmb z)$. Let $\Gamma$ be a path in $\mathbb{C}^2$ that intersects each projective line exactly once. Then define the scalar product over $\mathcal{D}^{(p,k)}[z_0,z_1]$:
\begin{align*}
(F,G) = \int_\Gamma \overline{F(\pmb z)} G(\pmb z) \Omega(\pmb z).
\end{align*}
Thus $\mathcal{D}^{(p,k)}[z_0,z_1]$ is a Hilbert space. Interestingly, the result does not depend on the path $\Gamma$ provided $p \in \mathbb{R}$, which we consider to be the case in the following. This scalar product is invariant for \SL: $(a \cdot F, a \cdot G) = (F,G)$. Thus the representation is unitary. It could be also shown to be irreducible. In the next subsection, we will see that the representation $\mathcal{D}^{(p,k)}[z_0,z_1]$ is equivalent to the representation $(\rho = 2 p,m = 2k)$ of the principal series described in section \ref{sec:infinite irreps}.

Using the language developed in chapter \ref{ch:geometric}, $\Gamma$ can be understood as a local section of the tautological bundle $\mathcal{O}(-1)$. Then $\bar{f} g \Omega$ is a homogeneous $2$-form of degree $0$. As explained in section \ref{sec:tautological}, it is no surprise that the integral does not depend on the choice of section $\Gamma$. Its computation can be made using what we call the \textit{Gel'fand section}
\begin{equation}
s : d \mapsto (d,(1, \zeta_0(d) )).
\end{equation}
Such a choice directly leads to the projective realisation of subsection \ref{sub:projective}.

	\subsection{Projective realisation}
	\label{sub:projective}
	
Consider the map
\begin{equation}
\iota : \left\{ \begin{array}{lll}
\mathbb{C} & \rightarrow &\mathbb{C}^2 \\
\zeta & \mapsto & (\zeta,1)
\end{array} \right.
\end{equation}
$\iota$ is a diffeomorphism from $\mathbb{C}$ to its range. It parametrises a horizontal straight line of $\mathbb{C}^2$. The projective construction consists in restricting the domain of definition of the homogeneous function to this line. If $F \in \mathcal{F}(\mathbb{C}^2)$, define $\iota^*F \in \mathcal{F}(\mathbb{C})$ as
\begin{equation}
\iota^* F (z) \overset{\text{def}}= F \circ \iota (z) = F(z,1).
\end{equation}
The $2$-form $\Omega$ becomes similarly
\begin{align*}
\iota^* \Omega (z) = \frac i 2 dz \wedge d\overline{z},
\end{align*}
which is nothing but the usual Lebesgue measure over $\mathbb{C}$. Thus we define the Hilbert space $L^2(\mathbb{C})$ with the scalar product
\begin{align*}
(f,g) = \frac i 2 \int_\mathbb{C} \overline{f(z)}g(z)  dz \wedge d\overline{z}.
\end{align*}
Thus we have a map $\iota^* :  \mathcal{D}^{(p,k)}[z_0,z_1] \rightarrow L^2(\mathbb{C})$. In fact $\iota^*$ is bijective: for all $f \in L^2(\mathbb{C})$, there exists a unique $F \in \mathcal{D}^{(p,k)}[z_0,z_1]$ such that $f = \iota^* F$. $F$ is given explicitly by
\begin{equation}
F (z_0,z_1) = z_1^{-1+ip+k} \bar{z}_1^{-1+ip-k} f \left( \frac{z_0}{z_1} \right).
\end{equation}

Importantly, $\iota^*$ induces naturally an action of \SL over $\mathcal{F}(\mathbb{C})$, such that $\iota^*$ becomes an intertwiner between two equivalent representations. After computation, we obtain:
\begin{equation}
a \cdot f(z) \overset{\text{def}}= (a_{12}z+a_{22})^{-1+ip +k} \overline{(a_{12}z + a_{22})}^{-1+ip -k} f\left( \frac{a_{11}z+a_{21}}{a_{12}z+a_{22}} \right).
\end{equation}
This formula is exactly the same formula as \eqref{eq:action Naimark}, with the indices $(p,k) = (\rho/2,m/2)$. Thus we have constructed explicitly the representations of the principal series, and we have shown the equivalence of the realisations $\mathcal{D}^{(p,k)}[z_0,z_1]$ and $L^2(\mathbb{C})$.

\subsection{\texorpdfstring{\SU}{\SU}-realisation}
\label{sub:principal-SU2}

Following Rühl (\cite{ruhl1970} p. 57), we are going to build another realisation of the unitary principal representations.

The first step is to observe the following diffeomorphism between manifolds:
\begin{equation}
\SL/K_+ \cong \SU / \U.
\end{equation}
Then, instead of constructing a space of functions over $\SL/K_+$ as was done originally (see subsection \ref{sub:principal-induced}), it is equivalent to consider functions $\phi$ over \SU satisfying a covariance condition for the group \U:
\begin{equation}
\phi \left(\begin{pmatrix}
e^{i \theta} & 0 \\ 0 & e^{-i \theta}
\end{pmatrix} u \right) = e^{i n \theta} \phi(u)
\end{equation}
with $n \in \mathbb{Z}$. The choice of the factor $e^{i n \theta}$ corresponds to uni-dimensional representations of \U (see chapter \ref{ch:harmonic}). 

Concretely, consider the map 
\begin{equation}
\kappa : \left\{ \begin{array}{lll}
\SU & \rightarrow & \mathbb{C}^2 \\
u & \mapsto & (u_{21},u_{22})
\end{array} \right.
\end{equation}
$\kappa$ is a diffeomorphism to its range. In some sense, \SU can be seen as the "unit circle" of $\mathbb{C}^2$, so that $\kappa$ can be seen as the injection of the circle in the plane $\mathbb{C}^2$. 

Then, define $\kappa^* : \mathcal{F}(\mathbb{C}^2) \rightarrow \mathcal{F}(\SU)$ such that
\begin{equation}\label{eq:kappa^*}
\kappa^* F (u) \overset{\text{def}}= F \circ \kappa (u) = F \left( u_{21}, u_{22} \right).
\end{equation}
If $F \in \mathcal{D}^{(p,k)}[z_0,z_1]$, then we show easily that $\kappa^*F$ satisfies the covariance property 
\begin{equation}\label{eq:covariance property}
\kappa^* F (e^{i\theta \sigma_3} u) =  e^{-2 i \theta k} \kappa^* F (u).
\end{equation}
We denote $\mathcal{D}^{(p,k)}[u] \overset{\text{def}}= \kappa^* \mathcal{D}^{(p,k)}[z_0,z_1]$. Thus $\kappa^*$ is a bijection from $ \mathcal{D}^{(p,k)}[z_0,z_1]$ to $\mathcal{D}^{(p,k)}[u]$. Its inverse is given explicitly by
\begin{equation}\label{eq:kappa*-1}
F (z_0 , z_1) = (|z_0|^2 + |z_1|^2)^{-1+ip} \phi \left( \frac{1}{\sqrt{|z_0|^2+|z_1|^2}} \begin{pmatrix}
z_1^* & -z_0^* \\
z_0 & z_1
\end{pmatrix} \right).
\end{equation}

We could also translate the measure $\kappa^* \Omega$, and thus endow $\mathcal{D}^{(p,k)}[u]$ with the structure of a Hilbert space. Interestingly, it is a subspace of $L^2(\SU)$. As previously, one can translate the action of \SL over $\mathcal{D}^{(p,k)}[u]$ such that $\kappa^*$ becomes a bijective intertwiner, and we obtain
\begin{multline}
a \cdot \phi(u) = (|\beta_{a,u}|^2 + |\alpha_{a,u}|^2)^{-1+ip} \\
\times \phi \left( \frac{1}{\sqrt{|\beta_{a,u}|^2+|\alpha_{a,u}|^2}} \begin{pmatrix}
\alpha_{a,u} & -\beta_{a,u}^* \\
\beta_{a,u} & \alpha_{a,u}^*
\end{pmatrix} \right),
\end{multline}
with $\alpha_{a,u} \overset{\text{def}}= (u_{21} a_{12} + u_{22} a_{22})^*$ and $\beta_{a,u} \overset{\text{def}}= u_{21} a_{11} + u_{22} a_{21}$. Thus $\mathcal{D}^{(p,k)}[u]$ is a third equivalent realisation of the unitary principal series. The equivalence with $L^2(\mathbb{C})$ is made through $(\kappa \circ \iota^{-1} )^*$ which gives explicitly
\begin{equation}
\phi (u)= u_{22}^{-1+ip+k} \overline{u_{22}}^{-1+ip-k} f\left(\frac{u_{21}}{u_{22}} \right) ,
\end{equation}
and conversely
\begin{equation}
f (z) =(1+|z|^2)^{-1+ip} \phi \left( \frac{1}{\sqrt{1+|z|^2}} \begin{pmatrix}
1 & -z^* \\
z & 1
\end{pmatrix} \right).
\end{equation}
Notice that this equivalence of representations supervenes on the Hopf bundle $\SU/\U \cong \mathbb{C}P^1$ (see section \ref{sec:hopf}).

\subsection{Canonical basis}

The advantage of the \SU-realisation is that we already know interesting functions over \SU, namely the coefficients of the Wigner matrix $D^j_{mn}$. Indeed, they are elements of $\mathcal{D}^{(p,k)}[u]$, provided that they satisfy the covariance property \eqref{eq:covariance property}. We compute easily
\begin{multline}
D^j_{mq}\left( e^{i\theta \sigma_3} u \right) = \sum_{n=-j}^j \bra{j,m} e^{i\theta \sigma_3} \ket{j,n} D^j_{nq}\left(  u \right) \\
= \sum_{n=-j}^j \delta_{mn} e^{2in\theta}  D^j_{nq}\left(  u \right) = e^{2im\theta}  D^j_{mq}\left(  u \right).
\end{multline}
Thus, the covariance property is satisfied if $m= -k$, and so
\begin{equation}\label{eq:Dbasis}
\forall j \in \{ |k|,|k|+1,... \}, \quad \forall q \in \{-j,..., j \} , \quad D^j_{-k,q} \in \mathcal{D}^{(p,k)}[u].
\end{equation}
Since the $D^j_{mn}(u)$ form a basis of $L^2(\SU)$, we show easily that the subset exhibited in \eqref{eq:Dbasis} form a basis of $\mathcal{D}^{(p,k)}[u]$. Another consequence is the following decomposition of $\mathcal{D}^{(p,k)}[u]$ into irreps of \SU:
\begin{equation}
\mathcal{D}^{(p,k)}[u] \cong \bigoplus_{j = |k|}^\infty \mathcal{Q}_j.
\end{equation}
We then call \textit{canonical basis} of $\mathcal{D}^{(p,k)}[u]$ the set of functions:
\begin{multline}\label{eq:canonical SU(2)}
\phi^{(p,k)}_{jm}(u) \overset{\text{def}}= \sqrt{\frac{2j+1}{\pi}} D^j_{-k,m}(u), \\
\text{with} \ j = |k|, |k|+1,... \ \text{and} \ -j \leq m \leq j.
\end{multline}	
From \eqref{eq:ortho D}, we see that they satisfy the orthogonality relations
\begin{equation}
\int_{\SU} \dd u \  \overline{\phi^{(p,k)}_{jm}(u)} \phi^{(p,k)}_{ln} (u)  = \frac 1 \pi \delta_{jl} \delta_{mn}.
\end{equation}
\begin{notabene}
In Rühl (\cite{ruhl1970} p. 59), the factor $\frac{1}{\sqrt{\pi}}$ is absent from the definition of the $\phi^{(p,k)}_{jm}$. Thus, the orthogonality relations do not show a factor $\frac{1}{\pi}$ on the \ac{RHS}. We have chosen this factor so that the canonical basis $f^{(p,k)}_{jm}$ of $L^2(\mathbb{C})$ (see below \eqref{eq:canonical complex}) is orthonormal for the usual scalar product with the Lebesgue measure $\dd z$ (for Rühl the measure is $\dd z / \pi$).

Moreover $\phi^{(p,k)}_{jm}$ could have been defined with a phase factor $e^{i\psi(p,j)}$. This is set to zero in some literature including \cite{ruhl1970, barrett2010}, and we follow that convention here. An alternative phase convention leading to real \SL-Clebsch-Gordan coefficients is obtained for the choice \cite{kerimov1978, speziale2017} $e^{i\psi(p,j)}  = (-1)^{-\frac{j}2} \frac{\Gamma(j+i\rho+1)}{|\Gamma(j+i\rho+1)|}$. 

An intermediate choice of phase is the one of \cite{dao1967, rashid2003}, which has the advantage of simplifying the recursion relations satisfied by the  Clebsch-Gordan coefficients \cite{anderson1970, anderson1970a}. The latter are now either real or purely imaginary.
\end{notabene}
The intertwiner $\kappa^*$ enables to translate this basis in  $\mathcal{D}^{(p,k)}[z_0,z_1]$, and we obtain the canonical basis:
\begin{multline}
F^{(p,k)}_{jm} (z_0,z_1) 
= \sqrt{\frac{2j+1}{\pi}} (|z_0|^2 + |z_1|^2)^{ip - 1 } \\
\times D^j_{-k,m} \left(  
\frac{1}{\sqrt{|z_0|^2+|z_1|^2}}
\begin{pmatrix}
z_1^* & -z_0^* \\
z_0 & z_1
\end{pmatrix} \right),
\end{multline}
where an explicit expression for $D^j_{-k,m}$ is given by equation \eqref{eq:Wigner matrix}.
The same is done with the intertwiner $\iota^*$ to $L^2(\mathbb{C})$, and we obtain the canonical basis:
\begin{equation}\label{eq:canonical complex}
f^{(p,k)}_{jm} (z)
= \sqrt{\frac{2j+1}{\pi}} (1+|z|^2)^{ip -1 -j } D^j_{-k,m} \begin{pmatrix}
1 & -z^* \\
z & 1
\end{pmatrix} 
\end{equation}
The constant factors of \eqref{eq:canonical SU(2)} have been chosen so that
\begin{equation}
\frac{i}{2} \int_\mathbb{C}  \overline{f^{(p,k)}_{jm}(z)} f^{(p,k)}_{ln} (z) dz d\overline{z} = \delta_{jl} \delta_{mn}.
\end{equation}
Finally, in ket notations, the canonical basis is denoted $\ket{p,k,jm}$.

\subsection{Action of the generators}

Similarly to equation \eqref{eq:generators homogeneous}, the action of the \SL-generators can be computed from the action of the group. The generators of the rotations "stay inside" the same \SU-irreps:
\begin{equation}
\begin{split}
&J_3 \ket{p,k,j,m} = m \ket{p,k,j,m} \\
&J_+ \ket{p,k,j,m} = \sqrt{(j+m+1)(j-m)} \ket{p,k,j,m+1} \\
&J_- \ket{p,k,j,m} = \sqrt{(j+m)(j-m+1)} \ket{p,k,j,m-1}.
\end{split}
\end{equation}
The generators of the boost spread over the neighbouring subspaces:
\begin{multline}
K_3 \ket{p,k,j,m} = \alpha_j \sqrt{j^2 - m^2} \ket{p,k,j-1,m} + \gamma_j m \ket{p,k,j,m} \\
- \alpha_{j+1} \sqrt{(j+1)^2 - m^2} \ket{p,k,j+1,m},
\end{multline}
\begin{multline}
K_+ \ket{p,k,j,m} = \alpha_j \sqrt{(j-m)(j-m-1)} \ket{p,k,j-1,m+1} \\
+ \gamma_j \sqrt{(j-m)(j+m+1)} \ket{p,k,j,m+1} \\
+ \alpha_{j+1} \sqrt{(j+m+1)(j+m+2)} \ket{p,k,j+1,m+1}
\end{multline}
\begin{multline}
K_- \ket{p,k,j,m} = - \alpha_j \sqrt{(j+m)(j+m-1)} \ket{p,k,j-1,m-1} \\
+ \gamma_j \sqrt{(j+m)(j-m+1)} \ket{p,k,j,m-1} \\
- \alpha_{j+1} \sqrt{(j-m+1)(j-m+2)} \ket{p,k,j+1,m-1}
\end{multline}
with $\gamma_j \overset{\text{def}}= \frac{k p}{j(j+1)}$ and $\alpha_j \overset{\text{def}}= i \sqrt{\frac{(j^2-k^2)(j^2+p^2)}{j^2(4j^2-1)}}$.
From these expressions, it is possible to compute the action of the two Casimir operators:
\begin{equation}
\begin{aligned}
&(\vec{K}^2 - \vec{J}^2) \ket{p,k,j,m}  = (p^2- k^2 + 1) \ket{p,k,j,m} \ , \\
&\vec K \cdot \vec J \ket{p,k,j,m} = pk \ket{p,k,j,m} \ .
\end{aligned}
\end{equation}

	\subsection{\texorpdfstring{\SL}{SL(2,C)} Wigner's matrix}

We define the \SL Wigner's matrix by its coefficients
\begin{equation}
D^{(p,k)}_{j_1q_1j_2q_2}(a) \overset{\text{def}}= \bra{p,k;j_1q_1} a \ket{p,k;j_2q_2}.
\end{equation}
These coefficients satisfy the orthogonality relations:
\begin{multline}
\int_{\SL} \dd h \, D^{(p_1,k_1)}_{j_1m_1l_1n_1}(h) D^{(p_2,k_2)}_{j_2m_2l_2n_2}(h) \\
= \frac{1}{4(p_1^2 + k_1^2)} \delta(p_1-p_2) \delta_{k_1k_2} \delta_{j_1j_2} \delta_{l_1l_2} \delta_{m_1m_2} \delta_{n_1n_2}.
\end{multline}
To compute it explicitly, it is useful to use Cartan decomposition, $g = u e^{r \sigma_3 /2} v^{-1}$, with $u,v \in \SU$ and $r \in \mathbb{R}_+$. Then, we have
\begin{equation}
D^{(p,k)}_{jmln}(g) = \sum_{q=-\min(j,l)}^{\min(j,l)} D^j_{mq}(u) d^{(p,k)}_{jlq}(r) D^l_{qn}(v^{-1}).
\end{equation}
with the \textit{reduced \SL Wigner's matrix} defined as
\begin{equation}
d^{(p,k)}_{jlm}(r) \overset{\text{def}}= D^{(p,k)}_{jmlm} (e^{r\sigma_3/2}).
\end{equation}
\begin{proof}
\begin{align*}
D^{(p,k)}_{jmln}(g) &= \sum_{j_1 \geq |k|} \sum_{m_1=-j_1}^{j_1} \sum_{j_2 \geq |k|} \sum_{m_2=-j_2}^{j_2} D^{(p,k)}_{jmj_1m_1}(u) D^{(p,k)}_{j_1m_1j_2m_2}(e^{r \sigma_3/2}) D^{(p,k)}_{j_2m_2ln}(v^{-1})\\ 
&= \sum_{j_1} \sum_{m_1} \sum_{j_2} \sum_{m_2} \delta_{jj_1} D^j_{mm_1}(u) D^{(p,k)}_{j_1m_1j_2m_2}(e^{r \sigma_3/2}) \delta_{j_2l} D^l_{m_2n}(v^{-1})\\ 
&= \sum_{m_1=-j}^j \sum_{m_2=-l}^{l} D^j_{mm_1}(u) \matrixel{p,k,jm_1 }{ e^{r \sigma_3/2}}{ p,k,lm_2} D^l_{m_2n}(v^{-1}) \\
&= \sum_{m_1=-\min(j,l)}^{\min(j,l)} D^j_{mm_1}(u) D^{(p,k)}_{jm_1lm_1} (e^{r\sigma_3/2}) D^l_{m_1n}(v^{-1})
\end{align*}
\end{proof}
We have the following symmetry properties:
\begin{multline}
d^{(p,k)}_{jlm}(r) = d^{(-p,k)}_{ljm}(-r) = d^{(p,-k)}_{jl,-m}(r) \\
 = (-1)^{j-l} d^{(-p,-k)}_{ljm}(r)=\overline{d^{(p,k)}_{ljm}(-r) }.
\end{multline}
It admits explicit formulae:

\vspace{\baselineskip}

\noindent
$\blacksquare$ \textbf{Integral formula 1.}
\begin{multline}
d^{(p,k)}_{jlm}(r) = \sqrt{(2j+1)(2l+1)} \left( \frac{(j-k)!(j+k)!}{(j+p)!(j-p)!} \right)^{1/2}  \left( \frac{(l-k)!(l+k)!}{(l+p)!(l-p)!} \right)^{1/2}  \\
\times \sum_{i= \max (0,p-k)}^{ \min(j-k,j+p)} \sum_{i'= \max (0,p-k)}^{ \min(l-k,l+p)} \Bigg[ (-1)^{j+l-2k-i-i'} e^{r(ip-1+p-k-2i')} \\
\times \binom{j+p}{i} \binom{j-p}{j-k-i} \binom{l+p}{i'} \binom{l-p}{l-k-i'} \\
\times \int^\infty_0 2 |\omega|  (1+|\omega|^2)^{-ip -1 -j }   (e^{-2r}+|\omega|^2)^{ip -1 -l } |\omega|^{2(j+l-i'-i+p-k)} d|\omega| \Bigg]
\end{multline}
\begin{proof}
\begin{align*}
D^{(p,k)}_{jplp} (e^{r\sigma_3/2})
&= \int_{\mathbb{C}} d\omega \overline{f^{(p,k)}_{jp} (\omega)}  e^{r(1-ip)} f^{(p ,k)}_{lp} (e^r \omega)  \\
&= \int_{\mathbb{C}} d\omega \sqrt{\frac{2j+1}{\pi}} (1+|\omega|^2)^{-ip -1 -j } \overline{D}^j_{-k,p} \begin{pmatrix}
1 & -\omega^* \\
\omega & 1
\end{pmatrix}  e^{r(1-ip)} \\
& \quad \times \sqrt{\frac{2l+1}{\pi}} (1+e^{2r}|\omega|^2)^{ip -1 -l } D^l_{-k,p} \begin{pmatrix}
1 & -e^r \omega^* \\
e^{r} \omega & 1
\end{pmatrix}.
\end{align*}
We conclude using equation \eqref{eq:Wigner matrix}.
\end{proof}

\vspace{\baselineskip}

\noindent
$\blacksquare$ \textbf{Integral formula 2.}
\begin{multline}
d^{(p,k)}_{jlm}(r) = \sqrt{(2j+1)(2l+1)} \sqrt{\frac{(j-k)!(j+k)!(l-k)!(l+k)!}{(j+m)!(j-m)!(l+m)!(l-m)!}} \\
\times \sum_{n_1,n_2} \Bigg[ (-1)^{j+l+2m-n_1-n_2} e^{r(ip -1-2n_2-k +m)}  \\
\times \binom{j+m}{n_1} \binom{j-m}{j-k-n_1} \binom{l+m}{n_2} \binom{l-m}{l-k-n_2} \\
\times \int_0^1 dt \left[1-(1-e^{-2r})t \right]^{ip -1-l} t^{n_1+n_2+k-m} (1-t)^{j+l-n_1-n_2-k+m}  \Bigg].
\end{multline}
\begin{proof}
Change of variables: $|\omega|^2 = \frac{1-t}{t} \Leftrightarrow t = \frac{1}{1+|\omega|^2}$ and $d\omega^2 = -\frac{1}{t^2} dt$.
\end{proof}

\vspace{\baselineskip}

\noindent
$\blacksquare$ \textbf{Hypergeometric formula.}
\begin{multline}
d^{(p,k)}_{jlm}(r) = \frac{\sqrt{(2j+1)(2l+1)} }{(j+l+1)!} \sqrt{\frac{(j-k)!(j+k)!(l-k)!(l+k)!}{(j-m)!(j+m)!(l-m)!(l+m)!}} \\
\times \sum_{n_1,n_2} (-1)^{n_1+n_2} \binom{j+m}{n_1} \binom{j-m}{n_1-m-k} \binom{l+m}{n_2} \binom{l-m}{n_2-m-k} \\
\times (j+l-n_1-n_2+m+k)!(n_1+n_2-m-k)!e^{r(m+k-ip-1 - 2n_1)} \\
\times {}_2F_1(n_1+n_2-m-k+1, j+ip+1, j+l+2;1-e^{-2r}).
\end{multline}
\begin{proof}
We have used the integral expression of the hypergeometric function ${}_2F_1$,
\begin{equation}
{}_2F_1 (a,b,c;z) = \frac{\Gamma (c)}{\Gamma(b) \Gamma(c-b)} \int_0^1 dt \ t^{b-1} (1-t)^{c-b-1} (1-zt)^{-a}.
\end{equation}
\end{proof}

\vspace{\baselineskip}

\noindent
$\blacksquare$ \textbf{Rühl's formula.} (\cite{ruhl1970} p. 64)
\begin{multline}
d^{(p,k)}_{jlq} (r) =  (2j+1)^{1/2} (2l+1)^{1/2} \int_0^1 dt  ((1-t) e^r + t e^{-r})^{-1 + i p}\\
\times d^j_{-k,q} \left( \arccos(2t-1) \right) d^l_{-k,q} \left( \arccos \left( \frac{t e^{-r} - (1-t)e^r }{t e^{-r} + (1-t)e^r } \right) \right) 
\end{multline}
\chapter{Recoupling of \texorpdfstring{\SL}{SL(2,C)}}
\label{ch:recoupling-SL2C}

\section{\texorpdfstring{\SL}{SL(2,C)}-Clebsch-Gordan coefficients}

Similarly to the \SU case, the tensor product of two irreps of \SL can be decomposed into a direct sum of irreps:
\begin{equation}
\mathcal{D}^{(p_1,k_1)} \otimes \mathcal{D}^{(p_2,k_2)} \cong \int_{\mathbb{R}} dp_3 \bigoplus_{\substack{k_3 \in \mathbb{Z}/2 \\ k_1 + k_2 + k_3 \in \mathbb{N}}} \mathcal{D}^{(p_3,k_3)}.
\end{equation}
Kerimov and Verdiev first got interested in the generalisation of the Clebsch-Gordan coefficients to the irreps of \SL \cite{kerimov1978}. The \SL-Clebsch-Gordan coefficients are defined by the relation 
\begin{multline}
\ket{p_3,k_3;j_3,m_3} = \int dp_1 dp_2 \sum_{k_1j_1m_1} \sum_{k_2j_2m_2} \\
C^{p_3 k_3 j_3 m_3}_{p_1 k_1j_1m_1, p_2 k_2j_2m_2} \ket{p_1,k_1;j_1,m_1} \otimes \ket{p_2,k_2;j_2,m_2}.
\end{multline}
The coefficients are non-zero only when $k_1 + k_2 + k_3 \in \mathbb{N}$, in addition to the usual triangle inequality $|j_1-j_2| \leq j_3 \leq j_1 + j_2$. 

We have an explicit expression for the \SL-Clebsch-Gordan coefficients but they are rather involved. First of all, remark that the magnetic part factorises as
\begin{multline}
C^{p_3 k_3j_3m_3}_{p_1 k_1j_1m_1, p_2 k_2j_2m_2} \\
= \chi (p_1, p_2, p_3, k_1, k_2,k_3;j_1,j_2,j_3) C^{j_3m_3}_{j_1m_1j_2m_2}.
\end{multline}
$\chi$ is a function of 9 variables which can be computed from the following expression (found initially in \cite{kerimov1978} but corrected slightly in \cite{speziale2017}):
\begin{multline}
\chi (p_1, p_2, p_3, k_1, k_2,k_3;j_1,j_2,j_3) \\
= \frac{\kappa \, N_{p_1}^{j_1} N_{p_2}^{j_2} \overline{N_{p_3}^{j_3}}}{4\sqrt{2\pi}} (-1)^{(j_1+j_2+j_3+k_1+k_2+k_3)/2} (-1)^{ - k_2 - k_1}   \\
\times  \sqrt{(2j_1+1)(2j_2+1)(2j_3+1)}  \left( \frac{(j_1-k_1)!(j_2+k_2)!}{(j_1+k_1)!(j_2-k_2)!}\right)^{1/2} \\
\times \Gamma(1 - \nu_3+\mu_3 ) \Gamma(1-\nu_3-\mu_3) \\
\times \sum_{n=-j_1}^{j_1} \left( \frac{(j_1-n)!(j_2+k_3-n)!}{(j_1+n)!(j_2-k_3+n)!} \right)^{1/2}  C^{j_3 k_3}_{j_1n;j_2,k_3-n}\\
\times \sum_{l_1 = \max(k_1,n)}^{\min(j_1,k_3+j_2)} \sum_{l_2 = \max(-k_2,n-k_3)}^{j_2}  \frac{(j_1+l_1)!(j_2+l_2)!}{(j_1-l_1)!(l_1-k_1)!(l_1-n)!} \\
\times \frac{(-1)^{l_1 - k_1 + l_2 + k_2}}{(j_2-l_2)!(l_2+k_2)!(l_2-n+k_3)!} \\
\times \frac{\Gamma(2 - \nu_1 -\nu_2 - \nu_3 + \mu_1 + l_1 + l_2 -n) \Gamma (1- \nu_1 + \mu_3 + l_1) }{\Gamma(2-\nu_1 - \nu_2 + l_1 + l_2) \Gamma (1- \nu_3 + \mu_1 - n)} \\
\times \frac{\Gamma (1- \nu_2 - \mu_3 + l_2)}{ \Gamma(2-\nu_1-\nu_3+l_1) \Gamma(2-\nu_3 - \nu_2 +l_2)}
\end{multline}
with 
\begin{equation}
\begin{array}{l}
\nu_1 = \frac{1}{2} (1 + ip_1 - i p_2 - ip_3) \\
\nu_2 = \frac{1}{2} (1 - ip_1 + i p_2 - ip_3) \\
\nu_3 = \frac{1}{2} (1 + ip_1 + i p_2 + ip_3) \\
\mu_1 = \frac{1}{2} (- k_1 + k_2 + k_3) \\
\mu_2 = \frac{1}{2} (k_1 - k_2 + k_3) \\
\mu_3 = \frac{1}{2} (- k_1 - k_2 - k_3) \\
\end{array}
\end{equation}
and a phase
\begin{multline}
\kappa = \frac{\Gamma (\nu_1 + \mu_1) \Gamma(\nu_2 + \mu_2) \Gamma(\nu_3 + \mu_3) }{|\Gamma(\nu_1+\mu_1)\Gamma(\nu_2+\mu_2) \Gamma(\nu_3+\mu_3)|} \\
\times \frac{\Gamma(-1+\nu_1+\nu_2+\nu_3+\mu_1+\mu_2+\mu_3)}{|\Gamma(-1+\nu_1+\nu_2+\nu_3+\mu_1+\mu_2+\mu_3)|}
\end{multline}
and 
\begin{equation}
N_p^j = \frac{\Gamma(1+j+ip)}{|\Gamma(1+j+ip)|},
\end{equation}
and the usual gamma function defined over $\mathbb{C}$ by analytic continuation of 
\begin{equation}
\Gamma ( z) = \int_0^{+\infty}  t^{z-1}\,e^{-t}\,\dd t, \quad {\rm with} \quad \Re z >0.
\end{equation}
\begin{notabene}
The phase $\kappa$ satisfying $|\kappa|=1$ was chosen to make the \SL-Clebsch-Gordan coefficients real (equivalent to the Condon-Shortley convention in the \SU case). Contrary to the usual \SU-Clebsch-Gordan coefficients, there is no consensual convention for this phase. The choice of Kerimov differs from that of Anderson \cite{anderson1970} or Speziale \cite{speziale2017}.
\end{notabene}
These seemingly intricate expressions have nevertheless been used very efficiently in \cite{speziale2017} to numerically compute spin-foam amplitudes. The formula is indeed interesting because it is expressed with only finite sums.

\section{Graphical calculus}

When one wishes to define a graphical calculus for \SL, one encounters the difficulty of finding a suitable \SL-analogue of the $3jm$-symbol of \SU recoupling theory, such that it would satisfy the appropriate symmetry relations to be well represented by a $3$-valent vertex. This issue is investigated in \cite{anderson1970}, but the symmetry relations are intricate and depend on the convention chosen for the phase $\kappa$. As a result there is no consensus about the definition of the rules of graphical calculus for \SL. Following the phase convention of \cite{speziale2017}, we then define  
\begin{multline}
\Wthree{(p_1,k_1)}{(p_2,k_2)}{(p_3,k_3)}{(j_1,m_1)}{(j_2,m_2)}{(j_3,m_3)} \\ \overset{\text{def}}= (-1)^{2j_1-j_2+j_3-m_3} C^{p_3 k_3j_3,-m_3}_{p_1 k_1j_1m_1, p_2 k_2j_2m_2}.
\end{multline}
Graphically it corresponds to the $3$-valent vertex
\begin{equation}
 \Wthree{(p_1,k_1)}{(p_2,k_2)}{(p_3,k_3)}{(j_1,m_1)}{(j_2,m_2)}{(j_3,m_3)}  = \begin{array}{c}
  \begin{overpic}[scale = 0.6]{figures/3CG-out.png}
 \put (-15,70) {$(p_1,k_1)$}
  \put (37,70) {$(p_2,k_2)$}
   \put (90,65) {$(p_3,k_3)$}
\end{overpic}
 \end{array}
\end{equation}
With the same rules of orientation and summation as those of section \ref{sec:graphical calculus}, we can then fully develop the graphical calculus of \SL. For instance, we can define \SL-invariant functions, like the $(6p,6k)$-symbol. The \SL-$15j$-symbol can be used to define the spin-foam amplitude (see section \ref{sec:spin-foam}).
\chapter{Loops and Foams in a nutshell}
\label{ch:loops}

\acf{LQG} is a good candidate theory for quantum gravity. It is obtained by the canonical quantisation of general relativity and describes the quantum states of space with the so-called spin-networks. Spin-foam theory is a later spinoff of both \ac{LQG} and the sum-over-histories approach to quantum gravity. It describes quantum spacetime, seen as the time evolution of spin-networks.

Most of the main textbooks provide a derivation of the theory, following more or less its historical developments through the process of quantisation \cite{CLQG, baez1999, dona2010}. Here we will only introduce the general mathematical framework of the theory, trying to be as concise as possible, since we believe that a full-fledged fundamental theory should come to a point where it stands on its own, with its mathematical framework and physical principles, without any reference to older approximate theories like general relativity or non-relativistic quantum mechanics.

\section{Spin-network}

Like any good quantum theory, \ac{LQG} comes with a Hilbert space. It is the mathematical space of the various possible states of physical space. A very convenient basis is parametrised by the so-called spin-networks that we first define.

\begin{figure}[h!]
\centering
\begin{overpic}[scale = 0.8]{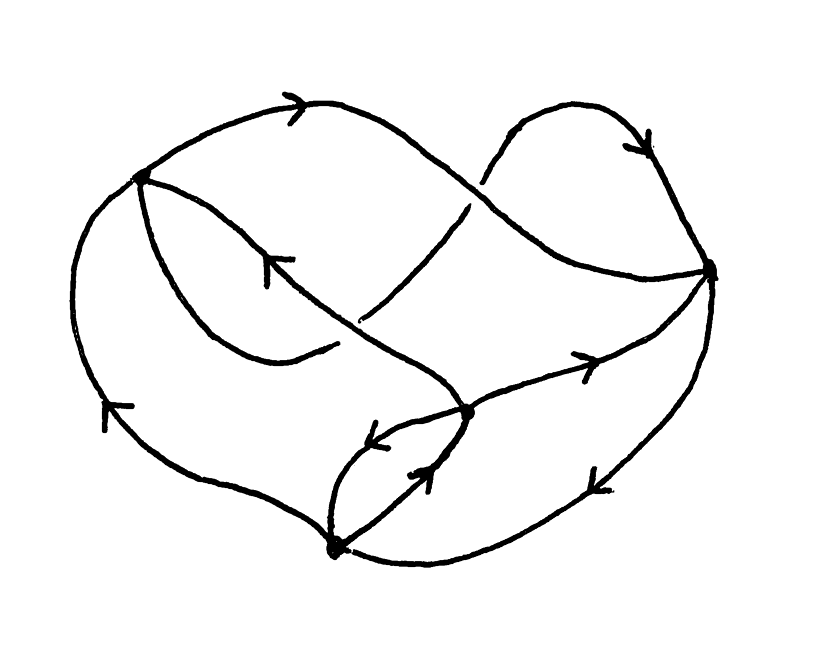}
 \put (45,70) {$j_1$}
  \put (85,65) {$j_2$}
   \put (33,53) {$j_3$} 
   \put (12,22) {$j_4$}
  \put (35,25) {$j_5$}
   \put (55,20) {$j_6$}
    \put (65,28) {$j_7$}
  \put (80,20) {$j_8$}
   \put (10,65) {$\ket{\iota_1}$}
  \put (90,45) {$\ket{\iota_2}$}
   \put (55,35) {$\ket{\iota_3}$} 
   \put (35,5) {$\ket{\iota_4}$}
\end{overpic}
\caption{A $4$-valent spin-network.}
\label{fig:spinnetwork}
\end{figure}

\paragraph{Spin-network.}
An \textit{abstract\footnote{Strictly speaking "\ac{LQG}" refers to the canonical approach for which the spin-networks are embedded in a space-like hypersurface. Here, we adopt a more abstract point of view, sometimes called "covariant \ac{LQG}", which is motivated by spin-foams. This alternative construction raises difficulties for defining the hamiltonian, but they are circumvented by the spin-foam formalism.} directed graph $\Gamma$} is an ordered pair $\Gamma = (\mathcal{N}, \mathcal{L})$, where $\mathcal{N} = \{ n_1,...,n_N\}$ is a finite set of $N$ nodes, and $\mathcal{L} = \{ l_1,...,l_L\}$ a finite set of $L$ links\footnote{Mathematicians usually say \textit{edge} or \textit{arrow}, but not "link", which has another meaning in knot theory. The terminology of \ac{LQG} keeps "edge" for spin-foams (see below), and uses "link" for spin-networks.}, endowed with a target map $t:\mathcal{L} \to \mathcal{N}$ and a source map $s:\mathcal{L} \to \mathcal{N}$, assigning each link to its endpoints (respectively the head or the tail, defined by the orientation). We denote $\mathcal{S}(n)$ (resp. $\mathcal{T}(n)$) the set of links for which the node $n$ is the source (resp. the target). The valency of a node $n$ is the number of links which have $n$ as an endpoint. A graph is said to be $p$-valent if the valency of each node is $p$. Given a directed graph $\Gamma$, we denote $\Lambda_\Gamma$ the set of labellings $j$ that assign to any link $l \in \mathcal{L}$, an \SU-irrep $j_l \in \mathbb{N}/2$.  Given a labelling $j \in \Lambda_\Gamma$,  we denote
\begin{equation}\label{eq:inter hilb node}
\text{Inv} (n, j) \overset{\text{def}}= \text{Inv}_{\SU} \left(\bigotimes_{l \in \mathcal{S}(n)} \mathcal{Q}_{j_l} \otimes \bigotimes_{l \in \mathcal{T}(n)} \mathcal{Q}_{j_l}^* \right).
\end{equation}
The tensor product above assumes the prescription of an ordering of the links around a node, \ie a sense of rotation and a starting link.
A \textit{spin-network} is a triple $\Sigma = (\Gamma, j, \iota)$, with $\Gamma$ a directed graph, $j \in \Lambda_\Gamma$ a labelling, and $\iota$ a map that assigns to any $n \in \mathcal{N}$ an intertwiner $\ket{\iota_n} \in \text{Inv}(n,j)$. Figure \ref{fig:spinnetwork} shows a pictorial representation of a $4$-valent spin-network.

\paragraph{Hilbert space.}
The Hilbert space of \ac{LQG} is given by
\begin{equation}
\mathcal{H}_{LQG} = \bigoplus_\Gamma \mathcal{H}_\Gamma
\end{equation}
where the direct sum is made over all possible directed $4$-valent graphs $\Gamma$, and $\mathcal{H}_\Gamma$ is 
\begin{equation}
\mathcal{H}_\Gamma \overset{\text{def}}= \bigoplus_{j \in \Lambda_\Gamma} \bigotimes_{n \in \mathcal{N}} \text{Inv} (n,j)
\end{equation}
It is spanned by the set of \textit{spin-network states}
\begin{equation}
\ket{\Gamma, j, \iota} = \bigotimes_{n \in \mathcal{N}} \ket{\iota_n}
\end{equation}
where $\Gamma$ ranges over all possible $4$-valent graphs, $j$ over $\Lambda_\Gamma$, and $\ket{\iota_n}$ over an orthonormal basis of $\text{Inv}(n,j)$. By definition of the invariant space $\text{Inv} (n, j)$, it is straightforward to see that "the action of any $g_n \in \SU$ on a node $n$", \ie on $\text{Inv} (n, j)$, leaves the spin-network states invariant:
\marginpar{The designation of "Gauss constraint" comes from an analogy with Maxwell theory of electromagnetism.}\begin{equation}\label{eq:Gauss constraint}
g_n \cdot \ket{\Gamma, j, \iota} = \ket{\Gamma, j, \iota}.
\end{equation}
With this property, the spin-network states are said to satisfy the \textit{Gauss constraint} at each node. 

Since we only consider $4$-valent graphs, an orthonormal basis of $\text{Inv}(n,j)$ is given by the states of equation \eqref{eq:intertwiner 4jm}. Thus, instead of writing the abstract states $\ket{\iota}$, it is equivalent to split each $4$-valent node (according to the prescribed ordering of the links around the nodes), like
\begin{equation}\label{eq:split}
\begin{array}{c}
\begin{overpic}[scale = 0.8]{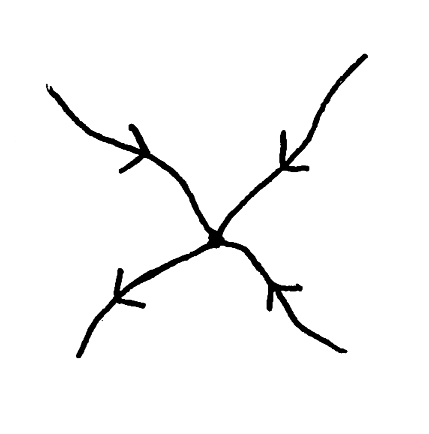}
 \put (20,80) {$j_1$}
  \put (60,80) {$j_2$}
   \put (60,15) {$j_3$} 
   \put (25,15) {$j_4$}
  \put (60,45) {$\ket{\iota}_{12}$}
\end{overpic}
\end{array}
=
\begin{array}{c}
\begin{overpic}[scale = 0.8]{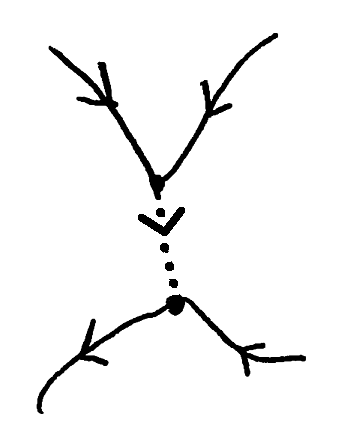}
 \put (5,75) {$j_1$}
  \put (55,75) {$j_2$}
   \put (50,5) {$j_3$} 
   \put (20,5) {$j_4$}
  \put (45,45) {$\iota$}
\end{overpic}
\end{array}
\end{equation}
and then associate to the virtual link the spin $\iota \in \{ \max (|j_1-j_2|,|j_3-j_4|) ,...,\min (j_1+j_2,j_3+j_4) \} $, which parametrises the basis $\ket{\iota}_{12}$ of equation \eqref{eq:intertwiner 4jm}. By metonymy the spin $\iota$ is also called an intertwiner. Thus the spin-network of figure \ref{fig:spinnetwork} becomes
\begin{equation}
\begin{overpic}[scale = 0.8]{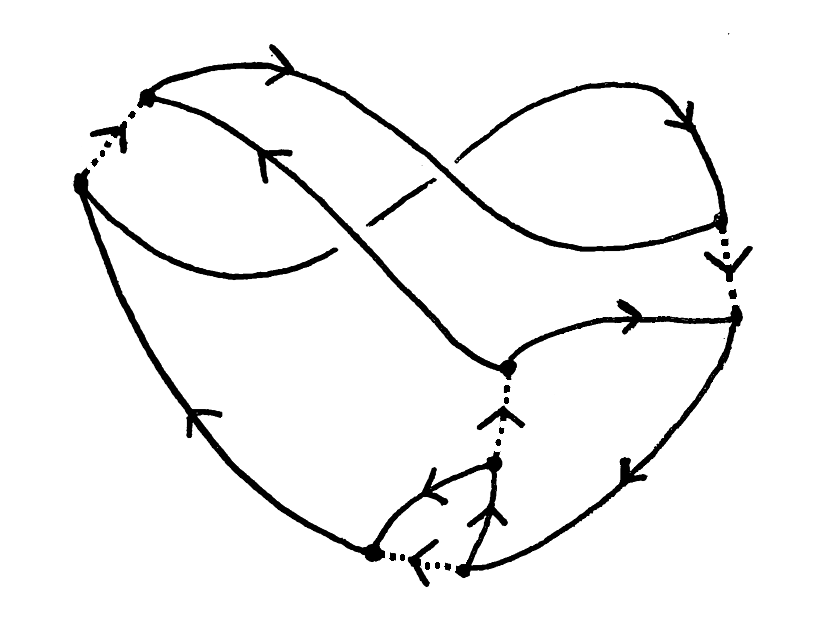}
 \put (45,65) {$j_1$}
  \put (85,65) {$j_2$}
   \put (45,35) {$j_3$} 
   \put (12,22) {$j_4$}
  \put (42,15) {$j_5$}
   \put (62,15) {$j_6$}
    \put (65,37) {$j_7$}
  \put (80,15) {$j_8$}
   \put (8,60) {$\iota_1$}
  \put (90,45) {$\iota_2$}
   \put (62,25) {$\iota_3$} 
   \put (50,0) {$\iota_4$}
\end{overpic}
\end{equation}

\paragraph{Spin-network wave function.}	
The isomorphism \eqref{eq:Peter L(SU(2))}, deduced from Peter-Weyl's theorem, offers another possible realisation of $\mathcal{H}_\Gamma$, as a subspace of $L^2(\SU^L)$, denoted\footnote{This subspace is sometimes denoted $L^2 \left(\SU^L / \SU^N \right)$, but this is not mathematically rigorous.} $L^2_\Gamma(\SU^L /\SU^N)$. A spin-network state $\ket{\Gamma, j, \iota}$ becomes a \textit{spin-network wave function}
\begin{equation}
\Psi_{(\Gamma,j,\iota)} (g_{l_1},...,g_{l_L}) \in L^2_\Gamma(\SU^L /\SU^N),
\end{equation}
obtained with the following procedure:
\begin{enumerate}
\setlength\itemsep{0 \baselineskip}
\item Associate to each link $l$
\begin{equation}\label{eq:line}
\begin{array}{c}
\begin{overpic}[scale = 0.8]{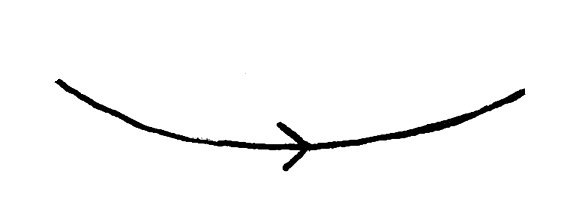}
   \put (40,20) {$j_l$} 
    \put (0,25) {$n_l$} 
     \put (95,25) {$m_l$} 
\end{overpic}
\end{array}
\cong D^{j_l}_{m_ln_l}(g_l)
\end{equation}
with the magnetic indices $m_l$ or $n_l$, depending on the orientation, and the variable $g_l \in \SU$.
\item Associate to each (splitted) node a $4jm$ symbol, like
\begin{equation}
\begin{array}{c}
\begin{overpic}[scale = 0.8]{figures/split.png}
 \put (5,75) {$j_1$}
  \put (55,75) {$j_2$}
   \put (50,5) {$j_3$} 
   \put (20,5) {$j_4$}
  \put (45,45) {$\iota$}
\end{overpic}
\end{array}
\cong (-1)^{j_4-n_4} \Wfour{j_1}{j_2}{j_3}{j_4}{m_1}{m_2}{m_3}{-n_4}{\iota}
\end{equation}
with an index $-n$ and a phase $(-1)^{j-n}$ for outgoing links.
\item Finally multiply all together, and sum over all the magnetic indices.
\end{enumerate}
For instance, the spin-network
\begin{equation}
\begin{array}{c}
\begin{overpic}[scale = 0.8]{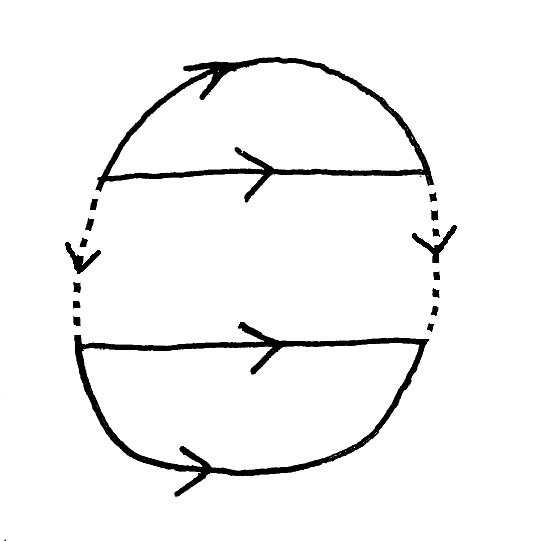}
 \put (15,75) {$j_1$}
  \put (55,70) {$j_2$}
   \put (55,40) {$j_3$} 
   \put (60,5) {$j_4$}
  \put (5,50) {$\iota$}
  \put (85,50) {$\kappa$}
\end{overpic}
\end{array}
\end{equation}
encodes the function
\begin{multline}
 \Psi(u_1,u_2,u_3,u_4) \\
 = \sum_{m_i,n_i} (-1)^{\sum_i (j_i - n_i)}  \Wfour{j_1}{j_2}{j_3}{j_4}{-n_1}{-n_2}{-n_3}{-n_4}{\iota} \\
 \times \Wfour{j_1}{j_2}{j_3}{j_4}{m_1}{m_2}{m_3}{m_4}{\kappa} \prod_{i=1}^4 D^{j_i}_{m_in_i}(u_i).
\end{multline}
From the isomorphism \eqref{eq:Peter L(SU(2))}, we can express the Gauss constraint \eqref{eq:Gauss constraint} as an invariance of the functions $\Psi_{(\Gamma,j,\iota)}(g_{l_1},...,g_{l_L})$: for all sets $(u_n) \in \SU^N$, parametrised by the nodes $n \in \mathcal{N}$, we have
\begin{equation}\label{eq:Gauss-constraint-functions}
\Psi_{(\Gamma,j,\iota)}(g_{l_1},...,g_{l_L}) =  \Psi_{(\Gamma,j,\iota)}(u_{t(l_1)} g_{l_1} u^{-1}_{s(l_1)} ,..., u_{t(l_L)} g_{l_L}u^{-1}_{s(l_L)})
\end{equation}
with $s$ and $t$ the source and target map of the graph. In fact, the space $L^2_\Gamma(\SU^L /\SU^N)$ can be characterized as the subspace of functions of $L^2(\SU^L)$ that satisfy this property.

Notice finally that evaluating the function at the identity on all links results in the graphical calculus previously defined in section \ref{sec:graphical calculus}.

\paragraph{Algebra of observables.} In fact, there is not much information in the Hilbert space itself. What really matters physically is the algebra of observables $\mathcal{A}$ acting upon it. The observables of \ac{LQG} are obtained by the principle of correspondence. Thus, they come with a geometrical interpretation: they correspond notably to measurements of area or measurements of volume. The Hilbert space $\mathcal{H}_{LQG}$ is built from the building block spaces $\mathcal{Q}_{j_l}$, where $j_l$ labels a link $l$. Similarly, the algebra of observables is built from the action of \su (the flux) and \SU (the holonomy) on $\mathcal{Q}_{j_l}$. Notice that an observable should not "go out" of $\mathcal{H}_{LQG}$: in other words, an observable should commute with the Gauss constraint. 

Given a graph $\Gamma$, the observable of area associated to a link $l$ is 
\begin{equation}
\hat A_l \overset{\text{def}}= 8 \pi \frac{\hbar G}{c^3} \gamma \sqrt{\vec{J}^2_l},
\end{equation}
where $\gamma$ is a real parameter called the \textit{Immirzi parameter}, and $\vec{J}_l$ are the generators of \SU acting on $\mathcal{Q}_{j_l}$. The spin-network basis diagonalises $\hat A_l$:
\begin{align}
\hat A_l  \ket{\Gamma, j, \iota} =  8 \pi \frac{\hbar G}{c^3} \gamma \sqrt{ j_l (j_l+1)} \ket{\Gamma, j, \iota}.
\end{align}
It also diagonalises the observable $(\vec{J}_1 + \vec{J}_2)^2$, acting on a node $n$,
\begin{equation}\label{eq:node}
\begin{array}{c}
\begin{overpic}[scale = 0.8]{figures/split.png}
 \put (5,75) {$j_1$}
  \put (55,75) {$j_2$}
   \put (50,5) {$j_3$} 
   \put (20,5) {$j_4$}
  \put (45,45) {$\iota_{12}$}
\end{overpic}
\end{array},
\end{equation} 
so that
\begin{align}
(\vec{J}_1 + \vec{J}_2)^2  \ket{\Gamma, j, \iota} = \iota_{12} (\iota_{12}+1) \ket{\Gamma, j, \iota}.
\end{align}
The latter observable encodes a notion of "angle" between the links $j_1$ and $j_2$. Given a graph $\Gamma$, the set of area observables associated to each link and the set of "angle operators" like $(\vec{J}_1 + \vec{J}_2)^2$ (one per each node), define a \acf{CSCO} on $\mathcal{H}_\Gamma$, diagonalised by the spin-network basis.

On each node like \eqref{eq:node}, we can also define the volume operator
\begin{equation}
\hat V_n = \frac{\sqrt{2}}{3} \left(\frac{8 \pi G \hbar \gamma}{c^3}\right)^{3/2} \sqrt{|\vec{J}_1 \cdot (\vec{J}_2 \times \vec{J}_3)|}.
\end{equation}
It is not diagonalised by the spin-network basis, but its eigenvalues can be computed numerically. It does not commute with $(\vec{J}_1 + \vec{J}_2)^2$ but it does with the areas, so that the areas $\hat A_l$ and the volumes $\hat V_n$ form another \ac{CSCO} (diagonalised by another basis than that of spin-networks).

These geometric operators of area, volume or angle, built from the principle of correspondence, suggest a vision of the "quantum geometry". It is obtained as the dual picture of a graph $\Gamma$: a tetrahedron is associated to each node, and they glue together along faces (whose area is given by  the eigenvalue of $\hat A_l$) dual to links.

\section{Spin-foam}
\label{sec:spin-foam}

\paragraph{Dynamics.} The latter mathematical framework of \ac{LQG} is obtained through the canonical quantisation of general relativity: the spin-network states represent quantum states of space. The time evolution of these states should be found by looking for the subspace formed by the solutions to the hamiltonian constraint $\hat H \ket{\Psi} = 0$, where $\ket{\Psi}$ is a superposition of spin-network states, $\hat{H}$ the quantized hamiltonian. This hard path of finding the dynamics was followed notably by Thiemann \cite{thiemann2007}. Below we present a way to short-circuit the issue, called spin-foams, which takes inspiration from former sum-over-histories approaches to quantum gravity. Spin-foams can be seen as the time evolution of spin-networks, or also as quantum states of spacetime.

\paragraph{Spin-foams.}
Spin-foams can be seen both as a higher dimensional version of Feynman diagrams propagating the gravitational field, and as the time evolution of spin-networks. Spin-foams are built out of combinatorial objects, which generalise graphs to higher dimensions, called \textit{piecewise linear cell complexes}, often abbreviated as \textit{complexes}. An \textit{oriented $2$-complex} is an ordered triple $\kappa = (\mathcal{E}, \mathcal{V}, \mathcal{F})$, with a finite set $\mathcal{E} = \{ e_1,...,e_E\}$ of edges, a finite set $\mathcal{V} = \{ v_1,...,v_V\}$ of vertices, and a finite set $\mathcal{F} = \{ f_1,...,f_F\}$ of faces, such that they all "glue consistently"\footnote{There is a way to give a precise meaning to this gluing, but it will be sufficient to keep it intuitive below, and to avoid these technicalities.}. The orientation is given on the edges by a target map $t:\mathcal{E} \to \mathcal{V}$ and a source map $s:\mathcal{E} \to \mathcal{V}$, and the orientation of each face gives a cyclic ordering of its bounding vertices.

\begin{figure}[h!]
\centering
\begin{overpic}[scale = 0.6]{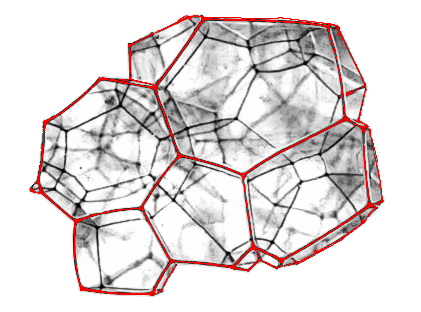}
\end{overpic}
\caption[A $2$-complex embedded in $3$-dimensional euclidean space.]{A $2$-complex embedded in $3$-dimensional euclidean space. Its boundary is a graph (in red).}
\label{fig:spin-foam}
\end{figure}

Given an oriented $2$-complex $\kappa$, we denote $\Lambda_\kappa$ the set of labellings $j$ that assign an \SU-irrep $j_f \in \mathbb{N}/2$ to any face $f \in \mathcal{F}$. Similarly we denote $I_\kappa$ the set of labellings $\iota$ that assign to each edge $e$ an intertwiner $\ket{\iota_e}$, 
\begin{equation}\label{eq:inter hilb vertex}
\ket{\iota_e} \in \text{Inv} (e, j) \overset{\text{def}}= \text{Inv}_{\SU} \left(\bigotimes_{f \in \mathcal{F}(e)} \mathcal{Q}_{j_f} \otimes \bigotimes_{f \in \mathcal{F}^*(e)} \mathcal{Q}_{j_f}^* \right),
\end{equation}
where $\mathcal{F}(e)$ and $\mathcal{F}^*(e)$ are the sets of faces adjacent to the edge $e$, whose orientation respectively matches and does not match that of $e$. 
A \textit{spin-foam} is a triple $F = (\kappa, j, \iota)$, where $\kappa$ is an oriented $2$-complex, $j \in \Lambda_\kappa$, and $\iota \in I_\kappa$. We can stick to a purely "abstract" combinatorial definition of $2$-complexes, but we can also adopt a geometrical "realisation" that represents "faces" as polygons. For instance, figure \ref{fig:spin-foam} shows a spin-foam embedded into $3$-dimensional euclidean space.
Notice that such a graphical representation is not always possible in $3$ dimensions, and sometimes a fourth dimension can be required. Interestingly, the boundary of a $2$-complex\footnote{The notion of boundary of an abstract $2$-complex requires a formal definition, but we keep it intuitive below for simplicity. We can admit that any $2$-complex comes with a boundary.} is a graph, as can be seen in figure \ref{fig:spin-foam}. Thus, the boundary of a spin-foam is a spin-network. The vertices and the edges of the boundary are called respectively nodes and links. Each link bounds an inside face, so that the spin of the link is also the spin of the face. Similarly, each node is an endpoint of an inside edge, so that the associated intertwiners match.

\paragraph{Spin-foam amplitude.} To each spin-foam we associate an amplitude, which is like the propagator associated to a Feynman diagram. Its interpretation is made precise below. Given a spin-foam $(\kappa, j , \iota)$, we define the spin-foam amplitude as
\begin{equation}
\mathcal{A}(\kappa, j , \iota) = \left(\prod_{f \in \mathcal{F}} (2j_f + 1) \right) \left(\prod_{e \in \mathcal{E}} (2\iota_e + 1) \right) \left(\prod_{v \in \mathcal{V}} A_v (j, \iota)\right).
\end{equation}
$A_v$ is called the vertex amplitude. In the short history of spin-foam amplitudes there have already been many different formulae proposed for the vertex amplitude. First, let us say that for quantum gravity, it is sufficient to consider spin-foams whose vertices are $5$-valent ($5$ edges attached to it) and whose edges are $4$-valent ($4$ faces attached to it). This restriction comes from the fact that the $2$-complexes of quantum gravity are built by dualising the triangulation of a $4$-dimensional manifold. Unfortunately there is no possible nice picture as figure \ref{fig:spin-foam} to visualise such a $2$-complex since it cannot be embedded into $3$-dimensional euclidean space. However it is sufficient to get an idea of the combinatorial structure of each vertex by representing the adjacent edges with dots and the faces with lines, so that we draw the \textit{vertex graph}
\begin{equation}\label{eq:vertex graph}
\begin{array}{c}
  \begin{overpic}[scale = 0.6]{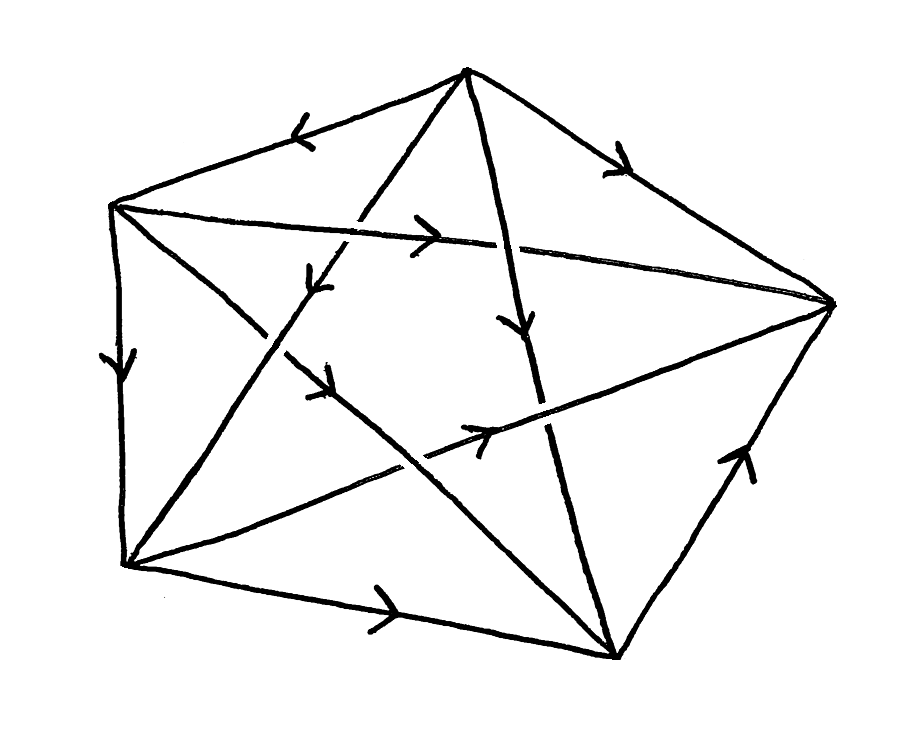}
 \put (25,70) {$j_1$}
  \put (38,47) {$j_2$}
   \put (60,45) {$j_3$}
   \put (75,60) {$j_4$}
   \put (45,60) {$j_5$}
   \put (55,25) {$j_6$}
    \put (75,15) {$j_7$}
  \put (30,30) {$j_8$}
   \put (35,5) {$j_9$}
   \put (0,35) {$j_{10}$}
   \put (50,80) {$\ket{\iota_1}$}
   \put (0,60) {$\ket{\iota_2}$}
      \put (10,10) {$\ket{\iota_3}$}
   \put (70,0) {$\ket{\iota_4}$}
   \put (95,50) {$\ket{\iota_5}$}
\end{overpic}
 \end{array}
 =
 \begin{array}{c}
  \begin{overpic}[scale = 0.6]{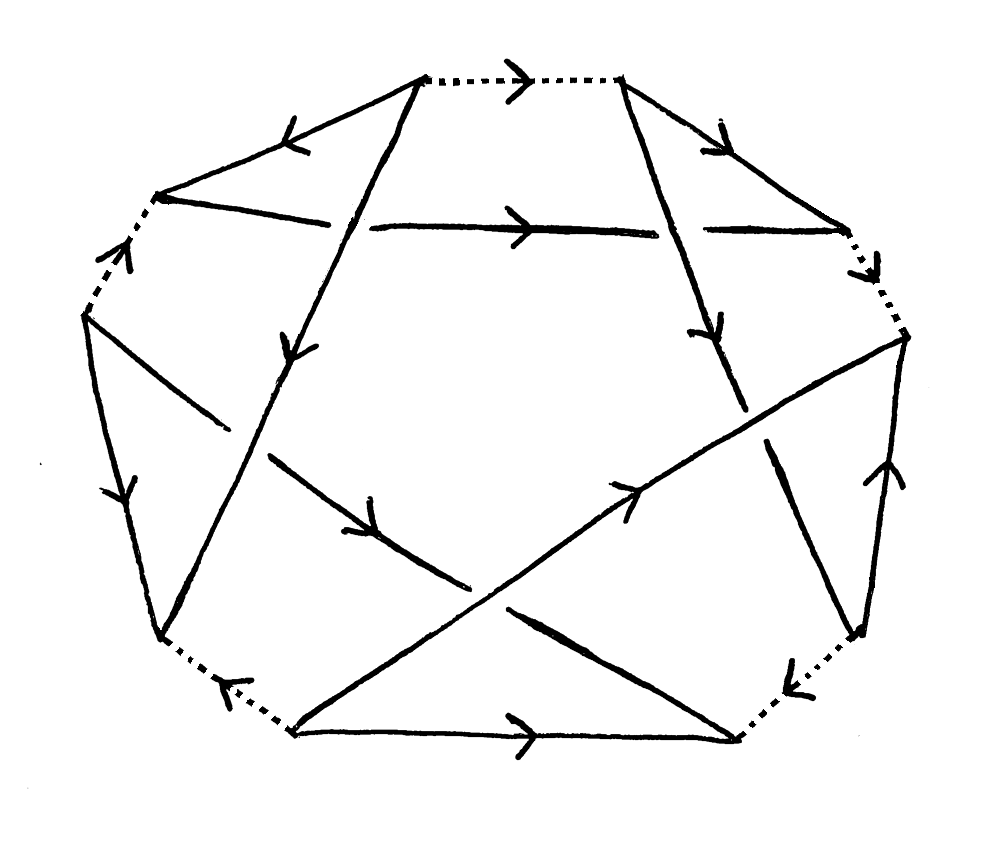}
 \put (25,75) {$j_1$}
  \put (33,47) {$j_2$}
   \put (60,50) {$j_3$}
   \put (80,70) {$j_4$}
   \put (50,55) {$j_5$}
   \put (55,35) {$j_6$}
    \put (95,35) {$j_7$}
  \put (37,37) {$j_8$}
   \put (55,5) {$j_9$}
   \put (0,35) {$j_{10}$}
   \put (45,81) {$\iota_1$}
   \put (0,60) {$\iota_2$}
      \put (15,5) {$\iota_3$}
   \put (80,10) {$\iota_4$}
   \put (93,60) {$\iota_5$}
\end{overpic}
 \end{array}
\end{equation}
The orientation and the spin of the links, and the intertwiners of the nodes are naturally inherited from the underlying spin-foam, so that the vertex graph is a spin-network.
\begin{notabene}
To avoid confusion, let us recap. Each spin-foam comes with a \textit{boundary spin-network}, and also with a \textit{vertex graph} for each of its vertices. If the spin-foam is made of only one vertex, then the boundary spin-network and the vertex graph coincide. Contrary to the boundary spin-network, there is in general no interpretation of the vertex graphs in terms of quantum states of space.
\end{notabene}
The combinatorial shape of each vertex suggests to define the amplitude $A_v$ as the value obtained with the rules of graphical calculus of \SU recoupling theory, defined in section \ref{sec:graphical calculus}. This is precisely what Ooguri did in \cite{ooguri1992} by defining the vertex amplitude as the $15j$-symbol, but it later appeared not to be a good candidate for quantum gravity. Since then many other models were suggested \cite{perez2013}. They all consist in finding other rules than that of \SU recoupling theory to assign a value to the vertex graph \eqref{eq:vertex graph}. 

The \acf{EPRL} model, introduced in \cite{engle2008}, is a model that is still considered a good candidate for quantum gravity. The vertex amplitude is computed from the vertex graph \eqref{eq:vertex graph} with the following rules:
\begin{enumerate}
\item Compute the spin-network wave function as shown in the previous section. We obtain a function of $L^2(\SU^{10})$ which satisfies the Gauss constraint \eqref{eq:Gauss-constraint-functions}:
\begin{equation}
\Psi_{(\Gamma,j,\iota)}(g_{l_1},...,g_{l_{10}}) 
\end{equation}
\item Apply the so-called \textit{$Y_\gamma$-map}, which is the linear map 
\begin{equation}
Y_\gamma : L^2(\SU^{10}) \to \mathcal{F}(\SL^{10})
\end{equation}
where $\mathcal{F}(\SL^{10})$ denotes the space of functions over $\SL^{10}$. It is defined over the canonical basis of Wigner matrix coefficients by
\begin{equation}
Y_\gamma \left( \prod_i D^{j_i}_{m_in_i} \right) = \prod_i D^{(\gamma j_i,j_i)}_{j_im_ij_in_i} ,
\end{equation}
where $\gamma$ is the Immirzi parameter. We thus obtain a function of $\mathcal{F}(\SL^{10})$
\begin{equation}
Y_\gamma\Psi_{(\Gamma,j,\iota)}(h_{l_1},...,h_{l_{10}}) .
\end{equation}
It still satisfies the invariance of the Gauss constraint \eqref{eq:Gauss-constraint-functions} for \SU action, but not for \SL.
\item Project down to the \SL-invariant subspace on each node with the projector $P_{\SL}$ acting as
\begin{multline}
P_{\SL} Y_\gamma \Psi_{(\Gamma,j,\iota)}(h_{l_1},...,h_{l_{10}}) 
=  \int_{\SL^5} \delta(a_{n_5}) \prod_{n \in \mathcal{N}} \dd a_n \\ \times \Psi_{(\Gamma,j,\iota)}\left(a_{t(l_1)} h_{l_1} a^{-1}_{s(l_1)} ,..., a_{t(l_{10})} h_{l_{10}} a^{-1}_{s(l_{10})} \right).
\end{multline}
with $n_5$ any of the 5 nodes. The delta function $\delta(a)$ (only non-vanishing when $a = \mathbb{1}$) is required to avoid the divergence of the integration, but the final result does not depend on the choice of node $n_5$. To put it differently the integration is only effective over (any) four nodes, while the fifth $a_{n_5}$ is fixed to the identity $\mathbb{1}$.
\item Evaluate all the variables $h_l$ to $\mathbb{1}$. So if $(\Gamma,j,\iota)$ is the vertex graph of a vertex $v$ in a spin-foam $(\kappa,j,\iota)$, we can finally write in a nutshell
\begin{equation}
A_v (j, \iota) = \left( P_{\SL} Y_\gamma \Psi_{(\Gamma,j,\iota)} \right) (\mathbb{1}).
\end{equation}
\end{enumerate}
Thus we have fully defined the spin-foam amplitude $\mathcal{A}(\kappa,j,\iota)$ of the \ac{EPRL} model. The specificity of this model is the $Y_\gamma$-map which selects only the irreps $(p=\gamma j, k=j)$ among the principal series of \SL. It implements the so-called \textit{simplicity constraints}, which reduce the topological BF theory to general relativity \cite{baez1999}. Besides, the apparently sophisticated procedure should not hide the fact that the value of $A_v(j,\iota)$ is the same as that obtained from the \SL graphical calculus, defined in chapter \ref{ch:recoupling-SL2C}, when the simplicity constraint is applied.
\begin{notabene}
For those only interested in the actual computation of the amplitude of a given vertex graph, we can summarise the previous procedure with the following algorithm:
\begin{enumerate}
\setlength\itemsep{0 \baselineskip}
\item Associate a variable $h_p \in \SL$ to each intertwiner $\iota_p$.
\item Associate to each link
\begin{equation}
\begin{array}{c}
\begin{overpic}[scale = 0.8]{figures/line.png}
   \put (40,20) {$j$} 
     \put (95,20) {$\iota_{p}$}  
       \put (0,20) {$\iota_{q}$} 
\end{overpic}
\end{array}
\cong D^{(\gamma j,j)}_{jmjn}(h^{-1}_p h_q).
\end{equation}
\item Associate a $3jm$-symbol to each node as in usual graphical calculus (eq. \eqref{eq:graphical 3jm}).
\item Multiply everything together and sum over all the magnetic indices $m$ and $n$.
\item Integrate over (any) four of the five \SL variables $h_p$, and fix the fifth to the identity $\mathbb{1}$.
\end{enumerate}
\end{notabene}

\paragraph{Interpretation.}

The interpretation of spin-foams relies on the general boundary formulation of quantum mechanics which was introduced by Oeckl \cite{oeckl2003, oeckl2008}. Consider a finite region of spacetime. Its boundary $\Sigma$ is a $3$-dimensional hypersurface which constitutes the quantum system under consideration. Its space of states is the Hilbert space of \ac{LQG}, $\mathcal{H}_{LQG}$, spanned by the spin-network states. An observer $\mathcal{O}$ may know some partial information about the state $\psi$ of $\Sigma$, which can be expressed by the fact that $\psi \in \mathcal{S}$, where $\mathcal{S}$ is a linear subspace of $\mathcal{H}_{LQG}$. Then $\mathcal{O}$ can carry out measurements with the operators of the algebra, to determine information about $\psi$. If $\mathcal{R}$ is a linear subspace of $\mathcal{S}$, then the probability to find $\psi \in \mathcal{R}$ is
\begin{equation}
P(\mathcal{R} | \mathcal{S}) = \frac{\sum_{i \in I} |\rho(\xi_i)|^2}{\sum_{j \in J} |\rho(\zeta_j)|^2},
\end{equation}
where $\xi_i$ (resp. $\zeta_j$) is an orthonormal basis of $\mathcal{R}$ (resp. $\mathcal{S}$). $\rho : \mathcal{H}_{LQG} \to \mathbb{C}$ is a linear map, called the \textit{transition amplitude} defined for a spin-network state $\Psi$ by
\begin{equation}
\rho(\Psi) \overset{\text{def}}= \sum_\sigma W_\sigma (\Psi)
\end{equation} 
where the sum is done over all possible $2$-complexes $\sigma$ which have $\Psi$ as a boundary, and $W_\sigma(\Psi)$ is the $2$-complex amplitude defined as
\begin{equation}
W_\sigma(\Psi) = \sum_j \sum_\iota \mathcal{A}(\sigma, j , \iota)
\end{equation}
where the sum is done over all the possible spin labellings $j \in \Lambda_\kappa$, and intertwiner labellings $\iota \in I_\kappa$, that are compatible with the spin-network $\Psi$ at the boundary. 

This completes the mathematical formulation of the theory and its probabilistic interpretation. Of course, much remains to be discovered. In particular, the theory has yet to meet the benchmark of experimental evidence!
\chapter{Commented bibliography}
\label{ch:commented}

This commented bibliography gathers the main textbooks that will provide more details than this primer. They are displayed in chronological order within each section.

\section*{Warmup}

\cite{knapp1986}
A. W. Knapp, \textit{Representation Theory of Semisimple Groups}, Princeton University Press, 1986.
\begin{quote}
This book is a clear and exhaustive introduction to the representation theory of semisimple groups, with specific focus on \SL and its subgroup. It nevertheless requires the reader to have followed a first-semester course on Lie groups and algebras.
\end{quote}

\vspace{\baselineskip}
\noindent
\cite{bernard2012} D. Bernard, Y. Laszlo and D. Renard, \textit{{\'{E}}l{\'{e}}ments de th{\'{e}}orie des groupes et sym{\'{e}}tries quantiques}, cours de l'{\'{E}}cole polytechnique, 2012.
\begin{quote}
This very pedagogical introduction to groups draws inspiration from physics. It covers a wide range of subjects in a concise manner. Unfortunately, there is only a French version.
\end{quote}

\vspace{\baselineskip}
\noindent\cite{hall2015}
Brian C. Hall, \textit{Lie Groups, Lie Algebras, and Representations}, Springer, 2015.
\begin{quote}
This book is a beautiful and modern introduction for physicists.
\end{quote}

\section*{Representation and recoupling of \SU}

\cite{yutsis1962}
A. P. Yutsis, I. B. Levinson and V. V. Vanagas, \textit{Mathematical Apparatus of the Theory of Angular Momentum}, Israel Program for Scientific Translations, 1962.
\begin{quote}
Although quite old, this book remains a reference for graphical calculus (Yutsis diagrams). Its conventions for the definition of the $nj$-symbols are widespread nowadays.
\end{quote}

\vspace{\baselineskip}
\noindent
\cite{moussouris1983}
John P. Moussouris, \textit{Quantum Models of Space-Time based on Recoupling Theory}, PhD thesis (Oxford), 1983.
\begin{quote}
This is a beautifully written PhD thesis by Moussouris, under the supervision of Roger Penrose. It deals notably with the recoupling theory of \SU, and its link to spacetime. Unfortunately, the document is not easily accessible.
\end{quote}

\vspace{\baselineskip}
\noindent
\cite{varshalovich1987}
D. A. Varshalovich, A. N. Moskalev and V. K. Khersonskii, \textit{Quantum theory of angular momentum}, World Scientific, 1987.
\begin{quote}
As the title suggests, this book could be regarded as the bible for the quantum aspects of angular momentum. It is supposed to be exhaustive in terms of formulae, so it is not really the kind of book you read, but rather something like a directory when you need something specific and not very memorable.
\end{quote}

\vspace{\baselineskip}
\noindent
\cite{sakurai2011}
J. J. Sakurai and Jim Napolitano, \textit{Modern Quantum Mechanics}, Addison-Wesley, 2011.
\begin{quote}
This classic book is an introduction to quantum mechanics. Chapter 3 deals with the theory of angular momentum. A number of basic formulae can be found there.
\end{quote}

\section*{Representation of \SL}

\cite{gelfand1963}
I. M. Gel'fand, R. A. Minlos and Z. Ya. Shapiro, \textit{Representations of the rotation and Lorentz groups and their applications}, Pergamon Press, 1963.
\begin{quote}
This book proposes a self-contained presentation of the representations of the rotation and Lorentz groups. However, its rudimentary page layout makes it a bit hard to read. In that respect, Naimark's book, one year later, is a better introduction (and is also probably more detailed in its content).
\end{quote}

\vspace{\baselineskip}
\noindent
\cite{naimark1964}
M. A. Naimark, \textit{Linear Representations of the Lorentz Group}, Pergamon Press, 1964.
\begin{quote}
This book introduces the subject to physicists. It is well-written, very introductory in the beginning, complete on the subject and quite rigorous (though not reaching the usual purely mathematical standards). Unfortunately the formalism and the notation start getting old and sometimes look a bit clumsy, which makes the reading a bit bumpy.
\end{quote}

\vspace{\baselineskip}
\noindent
\cite{gelfand1966}
I. M. Gel'fand, M. I. Graev and N. Ya. Vilenkin, \textit{Generalized Functions: Volume 5, Integral Geometry and Representation Theory}, Academic Press, 1966.
\begin{quote}
This book is the English translation of the Russian version, published in 1962. The chapters of interest for us are Chapter III, on the representations of \SL, and Chapter IV, on its harmonic analysis. 
\end{quote}

\vspace{\baselineskip}
\noindent
\cite{ruhl1970}
W. R{\"{u}}hl, \textit{The Lorentz Group and Harmonic Analysis}, W. A. Benjamin, Inc, 1970.
\begin{quote}
This old book was written by a physicist and is maybe too sloppy in its mathematical exposition. It is nevertheless a classic textbook with a lot of useful formulae. It focuses on the study of \SL and \textsc{SL$_2$($\mathbb{R}$)}.
\end{quote}

\section*{Loop Quantum Gravity and Spin-Foams}

\cite{quantum_gravity}
C. Rovelli, \textit{Quantum Gravity}, Cambridge University Press, 2004.
\begin{quote}
This major textbook is recommended for its insistence on underlying physical ideas. The mathematical formulae are also present but some of the tools of representation and recoupling theories are assumed to be already known.
\end{quote}

\vspace{\baselineskip}
\noindent
\cite{thiemann2007}
T. Thiemann, \textit{Modern Canonical Quantum General Relativity}, Cambridge University Press, 2007.
\begin{quote}
This quite technical book is one of the few main current textbooks in the field. It gives details on some of the representation theory results in this primer, including a proof of the Peter-Weyl theorem.
\end{quote}

\vspace{\baselineskip}
\noindent
\cite{CLQG}
C. Rovelli and F. Vidotto, \textit{Covariant Loop Quantum Gravity}, Cambridge University Press, 2014.
\begin{quote}
This book is a concise exposition of the covariant formulation of LQG, also known as the spin-foam formalism. It gathers all the main achievements of the theory. It can alternatively be used as a technical toolbox ready for use or as a general introduction that sketches the programme and the physical ideas upon which it relies. Nevertheless, the mathematics is not explained in detail (though lots of formulae are found) and it is sometimes a bit sloppy with the mathematical accuracy.
\end{quote}

\cleardoublepage\label{app:bibliography} 

\manualmark 
\markboth{\spacedlowsmallcaps{\bibname}}{\spacedlowsmallcaps{\bibname}} 
\refstepcounter{dummy}

\addtocontents{toc}{\protect\vspace{\beforebibskip}} 
\addcontentsline{toc}{chapter}{\tocEntry{\bibname}}

\sloppy 
\printbibliography

@incollection{penrose1971a,
  title = {Applications of {{Negative Dimensional Tensors}}},
  booktitle = {Combinatorial {{Mathematics}} and Its {{Applications}}},
  author = {Penrose, Roger},
  editor = {Welsh, D. J. A.},
  year = {1971},
  publisher = {{Academic Press}},
}

@article{east2021,
	title = {Spin-networks in the {ZX}-calculus},
	author = {East, Richard D. P. and Martin-Dussaud, Pierre and Van de Wetering, John},
	date = {2021},
	archiveprefix = {arXiv},
  	eprint = {2111.03114}
}

@incollection{wigner1993,
	title = {On the {{Matrices Which Reduce}} the {{Kronecker Products}} of {{Representations}} of {{S}}. {{R}}. {{Groups}}},
	booktitle = {The {{Collected Works}} of {{Eugene Paul Wigner}}: {{Part A}}: {{The Scientific Papers}}},
	author = {Wigner, E. P.},
	editor = {Wightman, Arthur S.},
	year = {1993},
	series = {The {{Collected Works}} of {{Eugene Paul Wigner}}},
	pages = {608--654},
	publisher = {{Springer}},
	address = {{Berlin, Heidelberg}},
	doi = {10.1007/978-3-662-02781-3_42},
	
	isbn = {978-3-662-02781-3},
	langid = {english}
}

@article{roberts1999,
  title = {Classical 6j-Symbols and the Tetrahedron},
  author = {Roberts, Justin},
  year = {1999},
  month = mar,
  doi = {10.2140/gt.1999.3.21},
  archiveprefix = {arXiv},
  eprint = {math-ph/9812013},
  journal = {Geometry \& Topology},
  number = {3},
  pages = {21–66}
}

@article{major1999,
  title = {A {{Spin Network Primer}}},
  author = {Major, Seth A.},
  year = {1999},
  volume = {67},
  pages = {972--980},
  issn = {0002-9505, 1943-2909},
  doi = {10.1119/1.19175},
  archiveprefix = {arXiv},
  eprint = {gr-qc/9905020},
  journal = {American Journal of Physics},
}

@book{feynman1964,
  title = {Lectures on {{Physics}}. {{Electromagnetism}}.},
  author = {Feynman, Richard P. and Leighton, Robert B. and Sands, Matthew},
  year = {1964},
  volume = {2},
  publisher = {{Addison-Wesley Publishing Company}},
}

@article{urbantke2003,
  title = {The {{Hopf}} Fibration--Seven Times in Physics},
  author = {Urbantke, H. K.},
  year = {2003},
  month = may,
  volume = {46},
  pages = {125--150},
  issn = {0393-0440},
  doi = {10.1016/S0393-0440(02)00121-3},
  journal = {Journal of Geometry and Physics},
  number = {2}
}

@article{hazewinkel1982,
  title = {A Short Elementary Proof of {{Grothendieck}}'s Theorem on Algebraic Vectorbundles over the Projective Line},
  author = {Hazewinkel, Michiel and Martin, Clyde F.},
  year = {1982},
  month = aug,
  volume = {25},
  pages = {207--211},
  issn = {0022-4049},
  doi = {10.1016/0022-4049(82)90037-8},
  journal = {Journal of Pure and Applied Algebra},
  number = {2}
}

@phdthesis{leaser2012,
author = {Leaser, Tyler},
school = {East Carolina University},
title = {{Fourier analysis on SU(2)}},
year = {2012}
}

@article{kerimov1978,
author = {Kerimov, G. A. and Verdiev, Yi. A. and Mathematical, O N},
journal = {Reports on Mathematical Physics},
number = {3},
pages = {315--326},
title = {{Clebsch-Gordan coefficients of the SL(2,C) group}},
volume = {13},
year = {1978}
}

@article{engle2008,
archivePrefix = {arXiv},
arxivId = {0711.0146},
author = {Engle, Jonathan and Livine, Etera and Pereira, Roberto and Rovelli, Carlo},
doi = {10.1016/j.nuclphysb.2008.02.018},
eprint = {0711.0146},
isbn = {0550-3213},
issn = {05503213},
journal = {Nuclear Physics B},
number = {1-2},
pages = {136--149},
title = {{LQG vertex with finite Immirzi parameter}},
volume = {B799},
year = {2008}
}

@article{perez2013,
archivePrefix = {arXiv},
arxivId = {1205.2019},
author = {Perez, Alejandro},
doi = {10.12942/lrr-2013-3},
eprint = {1205.2019},
issn = {14338351},
journal = {Living Reviews in Relativity},
pages = {3},
title = {{The Spin Foam Approach to Quantum Gravity}},
volume = {16},
year = {2013}
}

@inproceedings{baez1999,
archivePrefix = {arXiv},
eprint = {gr-qc/9905087},
author = {Baez, J C},
booktitle = {Geometry and quantum physics},
editor = {Gausterer, H and Grosse, H and Pittner, L},
pages = {25--94},
publisher = {Springer},
series = {Lecture Notes in Physics},
title = {{An Introduction to Spin Foam Models of Quantum Gravity and BF Theory}},
volume = {543},
year = {2000}
}

@article{Barrett2010,
author = {Barrett, John W and Dowdall, R J and Fairbairn, Winston J and Hellmann, Frank and Pereira, Roberto},
doi = {10.1088/0264-9381/27/16/165009},
issn = {0264-9381},
journal = {Classical and Quantum Gravity},
number = {16},
pages = {165009},
publisher = {IOP Publishing},
title = {{Lorentzian spin foam amplitudes: graphical calculus and asymptotics}},
volume = {27},
year = {2010},
eprint = {0907.2440},
archivePrefix = {arXiv}
}

@book{varshalovich1987,
author = {Varshalovich, D A and Moskalev, A N and Khersonskii, V K},
publisher = {World Scientific},
title = {{Quantum theory of angular momentum}},
year = {1987}
}

@article{freidel2010a,
archivePrefix = {arXiv},
arxivId = {1006.0199},
author = {Freidel, Laurent and Speziale, Simone},
doi = {10.1103/PhysRevD.82.084041},
eprint = {1006.0199},
issn = {15507998},
journal = {Physical Review D - Particles, Fields, Gravitation and Cosmology},
number = {8},
pages = {84041},
title = {{From twistors to twisted geometries}},
volume = {D82},
year = {2010}
}

@article{ooguri1992,
archivePrefix = {arXiv},
eprint = {hep-th/9205090},
author = {Ooguri, Hirosi},
journal = {Mod. Phys. Lett.},
pages = {2799--2810},
title = {{Topological lattice models in four-dimensions}},
volume = {A7},
year = {1992},
doi = {10.1142/S0217732392004171}
}

@book{thiemann2007,
author = {Thiemann, Thomas},
publisher = {Cambridge University Press},
title = {{Modern Canonical Quantum General Relativity}},
year = {2007}
}

@article{sarno2018,
author = {Sarno, Giorgio and Speziale, Simone and Stagno, Gabriele V.},
doi = {10.1007/s10714-018-2360-x},
issn = {0001-7701},
journal = {General Relativity and Gravitation},
number = {4},
pages = {43},
publisher = {Springer US},
title = {{2-vertex Lorentzian spin foam amplitudes for dipole transitions}},
volume = {50},
year = {2018},
eprint = {1801.03771},
archivePrefix = {arXiv}
}

@article{gelfand1947,
author = {Gel'fand, I. M. and Naimark, M. A.},
journal = {Izv. Akad. Nauk SSSR Ser. Mat.},
number = {5},
pages = {411--504},
title = {{Unitary representations of the Lorentz group}},
volume = {11},
year = {1947}
}

@inproceedings{ponzano1968,
author = {Ponzano, G and Regge, Tullio},
booktitle = {Spectroscopy and group theoretical methods in Physics},
editor = {Bloch, F},
publisher = {North-Holland},
title = {{Semiclassical limit of Racah coefficients}},
year = {1968}
}

@article{oeckl2003,
archivePrefix = {arXiv},
eprint = {hep-th/0306025},
author = {Oeckl, Robert},
journal = {Phys. Lett.},
pages = {318--324},
title = {{A 'general boundary' formulation for quantum mechanics and quantum gravity}},
volume = {B575},
year = {2003},
doi = {10.1016/j.physletb.2003.08.043}
}

@article{langvik2016,
archivePrefix = {arXiv},
arxivId = {1602.01861},
author = {L{\aa}ngvik, Miklos and Speziale, Simone},
doi = {10.1103/PhysRevD.94.024050},
eprint = {1602.01861},
issn = {24700029},
journal = {Physical Review D},
number = {2},
pages = {20},
title = {{Twisted geometries, twistors, and conformal transformations}},
volume = {94},
year = {2016}
}

@inproceedings{dona2010,
archivePrefix = {arXiv},
arxivId = {1007.0402},
author = {Don{\'{a}}, Pietro and Speziale, Simone},
booktitle = {TVC 79. Gravitation : th{\'{e}}orie et exp{\'{e}}rience},
editor = {Bounames, Abdelhafid and Makhlouf, Abnenacer},
eprint = {1007.0402},
issn = {{\textless}null{\textgreater}},
publisher = {Hermann},
title = {{Introductory lectures to loop quantum gravity}},
year = {2010}
}

@article{anderson1970a,
author = {Anderson, R. L. and Raczka, R. and Rashid, M. A. and Winternitz, P.},
doi = {10.1063/1.1665196},
issn = {00222488},
journal = {Journal of Mathematical Physics},
number = {3},
pages = {1050--1058},
title = {{Clebsch-Gordan coefficients for the coupling of SL(2, C) principal-series representations}},
volume = {11},
year = {1970}
}

@article{bargmann1947,
author = {Bargmann, V.},
doi = {10.2307/1969129},
journal = {Annals of Mathematics},
number = {3},
pages = {568--640},
title = {{Irreducible Unitary Representations of the Lorentz Group}},
volume = {48},
year = {1947}
}

@article{livine2011a,
archivePrefix = {arXiv},
arxivId = {1108.0369},
author = {Livine, Etera R. and Speziale, Simone and Tambornino, Johannes},
doi = {10.1103/PhysRevD.85.064002},
eprint = {1108.0369},
journal = {Physical Review D},
number = {6},
pages = {064002},
title = {{Twistor Networks and Covariant Twisted Geometries}},
volume = {85},
year = {2012}
}

@book{hall2015,
author = {Hall, Brian C.},
publisher = {Springer},
title = {{Lie Groups, Lie Algebras, and Representations}},
year = {2015}
}

@book{CLQG,
author = {Rovelli, Carlo and Vidotto, Francesca},
publisher = {Cambridge University Press},
title = {{Covariant Loop Quantum Gravity}},
year = {2014}
}

@article{bonzom2014b,
archivePrefix = {arXiv},
arxivId = {1402.2323},
author = {Bonzom, Valentin and Dupuis, Mait{\'{e}} and Girelli, Florian and Livine, Etera R.},
eprint = {1402.2323},
title = {{Deformed phase space for 3d loop gravity and hyperbolic discrete geometries}},
year = {2014}
}

@book{sakurai2011,
author = {Sakurai, J. J. and Napolitano, Jim},
edition = {2},
publisher = {Addison-Wesley},
title = {{Modern Quantum Mechanics}},
year = {2011}
}

@book{edmonds1957,
author = {Edmonds, A. R.},
booktitle = {The Mathematical Gazette},
publisher = {Princeton University Press},
title = {{Angular Momentum in Quantum Mechanics}},
year = {1957}
}

@book{ruhl1970,
author = {R{\"{u}}hl, W},
publisher = {W. A. Benjamin, Inc},
title = {{The Lorentz Group and Harmonic Analysis}},
year = {1970}
}

@book{Penrose1984,
author = {Penrose, Roger and Rindler, Wolfgang},
publisher = {Cambridge University Press},
title = {{Spinors and Space-time, Volume 1}},
year = {1984}
}

@article{Thiemann:2000bx,
archivePrefix = {arXiv},
eprint = {hep-th/0005234},
author = {Thiemann, Thomas and Winkler, Oliver},
journal = {Class. Quant. Grav.},
pages = {4629--4682},
title = {{Gauge field theory coherent states (GCS) III: Ehrenfest theorems}},
volume = {18},
year = {2001}
}

@book{maurin1997,
author = {Maurin, Krzysztof},
title = {{The Riemann Legacy}},
year = {1997},
publisher = {Kluwer Academic Publishers}
}

@article{Atiyah1974,
author = {Atiyah, M. F.},
journal = {Bull. IMA},
pages = {232--234},
title = {{How research is carried out}},
volume = {10},
year = {1974}
}

@article{Thiemann:2000by,
archivePrefix = {arXiv},
eprint = {hep-th/0005235},
author = {Thiemann, Thomas and Winkler, Oliver},
journal = {Class. Quant. Grav.},
pages = {4997--5054},
title = {{Gauge field theory coherent states (GCS). IV: Infinite tensor product and thermodynamical limit}},
volume = {18},
year = {2001}
}

@article{Dao1967,
author = {Dao, Vong Duc and NGuyen, Van Hieu},
journal = {Ann. Inst. Henri Poincar{\'{e}}},
number = {1},
pages = {17--37},
title = {{On the theory of unitary representations of the SL(2,C) group}},
volume = {VI},
year = {1967}
}

@article{Thiemann:2000bw,
archivePrefix = {arXiv},
eprint = {hep-th/0005233},
author = {Thiemann, Thomas},
doi = {10.1088/0264-9381/18/11/304},
issn = {02649381},
journal = {Classical and Quantum Gravity},
number = {11},
pages = {2025--2064},
title = {{Gauge field theory coherent states (GCS): I. General properties}},
volume = {18},
year = {2001}
}

@article{naimark1954,
author = {Naimark, M. A.},
journal = {Dokl. Akad. Nauk SSSR},
pages = {969--972},
title = {{On linear representations of the proper Lorentz group}},
volume = {97},
year = {1954}
}

@book{quantum_gravity,
author = {Rovelli, Carlo},
publisher = {Cambridge University Press},
title = {{Quantum Gravity}},
year = {2004}
}

@book{knapp1986,
author = {Knapp, Anthony W.},
publisher = {Princeton University Press},
title = {{Representation Theory of Semisimple Groups}},
year = {1986}
}

@article{harish-chandra1947,
author = {Harish-Chandra},
doi = {10.1098/rspa.1947.0047},
issn = {1364-5021},
journal = {Proceedings of the Royal Society A},
number = {1018},
pages = {372--401},
title = {{Infinite Irreducible Representations of the Lorentz Group}},
volume = {189},
year = {1947}
}

@article{anderson1970,
author = {Anderson, R. L. and Raczka, R. and Rashid, M. A. and Winternitz, P.},
doi = {10.1063/1.1665197},
issn = {00222488},
journal = {Journal of Mathematical Physics},
number = {3},
pages = {1059--1068},
title = {{Recursion and symmetry relations for the Clebsch-Gordan coefficients of the homogeneous Lorentz group}},
volume = {11},
year = {1970}
}

@book{yutsis1962,
author = {Yutsis, A. P. and Levinson, I. B. and Vanagas, V. V.},
publisher = {Israel Program for Scientific Translations},
title = {{Mathematical Apparatus of the Theory of Angular Momentum}},
year = {1962}
}

@article{rashid2003,
author = {Rashid, M. A.},
doi = {10.1063/1.524211},
issn = {0022-2488},
journal = {Journal of Mathematical Physics},
number = {7},
pages = {1514--1519},
title = {{Boost matrix elements of the homogeneous Lorentz group}},
volume = {20},
year = {2003}
}

@article{speziale2017,
archivePrefix = {arXiv},
arxivId = {1609.01632},
author = {Speziale, Simone},
doi = {10.1063/1.4977752},
eprint = {1609.01632},
journal = {Journal of Mathematical Physics},
pages = {032501},
title = {{Boosting Wigner's nj-symbols}},
volume = {58},
year = {2017}
}

@book{naimark1964,
author = {Naimark, M A},
publisher = {Pergamon Press},
title = {{Linear Representations of the Lorentz Group}},
year = {1964}
}

@article{Thiemann:2000aa,
archivePrefix = {arXiv},
eprint = {hep-th/0005233},
author = {Thiemann, Thomas and Winkler, Oliver},
journal = {Class. Quant. Grav.},
pages = {2561--2636},
title = {{Gauge field theory coherent states (GCS). II: Peakedness properties}},
volume = {18},
year = {2001}
}

@phdthesis{moussouris1983,
author = {Moussouris, John P.},
school = {Oxford},
title = {{Quantum Models of Space-Time based on Recoupling Theory}},
year = {1983}
}

@article{itzkowitz1991,
author = {Itzkowitz, Gerald and Rothman, Sheldon and Strassberg, Helen},
doi = {10.1016/0022-4049(91)90023-U},
issn = {00224049},
journal = {Journal of Pure and Applied Algebra},
number = {3},
pages = {285--294},
title = {{A note on the real representations of SU(2,C)}},
volume = {69},
year = {1991}
}

@book{gelfand1963,
author = {Gel'fand, I. M. and Minlos, R. A. and Shapiro, Z. Ya.},
publisher = {Pergamon Press},
title = {{Representations of the rotation and Lorentz groups and their applications}},
year = {1963}
}

@book{condon1959,
author = {Condon, E. U. and Shortley, G. H.},
publisher = {Cambridge at the University Press},
title = {{The Theory of Atomic Spectra}},
year = {1959}
}

@book{Penrose1986,
author = {Penrose, Roger and Rindler, Wolfgang},
publisher = {Cambridge University Press},
title = {{Spinors and Space-time, Volume 2}},
year = {1986}
}

@book{gelfand1966,
author = {Gel'fand, I. M. and Graev, M. I. and Vilenkin, N. Ya.},
isbn = {0-12-279505-9},
publisher = {Academic Press},
title = {{Generalized Functions: Volume 5, Integral Geometry and Representation Theory}},
year = {1966}
}

@article{oeckl2008,
archivePrefix = {arXiv},
eprint = {hep-th/0509122},
author = {Oeckl, Robert},
journal = {Adv. Theor. Math. Phys.},
pages = {319--352},
title = {{General boundary quantum field theory: Foundations and probability interpretation}},
volume = {12},
year = {2008},
doi = {10.4310/ATMP.2008.v12.n2.a3}
}

@book{bernard2012,
author = {Bernard, D and Laszlo, Y and Renard, D},
publisher = {Cours de l'{\'{E}}cole polytechnique},
title = {{{\'{E}}l{\'{e}}ments de th{\'{e}}orie des groupes et sym{\'{e}}tries quantiques}},
year = {2012},
url = {www.phys.ens.fr/~dbernard/Publications/PolyGroupSym2012.pdf}
}

\cleardoublepage

\pagestyle{empty}

\hfill

\vfill

\pdfbookmark[0]{Colophon}{colophon}

\section*{Colophon}
\label{colophon}

This document was typeset using the typographical look-and-feel \texttt{classicthesis} developed by Andr\'e Miede, released under the \textsmaller{GNU} General Public License as published by the Free Software Foundation, and available for both \LaTeX\ and \mLyX: 

\begin{center}
\url{https://bitbucket.org/amiede/classicthesis/}.
\end{center}

\noindent Photo of the front cover by Matteo Modica on Unsplash.
 
\bigskip

\noindent\emph{Final Version} as of \today. 

\end{document}